\documentclass[10pt,twocolumn,letterpaper]{article}
\usepackage[utf8]{inputenc}
\usepackage[T1]{fontenc}
\usepackage{lmodern}
\usepackage[margin=0.75in,columnsep=0.25in]{geometry}
\usepackage{microtype}
\usepackage{graphicx}
\usepackage{booktabs}
\usepackage{algorithm}
\usepackage[noend]{algpseudocode}
\usepackage[numbers,sort&compress]{natbib}
\usepackage{authblk}
\usepackage{url}
\usepackage[font=small,labelfont=bf,skip=5pt]{caption}
\usepackage{amsmath,amssymb,amsthm,mathtools}
\usepackage{bm, verbatim}
\usepackage{multirow}
\usepackage{xcolor}

\DeclareMathOperator{\E}{\mathbb{E}}
\DeclareMathOperator{\Var}{Var}
\DeclareMathOperator{\Cov}{Cov}

\newcommand{\dd}{\,\mathrm{d}}

\newcommand{\indep}{\perp\!\!\!\perp}

\newcommand{\Prob}{\mathbb{P}}
\newcommand{\R}{\mathbb{R}}
\DeclareMathOperator{\rank}{rank}

\newcommand{\Fcal}{\mathcal{F}}

\newcommand{\bb}{\mathbf{b}}

\newcommand{\btheta}{\boldsymbol{\theta}}

\newcommand{\dW}{\dd W_t}

\newcommand{\norm}[1]{\lVert #1 \rVert}

\newcommand{\convp}{\xrightarrow{\,\mathbb{P}\,}}
\newcommand{\convd}{\xrightarrow{\,d\,}}

\theoremstyle{plain}
\newtheorem{theorem}{Theorem}
\newtheorem{proposition}{Proposition}

\theoremstyle{definition}
\newtheorem{definition}{Definition}

\theoremstyle{remark}
\newtheorem{remark}{Remark}

\newcommand{\dt}{\dd t}
\newcommand{\transpose}{^{\!\top}}
\newcommand{\Filt}{\Fcal}

\newcommand{\Amat}{\mathbf{A}}
\newcommand{\Zmat}{\mathbf{Z}}
\newcommand{\bvec}{\bb}
\newcommand{\Wmat}{\widehat{\mathbf{W}}}
\newcommand{\Pmat}{\mathbf{P}}

\newcommand{\IV}{\textsc{iv}}
\newcommand{\GMM}{\textsc{gmm}}
\newcommand{\OLS}{\textsc{ols}}
\newcommand{\TSLS}{2\textsc{sls}}

\newcommand{\param}{\btheta}
\newcommand{\thetahat}{\hat{\btheta}}

\newcommand{\toprob}{\convp}
\newcommand{\todist}{\convd}

\makeatletter
\long\def\textcolor#1#2{{\color{#1}#2}}
\makeatother

\definecolor{accentgreen}{RGB}{0,110,60}

\definecolor{editblue}{rgb}{0.00,0.20,0.75}
\definecolor{editgold}{rgb}{0.55,0.45,0.00}
\definecolor{editdred}{rgb}{0.60,0.00,0.00}
\newcommand{\old}[1]{{\footnotesize\textcolor{editgold}{[old: #1]}}}
\newcommand{\reply}[1]{\textcolor{editdred}{\textbf{[Reply: #1]}}}

\usepackage[normalem]{ulem}

\newcommand{\del}[1]{}
\newcommand{\delblock}[1]{}

\newcommand{\hr}[1]{}

\newif\ifcleanmode
\cleanmodefalse

\usepackage[colorlinks=true,allcolors=blue]{hyperref}

\long\def\textcolor#1#2{#2}
\renewcommand{\hr}[1]{}

\renewcommand{\reply}[1]{}

\renewcommand{\old}[1]{}
\newcommand{\dropcap}[1]{#1}
\renewcommand{\eqref}[1]{Eq.~(\ref{#1})}

\makeatletter
\renewcommand{\@seccntformat}[1]{\csname the#1\endcsname.\quad}
\makeatother

\title{\textsc{WAIVES}: Weak-form Asymptotically-unbiased Instrumental-Variable Estimation for SDEs}
\author[1,*]{Hamza Ruzayqat}
\author[1]{David M.\ Bortz}
\author[1]{Vanja Duki\'c}
\affil[1]{Department of Applied Mathematics, University of Colorado Boulder, Boulder, CO 80309, USA}
\affil[*]{Corresponding author. E-mail: hamza.ruzayqat@colorado.edu}
\date{\today}

\begin{document}

\twocolumn[
\begin{@twocolumnfalse}
\maketitle
\begin{abstract}
Weak-form methods have transformed inference for deterministic dynamical
systems. By integrating the governing equations against smooth,
compactly supported test functions, they trade noise-amplifying
differentiation for noise-attenuating integration, yielding fast and robust
estimates. Extending this idea to stochastic differential equations (SDEs) has proven harder, and
a consistent single-trajectory formulation in the time domain has remained out of
reach. Applied naively in
the time domain, the weak form makes the regressors and the errors
functionals of the same noise, creating \emph{endogeneity} that leaves
weak-form least squares biased and inconsistent. We resolve it with
instrumental variables, a classical idea from econometrics. Weak-form
integrals over nearby disjoint time windows prove to be valid instruments,
correlated with the regressors yet uncorrelated with the error, restoring both
consistency and asymptotic normality. The resulting
generalized-method-of-moments estimator, which we term \textsc{WAIVES}
(Weak-form Asymptotically-unbiased Instrumental-Variable Estimation for
SDEs), requires only a linear solve, with no likelihood evaluations,
filtering, or simulation. It is therefore an order of magnitude
faster than exact-likelihood methods where those exist, and it applies
unchanged where they do not, with standard errors and calibrated confidence
intervals supplied directly by the moment conditions. It also handles
measurement noise, separating it from the intrinsic randomness of the
dynamics. Across nine example systems, spanning linear, bistable, chaotic,
excitable, state-dependent-diffusion, non-polynomial, hypoelliptic, and
high-dimensional regimes (up to 120 dimensions), \textsc{WAIVES} recovers
nonlinear drift and diffusion with negligible bias and near-nominal
interval coverage. We close with two real-data illustrations: a 167-year North Atlantic
sea-surface-temperature record and an optical-tweezer calibration.
\end{abstract}
\begin{center}
\begin{minipage}{0.9\textwidth}
\end{minipage}
\end{center}
\vspace{1.2em}
\end{@twocolumnfalse}
]
\begingroup\renewcommand{\thefootnote}{}\footnotetext{\textbf{Author contributions.} H.R.\ conceived the application of instrumental variables to weak-form SDE estimation. H.R., D.M.B., and V.D.\ developed the theory. H.R.\ and V.D.\ performed
the numerical experiments. H.R., D.M.B., and V.D.\ wrote the paper. The simulation,
generation of tables, and figure code, and manuscript editing, were
done with the assistance of large language models, as described in the
Acknowledgments; all results were verified by the authors.}\endgroup

\section{Introduction}
\dropcap{D}ata-driven inference of governing equations, from equation discovery with SINDy~\cite{Brunton2016} and WSINDy~\cite{MessengerBortz2021} to parameter estimation, has become a central
tool across the quantitative sciences~\cite{SchmidtLipson2009, Brunton2016,
RudyBruntonProctorKutz2017, BruntonKutz2022}. For deterministic
ordinary differential equations (ODEs), weak-form estimation has proven
especially powerful~\cite{MessengerBortz2021, MessengerBortz2021PDE,
Reinbold2020, BortzMessengerDukic2023}. Rather than estimating derivatives
pointwise from noisy data, the weak-form approach relies on multiplying
the governing equation by a smooth, compactly supported test function and
integrating by parts, thus converting differentiation of the data into
differentiation of the (known) smooth test function~\cite{MessengerBortz2021}.
The weak-form estimation of nonlinear dynamics (WENDy)
framework~\cite{BortzMessengerDukic2023, Rummel2025} exploits this to recover model
parameters by linear regression, resulting in inference that is ultra-fast and robust to additive and multiplicative measurement noise.

The extension of the WENDy framework to stochastic differential equations
(SDEs) has, to our knowledge, not previously been attempted. Many systems of scientific interest are not merely observed with noise but driven by it. Ion
channels in neurons, climate oscillators, ecological populations near
extinction, and molecular reaction networks are among the examples in which
randomness enters into the dynamics themselves, and not just into the
measurement process. For such systems, SDEs and stochastic partial differential equations (SPDEs) are more faithful models than ODEs and PDEs. Recovering the drift and diffusion of an SDE from sampled trajectories is the stochastic analogue of WENDy's goal and the objective of this paper. SPDEs are left for future work.

Inference for discretely observed diffusions is difficult because the transition density of the SDE is almost never available in closed form, so the exact likelihood of a discretely sampled path cannot be evaluated directly (see, e.g., \cite{Sorensen2004} and the references therein). A substantial body of work addresses this challenge. Some approaches use simulated-likelihood or data-augmentation schemes to account for the unobserved portions of the trajectory between observations~\cite{Pedersen1995, Roberts2001}. These methods, however, can be computationally demanding, as they require repeated simulation or sampling of the latent paths, and data-augmentation schemes can suffer from poor MCMC mixing due to strong dependence between the latent paths and the diffusion parameters. Other approaches construct closed-form approximations to the transition density~\cite{AitSahalia2002}. Although these expansions can be highly accurate, practical implementations rely on finite truncations, whose accuracy and computational cost depend on the order of the expansion and the complexity of the underlying model. Finally, exact-simulation methods eliminate discretization error entirely~\cite{BPRF2006, Beskos2009}. However, they typically require first transforming the SDE to one with unit diffusion coefficient (the Lamperti transform). Because this transformation generally exists only for scalar diffusions, and for a restricted class of multivariate ones, these methods do not extend to general multivariate systems. Despite their different approaches, these methods are generally developed for settings in which the state is observed exactly at discrete times, without additional measurement noise.

When the state is observed with measurement error, the unknown parameters
must be estimated within a state-space formulation that couples them to a
nonlinear filtering problem. This is handled
through sequential Monte Carlo~\cite{Golightly2008, Kantas2015, Beskos2021}, particle
MCMC~\cite{Andrieu2010}, iterated filtering~\cite{Ionides2006}, and
ensemble Kalman--Bucy schemes~\cite{ Crisan2022}, jointly
inferring the drift and diffusion parameters together with the latent path
and yielding consistent, and in some cases unbiased, estimators even in
high-dimensional or partially observed regimes. However, the performance varies with
the method and the model class; for example, ensemble Kalman--Bucy schemes are exact only for linear Gaussian dynamics and become
approximations without convergence guarantees once the drift or diffusion
is nonlinear. A parallel line of work instead parameterizes the
drift and diffusion by neural networks~\cite{Tzen2019, Li2020, Kidger2021};
such models are flexible and accommodate irregular sampling, but they
require large training sets and are not directly interpretable as governing
equations. In all but trivial cases, these methods are also highly
computationally intensive.

Within the governing equation discovery setting, sparse-regression and related methods have been adapted to learn the form of stochastic dynamics (and consequently the associated parameters), typically by fitting drift and diffusion fields to empirical estimates of conditional moments or to Fokker--Planck statistics~\cite{Boninsegna2018, Callaham2021, Wanner2024}. Weak-form ideas have recently entered this setting as well, for mean-field equations estimated from interacting-particle data~\cite{MessengerBortz2022} and for mean-field Fokker--Planck equations learned from sparse ecological experiments~\cite{Minor2026}. A weak-collocation approach integrates the Fokker--Planck equation against Gaussian kernels placed in state space and evaluates the resulting integrals as sums over observed samples. This has been done for aggregate snapshot data~\cite{lu2024wcr} and, through ergodic averaging, for a single stationary trajectory split into segments~\cite{jiang2026stwcr}. In all of these weak-form methods, the weak form acts on the density of the process, so the stochastic-integral term is removed by the averaging over samples before any regression is performed. In exchange, the temporal structure of the increments is not used. The strong-form approaches, in turn, rely on finite differencing of the sampled path, which amplifies measurement noise and makes parameter inference more challenging. Many of these approaches are also based on empirical estimates of conditional moments or densities, obtained from single-step increments, by binning, or by smoothing. Binning generally discards the within-bin dynamics, and the recovered parameters show bias that grows with the noise level and with coarser sampling~\cite{Ragwitz2001, Boettcher2006, Friedrich2011}. In contrast, \textsc{WAIVES} retains the weak form in the time variable, acting on a single trajectory. This preserves both the temporal information in the increments and the computational efficiency and speed of compactly supported test functions, the same properties behind the weak form’s success for deterministic systems. However, this formulation introduces a difficulty of its own, which we address next.

Applied naively to SDEs, weak-form regression in time fails. When the governing SDE is integrated against a test function in time, the resulting regression carries a stochastic-integral error that is correlated with its own regressors, because both are functionals of the same Brownian path over the test-function support. This is the \emph{endogeneity} problem of classical regression~\cite{Sargan1958}, and it makes the weak-form least-squares estimator biased and inconsistent, regardless of how many test functions are used or how long the trajectory grows. In fact, as we show in Section~\ref{sec:endogeneity}, for the Ornstein--Uhlenbeck process the least-squares estimate converges to zero regardless of the record length.

In this work, we propose to address the endogeneity issue directly, using
\emph{instrumental variables} (IV)~\cite{Sargan1958,Hansen1982}. IVs
are not new to system identification, where they have long been used to
obtain consistent estimates of dynamical models from noisy
data~\cite{Young1970, Soderstrom1983, Soderstrom1989}, with recent extensions to nonlinear continuous-time models through auxiliary-model instruments~\cite{Zeiringer2026} or sample-splitting of the record~\cite{KuangLin2026}. Both of these, however, address deterministic output-error models, where the only disturbance is measurement noise, and thus do not carry over to SDEs. Once the driving noise enters the dynamics itself, no noise-free auxiliary trajectory and no subsample independent of the Brownian path exist. Instrumental variable ideas have also appeared against measurement noise alone. For example, in strong-form stochastic SINDy, whose single-step regression is consistent in the absence of measurement noise, the drift term candidates evaluated one time step earlier can serve as instruments against that noise~\cite{Wanner2024}. That construction, however, does not address the endogeneity of the weak form, which is present even for perfectly observed paths, as we show in the next section. The instruments we construct below do.

In principle, an instrument can be any quantity that is correlated with the regressor, but
uncorrelated with the error. The difficulty lies in finding an
instrument for which both properties hold strongly. We divide the record into short sub-intervals (windows) and construct instruments from weak-form integrals over nearby windows disjoint from the target window on which the regressors are built. Because Brownian increments
over disjoint intervals are mutually independent, a weak-form integral computed on one window (against a test function supported on that window) is uncorrelated with the stochastic-integral error of a
non-overlapping window, while remaining correlated with the regressors
whenever the two windows lie within the memory of the process. The
resulting estimator, which we call \textsc{WAIVES}
(\textbf{W}eak-form \textbf{A}symptotically-unbiased
\textbf{I}nstrumental-\textbf{V}ariable \textbf{E}stimation for
\textbf{S}DEs), is shown to be consistent, asymptotically unbiased (with
the bias vanishing in the joint limit of many windows and fine sampling), asymptotically normal,
and it requires neither a path likelihood nor a Radon--Nikodym derivative.
\textsc{WAIVES} thus belongs to the moment-based tradition for diffusions, alongside martingale estimating functions~\cite{BibbySorensen1995} and generator-based generalized method of moments (GMM)~\cite{HansenScheinkman1995}, which derive their moment conditions from the model's transition or generator structure and presuppose exactly observed states. \textsc{WAIVES} builds them from the test functions and the noisy observations alone.

Because it
avoids the finite-difference estimates
on which strong-form methods
rely, it inherits the weak-form noise robustness and solver-free
efficiency, while still handling diffusions that are observed only at discrete times,
with noise, and possibly through a linear sensor map.
We demonstrate this on the nine example systems of Table~\ref{tab:systems}, from scalar diffusions to a $120$-dimensional Lorenz-96 model, and on two real data sets, the 167-year Atlantic Multidecadal Oscillation index (a North Atlantic sea-surface-temperature record) and an optical-tweezer calibration record.

\section{The weak form for SDEs}
\label{sec:weakform}

\subsection{Model}
We consider the state-space model in which a latent process obeys an
It\^o SDE and is observed discretely under additive Gaussian error,
\begin{align}
\text{State:}\quad & \dd X_t = f(X_t,\param)\dt + \sigma(X_t {,\bm{\nu}})\dW,
 \label{eq:state}\\
\text{Obs.:}\quad & Y_{t_n} = X_{t_n} + \eta_n,
 \qquad \eta_n \overset{\text{iid}}{\sim} \mathcal{N}(0,R),
 \label{eq:obs}
\end{align}
where $\{W_t\}$ is a standard Brownian motion in $\R^{d}$ and the state
$X_t\in\R^{d}$. The observations are taken at the discrete instants
$t_n=n\Delta t$ for $n=0,\dots,N$, so that the record spans $[0,T]$ with $T=N\Delta t$. The observation errors
$\eta_n\in\R^{d_y}$ have covariance $R\in\R^{d_y\times d_y}$, which is assumed
unknown.
Throughout we assume $d_y=d$, but Remark~\ref{rem:obsC} treats the more general linear observation model case $Y_{t_n}= C~X_{t_n}+\eta_n$ with
$d_y\geq d$.
The errors $\{\eta_n\}$ are assumed to be independent of
$\{W_t\}$, and the drift is assumed to be linear in the
unknown parameters:
\begin{equation}
f(x,\param) = \sum_{j=1}^{p}\theta_j\, g_j(x),
\label{eq:drift-lip}
\end{equation}
with known drift term candidates $\{g_j\}_{j=1}^p$ such as
monomials or trigonometric functions, and the unknown parameter vector
$\param=(\theta_1,\dots,\theta_p)^{\!\top}\in\R^p$.

The diffusion is parameterized in the same
way. We take $\sigma=\sigma(X_t,\bm{\nu})$ and, under the diagonal-noise
assumption used throughout (each state component driven by its own independent
Brownian motion), expand the squared diffusion of component $i$ in known diffusion term candidates $\{h_l\}_{l=1}^q$,
\begin{equation}
\sigma_i^2(x,\bm{\nu}) = \sum_{l=1}^{q}\nu_{il}\,h_l(x),
\label{eq:diff-lib}
\end{equation}
where $\bm{\nu}_i=(\nu_{i1},\dots,\nu_{iq})^{\!\top}\in\R^{q}$ collects the unknown diffusion parameters of component $i$ and $\bm{\nu}$ denotes their
collection, so that learning the SDE means recovering both $\param$ (drift)
and $\bm{\nu}$ (diffusion). The general non-diagonal case estimates the diffusion
matrix $D=\sigma\sigma^{\!\top}$; we use the diagonal form for clarity. The estimation of $\bm\nu$ is developed in
Section~\ref{sec:iv}.

\subsection{The deterministic weak form}
For the noise-free ODE $\dot x = f(x,\param)$, multiplying by a smooth,
compactly supported test function of time $\phi_k$ with
$\operatorname{supp}\phi_k\subset(a_k,b_k)$ and integrating by parts moves the
derivative onto $\phi_k$. We call the support interval $(a_k,b_k)$, a sub-interval of the record $[0,T]$, the
\emph{window} of the test function.
Because $\phi_k$ vanishes at the window
endpoints, the boundary terms drop and one obtains a linear system
$\bvec = \Amat\param$ with
\begin{equation}
\mathsf b_k = -\!\int_{a_k}^{b_k}\! x(t)\,\phi_k'(t)\dt,
\qquad
A_{kj} = \int_{a_k}^{b_k}\! g_j(x(t))\,\phi_k(t)\dt.
\label{eq:ode-weakform}
\end{equation}
Because $\phi_k$ and $\phi_k'$ vanish outside the window, the same integrals may equally be taken over the whole record $[0,T]$. Stacking $K$
such test functions, each supported on its own window, yields the WENDy
regression~\cite{BortzMessengerDukic2023}. Its integrals are computed from the observations $Y_{t_n}$ by quadrature, so their only error is the measurement noise $\eta_n$, which is independent of the latent path $x(t)$. WENDy accounts for the error
covariance this induces, treating $\bvec=\Amat\param+\varepsilon$ as an
errors-in-variables regression solved by iteratively reweighted generalized
least squares. As the noise is exogenous to the dynamics, this yields a consistent parameter estimator for ODE data with
additive measurement noise.

\subsection{The stochastic weak form}
Here and below we abbreviate
$\sigma(X_t)=\sigma(X_t,\bm{\nu})$. Integrating the SDE~\eqref{eq:state} against $\phi_k$ over its support,
applying semimartingale integration by parts to $\int\phi_k\dd X_t$, and
using the vanishing of the boundary terms, we obtain
\begin{equation}
\begin{split}
\underbrace{-\!\int_{a_k}^{b_k}\! X_t\,\phi_k'(t)\dt}_{\mathsf b_k}
={}&
\sum_{j=1}^p \theta_j
\underbrace{\int_{a_k}^{b_k}\! \phi_k(t)\,g_j(X_t)\dt}_{A_{kj}}\\
&{}+
\underbrace{\int_{a_k}^{b_k}\! \phi_k(t)\,\sigma(X_t)\dW}_{\varepsilon_k}.
\end{split}
\label{eq:sde-weakform}
\end{equation}
This is the ODE weak form~\eqref{eq:ode-weakform} with one new term, the residual $\varepsilon_k=\int_{a_k}^{b_k}\phi_k\sigma(X_t)\dW$, a stochastic integral, hence a martingale increment, with mean zero
$\E[\varepsilon_k]=0$. with mean zero. As it depends on the Brownian path, this residual is
correlated with the regressors, the design-matrix entries $A_{kj}$ of
\eqref{eq:sde-weakform}. The correlation arises because, within the
window, the path is built from the same Brownian increments that enter the
residual. For $a_k\leq s<t \leq b_k$, the state $X_t$ depends on the increment $\dd W_s$, so
$\E[g_j(X_t)\,\dd W_s]$ does not in general vanish. As a result, these contributions may accumulate
over the entire window, and result in a nonzero covariance between $A_{kj}$ and
$\varepsilon_k$.

\section{Inconsistency and bias of ordinary least squares}
\label{sec:endogeneity}

We now quantify this covariance. The regressor $A_{kj}$ and the residual $\varepsilon_k$
in~\eqref{eq:sde-weakform} are both functionals of the same Brownian path
$\{W_s:s\in[a,b]\}$ over the window, so their product does not in general
average to zero:
\begin{equation}
\E[A_{kj}\varepsilon_k]
= \E\Bigl[\int_{a_k}^{b_k}\!\phi_k g_j(X_t)\dt
 \int_{a_k}^{b_k}\!\phi_k\sigma(X_t)\dW\Bigr]\neq 0.
\label{eq:endogeneity}
\end{equation}
This covariance between regressors and error, termed
\emph{endogeneity}~\cite{Sargan1958}, violates the orthogonality condition
$\E[\Amat\transpose\varepsilon]=0$ and renders the ordinary least squares (OLS) estimator
$\thetahat_{\OLS}=(\Amat\transpose\Amat)^{-1}\Amat\transpose\bvec$ biased
and inconsistent~\cite{Wooldridge2010}.
The mechanism is most transparent for the stationary Ornstein--Uhlenbeck (OU)
process $\dd X_t=-\kappa X_t\dt+\sigma\dW$, $\kappa>0$, whose state admits
the moving-average representation
$X_t=\sigma\int_{-\infty}^{t}e^{-\kappa(t-s)}\dd W_s$~\cite{Pavliotis2014}.
With the drift term candidate $g(x)=x$, the regressor $A=\int_a^b\phi(t)X_t\dt$
contains every increment $\dd W_s$ with $s\in(a,b)$, each of which also
enters the residual $\varepsilon=\sigma\int_a^b\phi\,\dd W_t$, so that
$\E[A\varepsilon]
=\sigma^2\iint_{a<s<t<b}\!\phi(t)\phi(s)e^{-\kappa(t-s)}\dd s\dd t>0$.
Positivity is immediate for the nonnegative bump test functions used here,
but holds for any nonzero test function (Appendix~\ref{si:validity}).

On each window, the weak-form response is
$\mathsf b=\int_a^b\phi(t)\,\dd X_t=-\int_a^b\phi'(t)X_t\dt$, and
substituting the dynamics gives $\mathsf b=-\kappa A+\varepsilon$. Each window contributes one row $(A_k,\mathsf b_k,\varepsilon_k)$ of ~\eqref{eq:sde-weakform}; with the window span held fixed, a longer record adds more windows. As the
number of windows grows, the least-squares estimator converges in probability to the
population regression coefficient, obtained by replacing sample averages
with expectations:
\[
\theta^{\mathrm{pop}}_{\OLS}
:=\operatorname*{arg\,min}_{\theta\in\R}\;
\E\bigl[(\mathsf b-\theta A)^{2}\bigr]
=\frac{\E[A\,\mathsf b]}{\E[A^{2}]}
=-\kappa+\frac{\E[A\varepsilon]}{\E[A^{2}]}.
\]
The positive covariance $\E[A\varepsilon]$ thus attenuates the estimated
drift toward zero, away from its true value $-\kappa$. For OU, which is time-reversible, the attenuation is total. As we will see in Proposition~\ref{prop:ols-zero} below, $\E[A\varepsilon]=\kappa\,\E[A^{2}]$, so the population least-squares coefficient is exactly zero.

The same decomposition holds for a general set of drift term candidates. Writing
$\mathbf a_k\transpose$ for the $k$-th row of $\Amat$, the pairs
$(\mathbf a_k,\varepsilon_k)$ are identically distributed across windows
under stationarity, and the least-squares estimator converges to
$\param^{\mathrm{pop}}_{\OLS}
=\param+\E[\mathbf a\mathbf a\transpose]^{-1}\,
\E[\mathbf a\,\varepsilon]$,
provided $\E[\mathbf a\mathbf a\transpose]$ is
invertible~\cite{Wooldridge2010}. The bias is a fixed vector of per-window
expectations, independent of the number of windows, so it does not diminish asymptotically.

\begin{proposition}
\label{prop:ols-zero}
Let $X$ be a stationary It\^o diffusion
$\dd X_t=f(X_t)\dt+\sigma(X_t)\dW$ in $\R^d$ that is time-reversible,
that is, $(X_0,X_u)$ and $(X_u,X_0)$ have the same law for every $u$.
Let $g_1,\dots,g_p$ be drift term candidates, and assume $\E\|X_0\|^2$,
$\E\|f(X_0)\|^2$, $\E\|\sigma(X_0)\|^2$, and $\E[g_j(X_0)^2]$ are finite,
and that each cross-covariance
$C_{ji}(u)=\Cov\bigl(g_j(X_u),X^{(i)}_0\bigr)$ is continuously
differentiable for $u\neq0$ with bounded derivative. Let $\phi$ be any
bounded test function supported on $[a,b]$, and let
$a_j=\int_a^b\phi(t)g_j(X_t)\dt$ and
$\mathsf b_i=\int_a^b\phi(t)\,\dd X^{(i)}_t$. Then $
\Cov(a_j,\mathsf b_i)=0$ for all $j$ and $i$,
and, whenever the population Gram matrix of the drift term candidates is invertible,
the population regression coefficient of $\mathsf b_i$ on the drift term candidates,
which is the probability limit of the least-squares estimator as the
number of windows grows, is the zero vector, for every window and every test function.
\end{proposition}

The full proof is given in Appendix~\ref{si:validity}. Briefly, off the diagonal, stationarity gives
$\E[g_j(X_t)\,\dd X^{(i)}_s]=-C_{ji}'(t-s)\,\dd s$, and reversibility
makes $C_{ji}$ an even function of the lag, implying that the integrand
$\phi(t)\phi(s)\,C_{ji}'(t-s)$ is odd
and integrates to zero over the symmetric window square.
Also, note that every stationary ergodic scalar diffusion
is reversible, so for scalar systems (like OU and double-well) the entire drift is always erased.
Section~\ref{sec:amo} demonstrates this drift cancellation also on a real climate
record.
\section{Instrumental variables from disjoint windows}
\label{sec:iv}

\begin{figure}[!t]
\centering
\includegraphics[width=\linewidth]{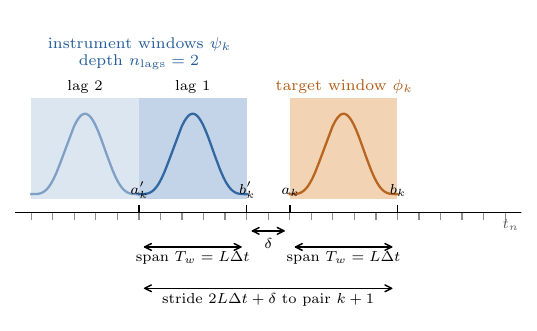}
\caption{Window layout for one moment pair. Each target window
$[a_k,a_k+T_w]$ of span $T_w=L\Delta t$ carries a test function $\phi_k$, and
its instrument is built from the $n_{\text{lags}}$ preceding blocks of the
same length, carrying $\psi_k$ and separated by a gap
$\delta$. These window pairs tile the record with stride $2L\Delta t+\delta$. The sketch shows two instrument windows (depth two), although the default used throughout is depth one.}
\label{fig:windows}
\end{figure}

The inconsistency of least squares discussed above stems from structural endogeneity, and consequently, the failure
of the orthogonality condition $\E[\Amat\transpose\bm{\varepsilon}]=0$. A
possible remedy is to impose orthogonality with respect to an instrumental variable instead. Writing the full weak-form system as
$\bvec=\Amat\param+\bm{\varepsilon}$, an instrument is a matrix
$\Zmat$ with one row per moment condition, like $\mathbf A$, and $p$ columns,
satisfying:
\begin{description}
\item[(IV1) Relevance.] $\E[\Zmat\transpose\Amat]$ has full column rank
 (the instrument is correlated with the regressors). When
 $\Zmat$ has as many columns as $\Amat$, full column rank makes
 $\E[\Zmat\transpose\Amat]$ square and invertible, and the estimator
 below is well defined.
\item[(IV2) Exogeneity.] $\E[\Zmat\transpose\bm{\varepsilon}]=0$
 (the instrument is uncorrelated with the residual).
\end{description}
Given a valid instrument, the IV estimator is defined by the empirical
moment condition $\Zmat\transpose(\bvec-\Amat\param)=\bm 0$, that is,
\begin{equation}
\thetahat_{\IV} = (\Zmat\transpose\Amat)^{-1}\Zmat\transpose\bvec.
\label{eq:iv-estimator}
\end{equation}
We make the condition (IV2) hold by building the instrument from a
test function supported away from the target window.
We tile the observation span into $N_w$ pairs of windows, each pair
holding a target window $[a_k,a_k+T_w]$ preceded by an instrument window
$[a_k',a_k'+T_w]$ with $a_k'+T_w\le a_k$, so distinct windows overlap at most at
the single boundary point $a_k'+T_w=a_k$ (Fig.~\ref{fig:windows}). Evaluating
$M_\phi$ test functions per pair (Remark~\ref{rem:testfns}) then gives
$M_\phi N_w$ stacked moment rows. Asymptotic statements are in the number
of window pairs, which we denote $K=N_w$.

\begin{definition}
\label{def:instrument}
For each window pair $k$, let $\phi_k$ be a target test function supported
on $[a_k,a_k+T_w]$ and let $\psi_k$ be an instrument test function supported on
$[a_k',a_k'+T_w)$.
Define
\begin{equation}
Z_{kj}:= \int_{a_k'}^{a_k'+T_w}\!\psi_k(t)\,g_j(X_t)\dt.
\label{eq:instrument-def}
\end{equation}
\end{definition}
Exogeneity holds because the residual $\varepsilon_k$ is a stochastic
integral of the Brownian increments over $(a_k,a_k+T_w]$, while $Z_{kj}$ is
measurable with respect to the increments up to time $a_k'+T_w\le a_k$. Past
and future increments are independent,
so
$\E[Z_{kj}\varepsilon_k]=0$, giving (IV2). Relevance (IV1) holds
provided the instrument window lies within the system's memory, so that the
state $X$ on $[a_k',a_k'+T_w]$ is informative about $g_j(X)$ on $[a_k,a_k+T_w]$. 

\subsection{Observable quantities}
States are observed via $Y_{t_n} = X_{t_n} + \eta_n$.
Each integral is evaluated as a grid sum of step $\Delta t$ over the
observation instants in its window,
$\mathcal{I}_k = \{\, n : t_n \in [a_k,\, a_k + T_w] \,\}$ for the
target and
$\mathcal{I}_k' = \{\, m : t_m \in [a_k',\, a_k' + T_w] \,\}$ for the
instrument. Because every test function and its derivative vanish at the window
endpoints, this sum coincides with both the left-endpoint Riemann sum
and the trapezoidal rule. Each resulting quantity decomposes into its
exact value plus an observation-noise perturbation:
\begin{align}
\mathsf b_k =& -\Delta t\!\sum_{n\in\mathcal{I}_k}\!\phi_k'(t_n)Y_{t_n}
 \;\approx\; \mathsf b_k^{X}
 \underbrace{-\,\Delta t\!\sum_{n\in\mathcal{I}_k}\!\phi_k'(t_n)\eta_n}_{\delta\mathsf b_k},\label{eq:obs-b}\\
A_{kj} &= \Delta t\!\sum_{n\in\mathcal{I}_k}\!\phi_k(t_n)g_j(Y_{t_n})
 \nonumber\\
 &\;\approx\; A_{kj}^{X}
 \underbrace{+\,\Delta t\!\sum_{n\in\mathcal{I}_k}\!\phi_k(t_n)g_j'(X_{t_n})\eta_n}_{\delta A_{kj}},
 \label{eq:obs-A}\\
Z_{kj} &= \Delta t\!\sum_{m\in\mathcal{I}_k'}\!\psi_k(t_m)g_j(Y_{t_m})
 \nonumber\\
 &\;\approx\; Z_{kj}^{X}
 \underbrace{+\,\Delta t\!\sum_{m\in\mathcal{I}_k'}\!\psi_k(t_m)g_j'(X_{t_m})\eta_m}_{\delta Z_{kj}}.
 \label{eq:obs-Z}
\end{align}
The approximations in \eqref{eq:obs-b}--\eqref{eq:obs-Z} contain two
remainders. The first is the $O(\Delta t)$ quadrature error, which
vanishes as $\Delta t\to0$. The second, present only in $A$ and $Z$ when
the drift term candidate $g_j$ is nonlinear, is the second-order Taylor remainder
of $g_j(X+\eta)$, which has nonzero mean of order $\norm{R}$, proportional to
the curvature $g_j''$ (removed by the debiasing of Section~\ref{sec:algorithm}; the higher-order
terms are absorbed in the remainder $\rho_k$ below). The response $\mathsf b$ is
linear in the data, so $\delta\mathsf b_k$ has
mean zero.
Using the identity $\mathsf b_k^X=A_k^X\param+\varepsilon_k$, the observable
regression is $\mathsf b_k=A_k\param+\varepsilon_k^{\mathrm{tot}}$, with the
observed $A_k$ instead of $A_k^{X}$ on the right, and with total residual
$\varepsilon_k^{\mathrm{tot}}=\varepsilon_k+\delta\mathsf b_k-\delta A_k\param+\rho_k$.
Here $\rho_k$ collects the $O(\Delta t)$ quadrature and $O(\|R\|^{2})$ Taylor
remainders of \eqref{eq:obs-b}--\eqref{eq:obs-Z}. The growth conditions of the
consistency theorem (Theorem~\ref{thm:cons}) ensure $\rho_k$ is asymptotically
negligible.

\subsection{Consistency and asymptotic normality}
The IV estimator~\eqref{eq:iv-estimator} has the exact error decomposition
\begin{equation}
\thetahat_{\IV}-\param
=\Big(\tfrac{1}{K}\Zmat\transpose\Amat\Big)^{-1}
 \Big(\tfrac{1}{K}\Zmat\transpose\bm{\varepsilon}^{\mathrm{tot}}\Big).
\label{eq:iv-error}
\end{equation}
Under (IV1)--(IV2) and $\E[Z_{kj}\varepsilon_k^{\mathrm{tot}}]=0$, the second
factor vanishes in probability and the first converges to an invertible limit, and thus $\thetahat_{\IV}\toprob\param$ as $K\to\infty$ with $\Delta t\to0$. Note that consistency
requires no rate coupling between the two, unlike asymptotic normality,
which requires $\sqrt K\,\Delta t\to0$ (Theorem~\ref{thm:clt}).
Because each instrument is built strictly before its target
window starts, the time-ordered summands $Z_k\varepsilon_k^{\mathrm{tot}}$ form
a martingale-difference sequence (the instruments themselves are dependent
across $k$; see Appendix~\ref{app:proofs}), and the martingale central limit theorem
gives
\begin{equation}
\sqrt{K}\,(\thetahat_{\IV}-\param)\ \todist\
 \mathcal{N}(\bm{0},\,\mathbf{G}),
\end{equation}
with covariance $\mathbf{G}$ determined by $Z_1$, $A_1$, and
$\varepsilon_1^{\mathrm{tot}}$. The precise statements follow, with proofs in Appendix~\ref{app:proofs}.

\begin{proposition}
\label{prop:exog}
Under model \eqref{eq:state}--\eqref{eq:obs} with $\{\eta_n\}\indep\{W_t\}$,
the process-noise residual satisfies $\E[Z_{kj}\,\varepsilon_k]=0$ exactly. The
observation-noise perturbations $\delta\mathsf b_k$ and $\delta A_{kj}$ are
orthogonal to $Z_{kj}$ to first order in the errors, so the total residual
satisfies $\E[Z_{kj}\,\varepsilon_k^{\mathrm{tot}}]=\E[Z_{kj}\,\rho_k]$.
After the debiasing this remainder is
$O(\Delta t)+O(\norm{R}^{2})$, and the rate conditions of the consistency
theorem make it negligible.
\end{proposition}

\begin{proposition}
\label{prop:rel}
Suppose $\{X_t\}$ is stationary and $\beta$-mixing with mixing coefficient
$\beta(u)\to0$, $u$ denoting the time separation, where
$\beta(u)=\E\sup_{B\in\sigma(X_s:\,s\ge u)}\bigl|\Prob(B\mid
\sigma(X_s:\,s\le0))-\Prob(B)\bigr|$ measures the dependence between the
past up to time $0$ and the future from time $u$ on, that the active
drift term candidates have finite moments of some order $\nu>2$ under the
stationary law, and that the gap
between instrument and target windows is $\delta\ge0$. Write
$\Gamma_{jj'}(u)=\Cov\bigl(g_j(X_0),g_{j'}(X_u)\bigr)$, and let
$\bm\Gamma(u)=[\Gamma_{jj'}(u)]_{j,j'=1}^{p}$ be the $p\times p$ matrix of these
covariances. Suppose further that the per-pair cross-moment matrix $\mathbf Q$
defined below is nonsingular; a sufficient condition, under which the
proof establishes positive-definiteness, is uniform positivity of the
cross-moment kernel: with $\bar{\bm g}=\E[\bm g(X_0)]$, there is $m>0$
with
$\bm\zeta\transpose\bigl[\bar{\bm g}\bar{\bm g}\transpose+\bm\Gamma(u)\bigr]\bm\zeta
\ge m\norm{\bm\zeta}^2$ for every $\bm\zeta\in\R^p$ and every separation
$\delta\le u\le\delta+2T_w$. (The centered matrix $\bm\Gamma(u)$ alone is
singular whenever the model contains a constant term, as every model
in this paper does, so the condition must be placed on the uncentered
kernel.) Let
$\mathbf Q\in\R^{p\times p}$ be the per-pair cross-moment matrix with
entries $\mathbf Q_{jj'}=\E\bigl[Z_{kj}A_{kj'}\bigr]$, the same for
every pair $k$ by stationarity, so that $\mathbf Q=\E[Z_kA_k\transpose]$.
Then $\mathbf Q$ is positive definite, and
$\E[\Zmat\transpose\Amat]$, a positive multiple of $\mathbf Q$, has
full column rank (IV1).
\end{proposition}

The following results establish consistency and asymptotic normality, and show that plugging in the estimated noise covariance and the estimated GMM weight leaves the first-order theory intact. All limits are in the
number of window pairs $K$, with the conditions on $\Delta t$ and on the
observation noise stated in each result. Throughout, the drift term
candidates are assumed to admit exact Gaussian deconvolution, as the
polynomial and trigonometric terms used in this paper do; general terms
corrected by the regularized Gauss--Hermite rule of
Appendix~\ref{si:numerics} incur an additional $O(\norm{R}^2)$ approximation
error, which must satisfy $\sqrt K\,\norm{R}^2\to0$ for the central
limit theorem to carry over.

\begin{theorem}\label{thm:cons}
Under Propositions~\ref{prop:exog}--\ref{prop:rel}, with $\sum_{u\ge1}\beta(u)^{1-2/\nu}<\infty$ for the moment order $\nu>2$ of Proposition~\ref{prop:rel} (exponential mixing, as in Remark~\ref{rem:mixing}, suffices) and with $K\to\infty$ and
$\Delta t\to0$ such that $K\Delta t^2\to0$ (equivalently $\sqrt K\,\Delta t\to0$, i.e.\ $\sqrt T\,\Delta t\to0$ at
fixed window geometry, since $K\asymp T/(2T_w+\delta)$; this single rate
serves both results below, and is sharp only for the CLT, since for consistency alone $\Delta t\to0$ suffices for the $O(\Delta t)$ quadrature remainder) and $\norm{R}_{\mathrm{op}}\Delta t/(\sigma_{\min}^2\,T_w^2)\to0$, with
$T_w=L\Delta t$ the window span, $\norm{R}_{\mathrm{op}}$ the spectral norm
of the observation-noise covariance, and $\sigma_{\min}^2$ the smallest
eigenvalue of $\sigma\sigma^{\!\top}$ over the support of the stationary
law (controlling the
observation-noise term, cf.\ \eqref{eq:noise-ratio}),
$\thetahat_{\IV}\toprob\param$.
\end{theorem}

\begin{remark}
The window width enters the two rate conditions in competing
directions. The observation-noise condition
$\norm{R}_{\mathrm{op}}\Delta t/(\sigma_{\min}^2T_w^2)\to0$ improves as
the span grows, since wider windows suppress measurement noise
quadratically (\eqref{eq:noise-ratio}), whereas relevance
(Proposition~\ref{prop:rel}) requires $T_w$ to stay within the
characteristic time $\tau$, favoring short windows. The two are reconciled by fixing the physical
span inside the memory, $T_w\lesssim\tau$, as the design rule of Appendix~\ref{si:choice} does, and
refining the sampling; as $\Delta t\to0$ at fixed $T_w$ the window holds
$L=T_w/\Delta t$ samples, the noise ratio vanishes as $\Delta t/T_w^2$,
and relevance is untouched, so consistency holds along every such
refinement.
\end{remark}

\begin{theorem}\label{thm:clt}
Under the same conditions, with $\mathbf Q=\E[Z_1 A_1\transpose]$ and
$\bm\Sigma=\Var(Z_1\varepsilon_1^{\mathrm{tot}})$ finite and positive definite,
\begin{equation}
\sqrt{K}\,(\thetahat_{\IV}-\param)\todist
\mathcal{N}\!\big(\bm0,\;\mathbf G\big),
\qquad \mathbf G=\mathbf Q^{-1}\bm\Sigma\,\mathbf Q^{-\top}.
\end{equation}
The over-identified GMM estimator with weight $\Wmat=\bm\Sigma^{-1}$ attains the
minimal covariance $(\mathbf Q\transpose\bm\Sigma^{-1}\mathbf Q)^{-1}$ among
GMM estimators based on these moment conditions.
\end{theorem}

The theorems above concern the estimator with known $R$ and a fixed
weight. The implemented estimator plugs in the variogram $\widehat R$, debiases
the drift term candidates with the same $\widehat R$, and estimates the GMM weight
from first-step residuals. The next results close this gap.

\begin{proposition}\label{prop:plugin}
Assume the conditions of the consistency theorem, including the noise
rate $\norm{R}_{\mathrm{op}}\Delta t/(\sigma_{\min}^2T_w^2)\to0$, and
let $\widehat R$ be the multilag variogram estimator of
Section~\ref{sec:iv} computed from the same record of length $N$.
Write $\thetahat(\widehat R)$ for the Algorithm~\ref{alg:waives} estimator in
which $\widehat R$ enters the response subtraction and the
debiasing, and suppose the debiasing map $R\mapsto\widetilde g_R$ is
continuously differentiable in $R$ near the truth, as it is for polynomial
and trigonometric terms. Then, at fixed window span $T_w$, as in the consistency theorem,
$\thetahat(\widehat R)$ is consistent,
$\|\thetahat(\widehat R)-\thetahat(R)\|=O_p(\|\widehat R-R\|)
=O_p(N^{-1/2})$, and
$\sqrt K\,(\thetahat(\widehat R)-\param)\todist\mathcal N(\bm0,\mathbf G_{\widehat R})$
at the same rate $\sqrt K$ as with $R$ known, with
$\mathbf G_{\widehat R}=\mathbf G+\mathbf G_{\mathrm{pl}}$, where
$\mathbf G_{\mathrm{pl}}$ collects the variance of $\widehat R$ propagated
through the debiased columns and its covariance with the moment error, both
of relative order $K/N\asymp1/L$. When $R$ enters no
drift column, as for the affine drift of Sections~\ref{sec:amo} and~\ref{sec:trap},
$\thetahat(\widehat R)=\thetahat(R)$ exactly. Estimating the weight leaves
the first-order theory unchanged whenever $\Wmat\toprob\bm\Sigma^{-1}$,
which holds for the first-step residual plug-in~\cite{Hansen1982}.
\end{proposition}

\begin{remark}
\label{rem:obsC}
The estimator applies verbatim to the observation model
\begin{equation}\label{eq:obs-C}
 Y_{t_n}\;=\;C\,X_{t_n}+\eta_n,\quad
 \eta_n\overset{\text{iid}}{\sim}\mathcal N(0,R),\quad
 C\in\R^{d_y\times d},
\end{equation}
with $C$ known, $d_y\ge d$ and $\rank C=d$. The deconvolution
$\widehat Y_{t_n}=(C^\top C)^{-1}C^\top Y_{t_n}$ recovers the state up to
additive Gaussian error with covariance
\begin{equation}\label{eq:R-eff}
 R_{\mathrm{eff}}\;=\;(C^\top C)^{-1}C^\top R\,C\,(C^\top C)^{-1},
\end{equation}
where the feasible estimator uses $\widehat Y$ and
$\widehat R_{\mathrm{eff}}$ in place of $\widehat R$. Note that for $d_y<d$ the state
is not identified pointwise.
\end{remark}

\subsection{Quadratic suppression of observation noise}
The relative size of measurement-noise and process-noise contributions to the
response is governed by the window length $L$, the number of observation
steps in each window, whose support span is $T_w=L\Delta t$. The instrument and target windows are designed to share the same length
$L$.

Let $\|f\|_{L^2}^2=\int_{a_k}^{a_k+T_w}f(t)^2\dt$. For $\sigma$ roughly
constant per window, the It\^o
isometry gives
$\Var[\varepsilon_k]\approx\sigma^2\|\phi_k\|_{L^2}^2$. Similarly, the noise sum gives
$\Var[\delta\mathsf b_k]\approx R\,\Delta t\,\|\phi_k'\|_{L^2}^2$. Each $\phi_k$
is a fixed reference bump stretched to its window, so
$\|\phi_k\|_{L^2}^2\propto T_w$ and $\|\phi_k'\|_{L^2}^2\propto T_w^{-1}$, and hence
\begin{equation}
\frac{\Var[\delta\mathsf b_k]}{\Var[\varepsilon_k]}
\;\approx\;\frac{R\,\Delta t}{\sigma^2 T_w^2}.
\label{eq:noise-ratio}
\end{equation}
Wider windows suppress observation noise quadratically but weaken the instrument. Choosing $L$ is the main practical decision in \textsc{WAIVES}; Section~\ref{sec:numerics} does it empirically.

\subsection{Joint diffusion estimation}
The same weak-form construction can recover the diffusion in addition to the drift. Writing the diagonal
ansatz $\sigma_i^2(x)=\sum_l\nu_{i,l}h_l(x)$ for diffusion term candidates
$\{h_l\}$, the quadratic variation of $X^i$, the $i$th component of
the state, is absolutely continuous, with
$\dd[X^i,X^i]_t=\sigma_i^2(X_t)\dt$. For each component $i$ and window
$k$, testing against $\phi_k$ and discretizing gives
\begin{equation}
\begin{aligned}
\mathsf b^{\sigma}_{k,i}&=\sum_{n\in\mathcal{I}_k}\bar\phi_k(t_n)\,\Bigl[(e^i_n)^2
-2\widehat R_{ii}\Bigr],\\
\mathsf A^{\sigma}_{k,l}&=\Delta t\!\sum_{n\in\mathcal{I}_k}\phi_k(t_n)\,\widetilde h_l(Y_n).
\end{aligned}
\label{eq:qv-system}
\end{equation}
Here,
$e^i_n=\Delta Y^i_n-\tfrac12(\widehat f_i(Y_n)+\widehat
f_i(Y_{n+1}))\Delta t$, with $\Delta Y^i_n=Y^i_{t_{n+1}}-Y^i_{t_n}$ the
lag-one increment, $\bar\phi_k(t_n)=\tfrac12(\phi_k(t_n)+\phi_k(t_{n+1}))$
is the trapezoidal step weight, $\widehat f$ is the
drift estimate, and $\widetilde h_l$ is the diffusion term candidate
debiased for observation noise as described in Section~\ref{sec:algorithm}.
Stacking the rows over the window pairs and test functions gives, for
each component $i$, the linear system
$\mathsf b^{\sigma}_{i}=\mathsf A^{\sigma}\bm\nu_i+\bm\varepsilon^{\sigma}_i$,
solved for $\bm\nu_i$ by the same disjoint-window IV/GMM step as the drift,
with instruments
$\mathsf Z^{\sigma}_{k,l}=\Delta t\sum_{m\in\mathcal I_k'}\psi_k(t_m)\,\widetilde h_l(Y_m)$
on the instrument window (Section~\ref{si:qv-derivation}).

Differencing the
observed series injects a deterministic measurement-noise variance
$2\widehat R_{ii}\sum_n\bar\phi_k(t_n)$ into the raw quadratic variation, which we
subtract directly. Additionally, the increment $\Delta Y^i_n$ contains the drift
$\int_{t_n}^{t_{n+1}} f_i(X_t)\,\mathrm{d}t$, which $e^i_n$ removes by
subtracting its trapezoidal approximation
$\tfrac12[\widehat f_i(Y_n) + \widehat f_i(Y_{n+1})]\Delta t$.
The trapezoidal step also cancels the leading bias that a left-endpoint
drift subtraction would leave, a cross term between the Brownian increment
and the drift's change within the step, of relative size $f'\Delta t$, and
it retains the $O(\Delta t^2)$ deterministic correction, without which the
$|J_f f|^2\Delta t^4$ bias of an uncorrected increment can swamp the
$\sigma^2\Delta t$ signal in strongly driven systems (Section~\ref{si:qv-derivation}).

The noise covariance $\widehat R$ entering \eqref{eq:qv-system},
and a second, independent diffusion estimate, both come from one further
construction: a multilag variogram of the observed increments, where the lag
is the number of steps $k$ an increment spans. For
$k=1,\dots,n_{\mathrm v}$ lags, let
\begin{equation}\label{eq:R-multilag}
 V_k=\frac{1}{N-k}\sum_{n}\bigl(Y_{t_{n+k}}-Y_{t_n}\bigr)
 \bigl(Y_{t_{n+k}}-Y_{t_n}\bigr)^{\!\top},
\end{equation}
whose expectation satisfies, for $k\Delta t$ small,
\begin{align}\label{eq:Vk-expansion}
 \E\bigl[(V_k)_{ij}\bigr]=2R_{ij}&+\E_\pi\bigl[(\sigma\sigma^\top)_{ij}\bigr]\,(k\Delta t) \nonumber
 \\
 &+c_{2,ij}\,(k\Delta t)^2+O\bigl((k\Delta t)^3\bigr),
\end{align}
where $c_{2,ij}$ is a drift-dependent stationary constant (identified for
the OU case in Section~\ref{si:slope}). Stationarity of
$X$ is assumed throughout, so each expectation is affine in $(k,k^2)$. For each pair
$(i,j)$, the $n_{\mathrm v}$ numbers $(V_1)_{ij},\dots,(V_{n_{\mathrm v}})_{ij}$ are then regressed
on the design $(1,k,k^2)$ by ordinary least squares,
$\widehat{(V_k)_{ij}}=\tilde{c}_{ij}+\tilde{d}_{ij}k+\tilde{e}_{ij}k^2$. The
intercepts of that regression give
\begin{equation}\label{eq:R-hat}
 \widehat R=\Pi_{\succeq0}\bigl(\tfrac12 \widetilde{\mathbf c}\bigr),\qquad \widetilde{\mathbf c}=(\tilde{c}_{ij}),
\end{equation}
with $\Pi_{\succeq0}$ the Frobenius-norm projection onto the positive-semidefinite cone, computed by zeroing negative eigenvalues. This projection is unique, and
the fit is well posed for any $n_{\mathrm v}\ge3$. We use $n_{\mathrm v}=6$,
and show that sensitivity to $n_{\mathrm v}\in\{3,\dots,10\}$ is negligible (Appendix~\ref{si:numerics}).

Reading $\widehat R$ off
the intercept requires no knowledge of the drift, and in particular of $c_{2,ij}$,
whose value is absorbed by the quadratic regressor. Off-diagonal entries follow from the polarization
identity, applied to the scalar variograms of the sums and differences
$Y^i\pm Y^j$ (Appendix~\ref{si:numerics}). Under the
observation model \eqref{eq:obs-C} of Remark~\ref{rem:obsC}, the same
construction, run on the deconvolved observations, returns
$\widehat R_{\mathrm{eff}}$ for the $R_{\mathrm{eff}}$ of
\eqref{eq:R-eff} directly.

The slope $\tilde{d}_{ij}$ of the same fit identifies the stationary
mean $\E_\pi[(\sigma\sigma^\top)_{ij}]\,\Delta t$ of \eqref{eq:Vk-expansion},
the diffusion matrix itself when $\sigma$ is constant, and so supplies a
second, independent estimate of that mean,
$\widehat{(\sigma\sigma^\top)}_{ij}=\tilde{d}_{ij}/\Delta t$.
We refer to this estimator as the multilag estimator. Its per-window,
weak-form version replaces the lag-one response of \eqref{eq:qv-system}
by
\begin{equation}\label{eq:qv-vgram}
 \mathsf b^{\sigma,\mathrm v}_{k,i}
 =\sum_{j=2}^{j_{\max}} w_j\,S_{k,i}(j); \quad
 S_{k,i}(j)=\sum_{n\in\mathcal{I}_k}\phi_k(t_n)\bigl(Y^i_{n+j}-Y^i_n\bigr)^2
\end{equation}
with $(w_j)$ the fixed weights that read the slope off a $(1,j,j^2)$ fit
across lags $j=2,\dots,j_{\max}$. Every lag uses the same anchors $n\in\mathcal I_k$, so the total test-function weight does not depend on $j$ and the slope carries no lag-dependent truncation, while increments anchored near the end of the window reach up to $j_{\max}$ steps past it. The design $\mathsf A^\sigma$ is
unchanged, and the derivation and defaults are given in
Appendix~\ref{si:numerics}. The multilag estimator needs no noise subtraction and no drift correction,
and it is insensitive to additive distortions of the increment confined to lags below the first fitted one, $j_{\min}\Delta t$ (Section~\ref{si:slope}). Its variance is comparable to the lag-one estimator's at
moderate noise and larger at very low noise.
The multilag estimator can also serve as an internal model check, since under
\eqref{eq:obs} the two diffusion estimators agree up to finite-lag, discretization, and sampling error.

Both diffusion estimators are governed by the per-component noise-to-signal ratio
$\varrho_i=2R_{ii}/(\sigma_i^2\Delta t)$ of the increment variance. (When
every component of a system shares a common scale, as in all our examples, we
drop the subscript and write a single $\varrho$.) As we will see, the multilag
fit recovers $\widehat R$ accurately in all our regimes, because the
quadratic coefficient absorbs the drift bias even for the chaotic flow at
$\varrho\approx50$. We also show that the diffusion estimators are accurate to within Monte
Carlo error in regimes with $\varrho\lesssim20$. Beyond that
noise-to-signal ratio, however, their spread and their bias grow with $\varrho$, and for chaotic systems such as Lorenz-63 at
$\varrho\approx50$ the spread can reach $20\%$ per component and the $y,z$ estimates are
biased high (Section~\ref{sec:numerics}).
Sub-sampling from a finer grid, when available, lowers $\varrho$ and can restore precision, provided the
coarsened step still resolves the drift. Several of the systems of Section~\ref{sec:numerics} illustrate performance on thinned grids.

\subsection{Derivation of the quadratic-variation system}
\label{si:qv-derivation}

For component $i$ of \eqref{eq:state}, $\dd X^i_t=f_i(X_t)\dt+\sigma_i(X_t)\dd W^i_t$,
the quadratic variation is $\dd[X^i,X^i]_t=\sum_l\nu_{il}h_l(X_t)\dt$, and
testing against $\phi_k$ gives
\begin{equation}\label{eq:qv-weak}
\int_{a_k}^{b_k}\phi_k(t)\,\dd[X^i,X^i]_t
=\sum_{l}\nu_{il}\int_{a_k}^{b_k}\phi_k(t)\,h_l(X_t)\dt .
\end{equation}
The right-hand side is $\sum_l\nu_{il}\mathsf A^{\sigma}_{k,l}$ up to the
$O(\Delta t)$ quadrature error, with $\widetilde h_l(Y_n)$ in place of
$h_l(X_{t_n})$ to remove the observation-noise blur. The left-hand side is
the limit of $\sum_{n\in\mathcal I_k}\bar\phi_k(t_n)(\Delta X^i_n)^2$, the
increment over $[t_n,t_{n+1}]$ carrying the step average $\bar\phi_k(t_n)$
of the test function. Three corrections turn $(\Delta X^i_n)^2$ into the
observable $(e^i_n)^2$. The increment contains the drift, which the
trapezoidal step $\tfrac12[f_i(X_{t_n})+f_i(X_{t_{n+1}})]\Delta t$ removes
to $O(\Delta t^{2})$ in mean, leaving the martingale part with conditional
second moment $\sigma_i^2(X_{t_n})\Delta t+O(\Delta t^2)$ (the left-endpoint
rule would leave a cross term between the Brownian increment and the
within-step change of the drift, of mean $f_i'\sigma_i^2\Delta t^2$, that is,
of relative size $f_i'\Delta t$, which the trapezoidal step cancels, as well
as the deterministic term $\tfrac14|J_ff|^2\Delta t^4$ above). The observations replace $X$ by $Y=X+\eta$, so $e^i_n$ carries
$\eta^i_{n+1}(1-\tfrac12 f_i'\Delta t)-\eta^i_n(1+\tfrac12 f_i'\Delta t)$ to
first order, of variance $2R_{ii}+O(R\Delta t^2)$, the term
$2\widehat R_{ii}$ subtracted in \eqref{eq:qv-system}; the symmetric
weights cancel the cross term with $\eta^i_n$, which a left-endpoint drift
would leave as a bias of $+2\Delta t\,f_i'R_{ii}$ per step. The cross terms
between the martingale part and the observation errors have mean zero, and
the drift-estimation error contributes $O(\Delta t\,\|\thetahat-\param\|)$.
Collecting terms,
$\E[(e^i_n)^2\mid X_{t_n}]=\sigma_i^2(X_{t_n})\Delta t+2R_{ii}+O(\Delta t^2)+O(R\Delta t^2)$,
and summing with the weights $\bar\phi_k(t_n)$ yields
\begin{equation}\label{eq:qv-rows}
\mathsf b^{\sigma}_{k,i}
=\sum_{l}\nu_{il}\,\mathsf A^{\sigma}_{k,l}+\varepsilon^{\sigma}_{k,i}+\rho^{\sigma}_{k,i},
\end{equation}
with $\varepsilon^{\sigma}_{k,i}$ a sum of mean-zero terms, martingale differences and a one-dependent noise part, and
$\rho^{\sigma}_{k,i}$ the $O(\Delta t)$ remainders, the error of
$\widehat R$ and the debiasing remainder. Stacked over window pairs and
test functions this is the system \eqref{eq:qv-system}. For a constant
diffusion term candidate the regressor $\mathsf A^{\sigma}$ is deterministic; for a
state-dependent term such as $h(x)=x$ of the Cox--Ingersoll--Ross model it
is built from the same increments as $\varepsilon^{\sigma}_{k,i}$, the
endogeneity of the drift case, and the disjoint-window instrument
$\mathsf Z^{\sigma}_{k,l}$ restores exogeneity by Proposition~\ref{prop:exog}.
The subtraction of $2\widehat R_{ii}$ is where the observation noise enters
at order one, the variance of $(e^i_n)^2-2\widehat R_{ii}$ being of order
$(\sigma_i^2\Delta t+2R_{ii})^2$ against a signal $\sigma_i^2\Delta t$, which
is why the precision of the lag-one form is governed by
$\varrho_i=2R_{ii}/(\sigma_i^2\Delta t)$.

\subsection{The multilag slope estimator}
\label{si:slope}

The slope response \eqref{eq:qv-vgram} regresses the test-function-weighted
squared $j$-increments on $(1,j,j^2)$ and keeps the linear coefficient. Its
properties follow from the small-time expansion of the variance of an observed increment.
For a stationary diffusion observed with the noise of
\eqref{eq:obs}, and for a lag $u=j\Delta t$,
\begin{equation}\label{eq:slope-expansion}
\E\bigl[(Y^i_{t+u}-Y^i_t)^2\bigr]
= 2R_{ii} + \bar\sigma_i^2\,u + c_{2,i}\,u^{2} + c_{3,i}\,u^{3} + O(u^4),
\end{equation}
where $\bar\sigma_i^2=\E[\sigma_i^2(X_t)]$ is the stationary mean of the
$i$th diagonal diffusion coefficient of Section~\ref{sec:iv}, the quantity
the slope estimates, and $c_{2,i}$ and $c_{3,i}$ are stationary moments of
the drift and its derivatives. The constant term is the only place where
the observation noise enters, since the two errors in the increment are
independent of the path and of each other. For the OU model the expansion is exact,
$\E[(Y_{t+u}-Y_t)^2]=2R+2v(1-e^{-\kappa u})$ with $v=\sigma^2/2\kappa$, so
$c_2=-\sigma^2\kappa/2$ and $c_3=\sigma^2\kappa^2/6$. Observation noise enters \eqref{eq:slope-expansion} only through the $u$-independent level, which the intercept absorbs, and so does any additive distortion whose correlation vanishes beyond the first fitted lag, $\Cov(\eta_{t+u},\eta_t)=0$ for $u\ge j_{\min}\Delta t$. An error correlation $c(u)$ of longer range adds $2R_{ii}[1-c(u)]$, which varies across the fitted lags and enters the slope, so the immunity is to distortions confined below $j_{\min}\Delta t$, a condition that Section~\ref{sec:trap} checks against the Einstein value rather than assumes.

The fit on $(1,j,j^2)$ is exact for the first three terms of
\eqref{eq:slope-expansion}, so the intercept returns $2R_{ii}$, the linear
coefficient returns $\bar\sigma_i^2\Delta t$, and the quadratic coefficient
absorbs $c_{2,i}u^2$, the drift contamination. The only bias of the slope is therefore the part of the cubic term $c_{3,i}u^3$ that the fit assigns to
the linear coefficient. Regressing $j^3$ on $(1,j,j^2)$ over
$j\in\{2,\dots,j_{\max}\}$ gives that part in closed form, and the relative
bias of $\widehat{\sigma^2}$ is $c\,(c_{3,i}/\bar\sigma_i^2)(j_{\max}\Delta t)^2$ with a design constant $c$ of order one. For OU this is $c\,(\kappa\,j_{\max}\Delta t)^2/6$, about
$3\times10^{-3}$ at the defaults, and
the rule $j_{\max}\Delta t\lesssim0.1\tau$ keeps it negligible in general,
while $j_{\max}\le L/2$ keeps the increments anchored in the window from reaching far beyond it. For the
double-well benchmark the two drift term candidates can be written explicitly. The
noise adds only the constant $2R$ to \eqref{eq:slope-expansion}, so
$c_2$ and $c_3$ are those of the path itself, and for a reversible scalar
diffusion with constant $\sigma$ they are stationary averages of the drift,
$c_2=-\E[f(X)^2]$ and
$c_3=-\tfrac13 \E\bigl[f(X)\bigl(ff'+\tfrac12\sigma^2f''\bigr)(X)\bigr]$.
On the double-well stationary law both equal $0.195$, so at $j=j_{\max}$
the quadratic term is $|c_2|j_{\max}\Delta t/\sigma^2\approx6\%$ of the
linear signal and the cubic term $0.5\%$; the fit removes the first exactly
and leaves a slope bias of $0.5\%$ from the second. 

Per window the same expansion holds conditionally on the path, with $\sigma_i^2(X_{t_n})$ at each anchor in place of $\bar\sigma_i^2$, so the linear coefficient of $S_{k,i}(j)$ is $\Delta t\sum_{n\in\mathcal I_k}\phi_k(t_n)\sigma_i^2(X_{t_n})=\sum_l\nu_{il}\mathsf A^\sigma_{k,l}$ up to the quadrature remainder. The slope response therefore obeys the moment equation \eqref{eq:qv-rows} and recovers the state-dependent coefficients $\bm\nu$, while the pooled fit \eqref{eq:R-multilag} gives only the stationary mean $\bar\sigma_i^2$. Since increments anchored near the end of the target window reach up to $j_{\max}$ steps into the next instrument window, consecutive pair scores $\bm\xi_k=Z_k\varepsilon_k$ are correlated at lag one. This does not affect exogeneity within a pair, because the instrument still precedes its own target. It changes only the variance. For a stationary sequence with correlation at lag one the central limit theorem still holds, with $\bm\Sigma$ replaced by the long-run covariance $\bm\Sigma+\bm\Sigma_1+\bm\Sigma_1^{\!\top}$, $\bm\Sigma_1=\Cov(\bm\xi_k,\bm\xi_{k+1})$, and this long-run covariance is exactly what the Newey--West plug-in of Appendix~\ref{si:coverage} estimates.

The slope form carries a variance cost, since a slope fitted through $j_{\max}-1$ correlated points is noisier than a single lag-one level. The
lag-one estimator, however, must subtract $2\widehat R$, and that
subtraction is itself noisy when $2R_{ii}$ is large against
$\bar\sigma_i^2\Delta t$. The two are therefore about equally precise on
the noisy synthetic benchmarks, while the slope estimate is the noisier one
on the low-noise optical-trap record (Table~\ref{tab:sigma},
Section~\ref{sec:trap}). What the slope provides is immunity to the short-lag
distortions of a real detector, which is why Section~\ref{sec:trap} uses it.

\section{Generalized method of moments}
\label{sec:efficiency}

The estimator~\eqref{eq:iv-estimator} is exactly identified. It uses one
instrument column per regressor, so the number of moment conditions equals the number of
unknowns, and the solution exists and is unique whenever
$\Zmat\transpose\Amat$ is invertible. One can instead use \emph{more} instruments than
parameters, so that $\Zmat$ has more columns than $\Amat$ and the moment
vector $\bm m(\param)=\Zmat\transpose(\bvec-\Amat\param)$ has more
conditions than parameters. This can happen when we create multiple disjoint
instrument windows per target window, or when more drift term candidates enter on the instrument side than on the target side. 
The system is then
\emph{over-identified}, and the efficient combination of moment conditions is
the generalized method of moments (GMM)~\cite{Hansen1982}:
\begin{equation}
\thetahat_{\GMM}
=(\Amat\transpose\Zmat\,\Wmat\,\Zmat\transpose\Amat)^{-1}
 \Amat\transpose\Zmat\,\Wmat\,\Zmat\transpose\bvec.
\label{eq:gmm}
\end{equation}
The optimal weight matrix
\begin{equation}
\Wmat=\Big(\tfrac{1}{K}\textstyle\sum_k
 Z_k\,\widehat\varepsilon_k\,\widehat\varepsilon_k\transpose Z_k\transpose
 \Big)^{-1}
\label{eq:gmm-weight}
\end{equation}
is estimated from first-step residuals
$\widehat\varepsilon_k=\mathsf b_k-A_k\thetahat_{\TSLS}$, obtained from the
two-stage least-squares estimator
\begin{equation}
\thetahat_{\TSLS}=(\Amat\transpose\Pmat_Z\Amat)^{-1}\Amat\transpose\Pmat_Z\bvec;
\quad \Pmat_Z=\Zmat(\Zmat\transpose\Zmat)^{-1}\Zmat\transpose.
\label{eq:tsls}
\end{equation}
The resulting estimator has the smallest asymptotic variance among GMM
estimators based on the given moment conditions~\cite{Hansen1982}. At
the default depth $n_{\text{lags}}=1$, meaning one past instrument block
per target window (Remark~\ref{rem:testfns}), the system is exactly
identified provided the instrument window carries the same drift term candidates and
test-function bank as the target (\eqref{eq:gmm} then reduces to
\eqref{eq:iv-estimator}). The
two-step estimator \eqref{eq:gmm} operates whenever the depth rule of
Remark~\ref{rem:testfns} raises $n_{\text{lags}}$ to two.%

\begin{figure}[t]
\centering
\includegraphics[width=\linewidth]{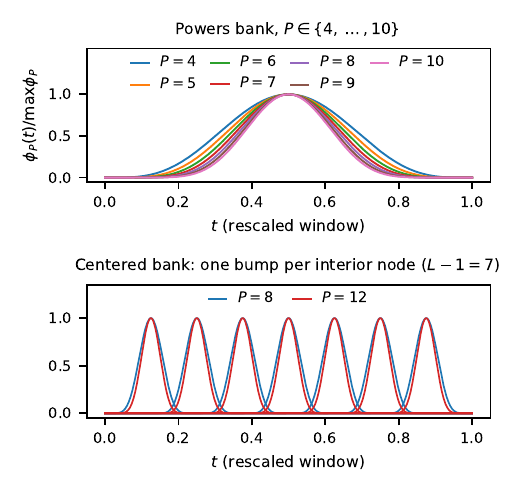}
\caption{Test-function banks (shown on a rescaled window). \emph{Top:} the
powers bank $\phi_P(t)\propto[(t-a)(b-t)]^P$ for $P\in\{4,\dots,10\}$, with
larger $P$ concentrating the bump toward the window center. \emph{Bottom:}
the centered bank, one compactly supported bump per interior node, for two
smoothness settings $P\in\{8,12\}$. Both banks supply the $M_\phi$ test
functions per window pair (Remark~\ref{rem:testfns}). The main results use
the powers bank with $P\in\{4,6,8\}$.}
\label{fig:banks}
\end{figure}

\begin{remark}
\label{rem:testfns}
For the test functions in both roles, $\phi_k$ on the target window and
$\psi_k$ on the instrument window, we use the compactly supported bump
functions of Messenger and Bortz~\cite{MessengerBortz2021},
$w_p(t)\propto(t-a_k)^p(a_k+T_w-t)^p\,\mathbf{1}_{[a_k,a_k+T_w]}(t)$. We evaluate a bank of $M_\phi=3$ test functions per window pair, with smoothness exponents $p\in\{4,6,8\}$; exponents up to $p=16$ perform comparably. To satisfy exogeneity we tile the observation grid into
consecutive disjoint blocks of $L$ observations (span $L\Delta t$)
and pair them sequentially. With $n_{\text{lags}}=1$ the instrument is
integrated over one block and the target over the next; with
$n_{\text{lags}}=2$ the instrument spans the two preceding
blocks. At $n_{\text{lags}}=1$ the moment system
is exactly identified, and when relevance (IV1) is marginal the estimator
inherits the usual weak-instrument bias toward the least-squares
limit~\cite{StaigerStock1997}, which for the reversible systems of
Proposition~\ref{prop:ols-zero} is zero. Doubling the instrument set makes
the system over-identified and brings in the GMM weighting.
We use $n_{\text{lags}}=1$ by default, and raise the depth to two only
when the diagnostics defined in Section~\ref{sec:numerics} flag
a weak or contaminated first stage. Instrument and target are further
separated by a gap $\delta\ge0$, an integer multiple of $\Delta t$. A positive
gap is not required for exogeneity, which holds even at $\delta=0$, the
setting used on the short climate record of Section~\ref{sec:amo}.
Rather, it buffers the instrument against short-range correlation in the
measurement noise. The guiding principle is to take $\delta$ beyond the
noise correlation length while keeping it far below $\tau$. A gap of a few
steps covers short-range noise correlation at negligible variance cost;
longer noise correlation calls for a longer gap. We use
$\delta=4\Delta t$ on the benchmarks, lengthened to $8\Delta t$ on the
low-pass-filtered trap record, following the design rule of
Appendix~\ref{si:choice} (see Fig.~\ref{fig:windows}). The test-function
banks are illustrated in Fig.~\ref{fig:banks}.
\end{remark}

\section{The WAIVES algorithm}
\label{sec:algorithm}

Algorithm~\ref{alg:waives} assembles the preceding constructions into a
complete procedure for estimating drift and diffusion parameters $\param$ and $\bm\nu$ from the data
$\{Y_{t_n}\}_{n=0}^N$, which carry both process and measurement noise.

\begin{algorithm}[!t]
\caption{\textsc{WAIVES}, weak-form IV/GMM estimation for SDEs}
\label{alg:waives}
\begin{algorithmic}[1]
\Require Observations $\{Y_{t_n}\}_{n=0}^N$; drift term candidates $\{g_j\}_{j=1}^p$
 and diffusion term candidates $\{h_l\}_{l=1}^{q}$; target test functions
 $\{\phi_k\}$ and instrument test functions $\{\psi_k\}$, shared by the drift
 system \eqref{eq:obs-b}--\eqref{eq:obs-Z} and the diffusion system
 \eqref{eq:qv-system}.
\State \textbf{Noise covariance.} Estimate $\widehat R$ from the multilag
 variogram of Section~\ref{sec:iv} (default $n_{\mathrm v}=6$), then project onto
 the positive-semidefinite cone.
\State \textbf{Weak-form system from raw $Y_{t_n}$.} Build $\mathsf b_k$, $A_{kj}$ over
 $[a_k,a_k+T_w]$ and $Z_{kj}$ over the disjoint window $[a_k',a_k'+T_w]$ via
 \eqref{eq:obs-b}--\eqref{eq:obs-Z}; debias polynomial drift term candidates for
 observation noise (rules in the text below).
\State \textbf{Drift estimate.} If exactly identified, compute
 $\thetahat_{\IV}$ by \eqref{eq:iv-estimator}. If over-identified, run GMM,
 (i) $\thetahat_{\TSLS}$ by \eqref{eq:tsls};
 (ii) residuals $\widehat\varepsilon_k=\mathsf b_k-A_k\thetahat_{\TSLS}$;
 (iii) $\Wmat$ by \eqref{eq:gmm-weight};
 (iv) $\thetahat_{\GMM}$ by \eqref{eq:gmm}.
\State \textbf{Iterative reweighting.} Recompute residuals, update the
 covariance, re-solve step 3; repeat to convergence (at most $10$ iterations). In the
exactly identified configuration used for the main-text tables the
solution does not depend on the weighting, so this step leaves the
estimate unchanged; it is active in the over-identified runs.
\State \textbf{Diffusion estimate.} With $\thetahat$ fixed, build the
 quadratic-variation system \eqref{eq:qv-system} on the same windows and
 instruments, using the multilag slope response \eqref{eq:qv-vgram} whenever the two responses disagree
 beyond their standard errors or the variogram intercept of
 Section~\ref{sec:iv} is fitted negative, and solve for the diffusion
 coefficients by the same IV/GMM steps.
\Ensure Drift parameters estimate $\thetahat$ and diffusion parameters estimate
 $\widehat{\bm\nu}$.
\end{algorithmic}
\end{algorithm}

Note that a nonlinear term evaluated at a
noisy observation is biased even for symmetric noise, since
$\E[g(Y)\mid X]$ represents $g$ averaged over the noise density. This
averaging of $g$ results in a bias (the classical
errors-in-variables bias of nonlinear regression) proportional to $R$ and to the
curvature of $g$ (for instance $\E[Y^2\mid X]=X^2+R$). 
The debiased term is the corrected score ~\cite{Stefanski1989,Nakamura1990,Carroll2006}, computed via the deconvolution $\widetilde g=e^{-\frac12\widehat R\,\partial^2}g$, resulting in adjustment to monomials ($Y^2-\widehat R$, $Y^3-3\widehat R\,Y$, and so on). General terms, including trigonometric ones, are handled by a regularized Gauss--Hermite rule (Appendix~\ref{si:numerics}).
 
\section{Numerical experiments}
\label{sec:numerics}

\begin{figure}[t]
\includegraphics[width=\linewidth]{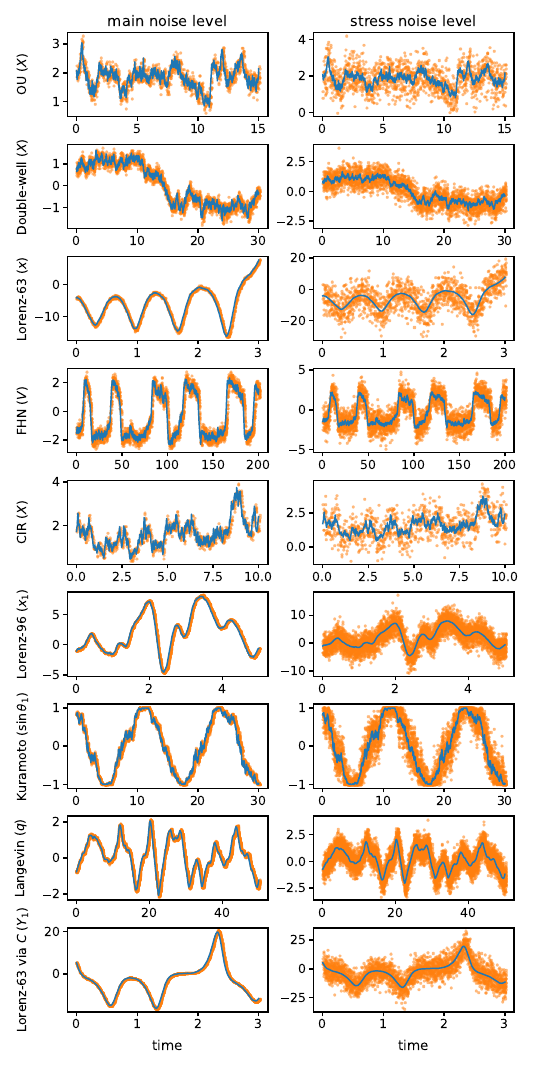}
\caption{Representative latent trajectory (blue) and noisy observations (orange) for the nine systems of Table~\ref{tab:systems}, at the main-analysis noise level (left) and at the stress level (right). The two levels, a fixed noise variance $R$ per system and a noise standard deviation of $15\%$ of the component’s range, are listed in Table~\ref{tab:noiselevels}. Kuramoto is shown as $\sin\theta_1$ with the noise applied to the phase.}
\label{fig:noise}
\end{figure}

In this section we test the performance of \textsc{WAIVES} on nine
systems. Table~\ref{tab:systems} summarizes each system's governing
equations and the difficulty it probes. Each system is described below, and the simulation protocol is given in Appendix~\ref{si:numerics}. Example trajectories at two noise levels of every system are shown in Fig.~\ref{fig:noise}. 

To test the performance of the \textsc{WAIVES} estimator over
replications, we simulate $M=1000$ independent trajectories for each model
by Euler--Maruyama on a grid finer than the observation step
(Appendix~\ref{si:numerics}), producing $M$ sets of observations
$\{Y_{t_n}\}_{n=0}^N$, with added i.i.d.\ Gaussian measurement noise with
covariance $R$. $R$ is fixed per system rather than set relative to the signal, so relative to each observed component's stationary standard deviation the main analyses span noise levels from $0.1\%$ to $30\%$ (Table~\ref{tab:noiselevels}). Measurements lengths differ across systems, each spanning a multiple of the system's characteristic time. Every experiment shares the bank and depth rule of Remark~\ref{rem:testfns}
and the gap $\delta=4\Delta t$ (Appendix~\ref{si:numerics}). 
Reported estimates are by default based on the exactly identified IV
($n_{\text{lags}}=1$), except where the depth rule raises
$n_{\text{lags}}$ to two. In the latter case the moment system is
over-identified, and the two-step GMM estimator \eqref{eq:gmm} is used
instead (Section~\ref{sec:efficiency}).

It is important to note that every drift and diffusion estimate is a
single-trajectory estimate, and each of the $M$ trajectories yields one
estimate from its own record.
The tables report the average and
standard deviation of the $M$ estimates, as estimator mean and
standard error, respectively. 
Single-trajectory uncertainty is quantified by Wald intervals built on
the plug-in GMM covariance (Appendix~\ref{si:coverage}), with a
Newey--West long-run correction across window pairs. On OU, the double-well, Lorenz-63 and FitzHugh--Nagumo these intervals attain near-nominal coverage at the operative
record lengths, and on OU and the double-well also at records one-eighth
as long; bootstrap intervals do not improve on them (Appendix~\ref{si:coverage}).

\begin{figure*}[t]
\centering
\includegraphics[width=\textwidth]{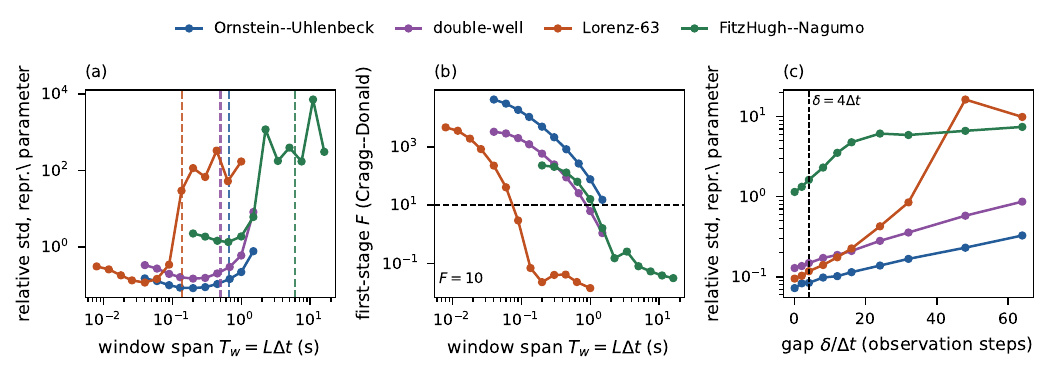}
\caption{Window and gap diagnostics, one representative parameter per
system ($M=1000$). (\emph{a}) Estimator standard deviation versus span
$T_w=L\Delta t$. Each curve turns sharply upward as the span approaches
the system's characteristic time $\tau$ (dashed vertical lines, values in
Appendix~\ref{si:choice}). (\emph{b}) The companion first-stage
Cragg--Donald $F$, which rises as the window shrinks. The dashed line
marks the $F=10$ threshold. (\emph{c}) Standard deviation versus the
instrument--target gap $\delta$ at $L=20$. Accuracy degrades as the gap
grows and the instrument leaves the system's memory (the dashed line
marks the default $\delta=4\Delta t$). FitzHugh--Nagumo is shown at depth
two (two instrument windows per target window), as selected by the depth rule, since its depth-one first stage falls below $F=10$
once the gap exceeds $6\Delta t$. Together the panels depict the operating
logic behind the design rule of Appendix~\ref{si:choice}.}
\label{fig:variance}
\end{figure*}

When the support span $T_w=L\Delta t$ exceeds the characteristic time
$\tau$ of the dynamics (the local relaxation time for the dissipative models, and for the
chaotic models the $1/e$ decorrelation time, computed here from long
simulated trajectories and in applications estimated from the record's
autocorrelation),
the state decorrelates across the
disjoint windows, relevance (IV1) degrades, and the variance of
$\thetahat$ increases sharply. For this reason, we compare $L=100$, $20$, and $5$
throughout  
(the full sweep is given in
Fig.~\ref{fig:variance}.)
Panel (c) of Fig.~\ref{fig:variance} shows the standard
deviation rising as the gap $\delta$ grows,
underscoring the need for the disjoint windows to stay close enough to remain
mutually informative. However,
even though   $\delta=0$ would be acceptable
(Section~\ref{sec:efficiency}), in practice a small but positive $\delta$ can guard against
short-range correlation in the measurement noise, with a relatively low
effect on variance.

Instrument strength is monitored by the first-stage Cragg--Donald $F$~\cite{CraggDonald1993} (Appendix~\ref{si:choice}), with $F\ge10$ treated as trustworthy relevance~\cite{StaigerStock1997,StockYogo2005}; below $10$ at depth one ($n_{\text{lags}}=1$) the depth rule of Remark~\ref{rem:testfns} is applied. This rule picks the depth from the data itself, a step the theorems do not cover, so we test its effect directly in the coverage study (Appendix~\ref{si:coverage}). As $F$ measures strength only, we also recompute the estimate at depth two. Under valid instruments both depths are consistent for the same parameters and agree to within their standard errors; a large disagreement indicates a residual correlation between the instrument and the error and calls for instruments beyond the contamination's range (Section~\ref{sec:trap}; Appendix~\ref{si:choice}).

\begin{table*}[t]
\centering
\caption{The nine example systems, each chosen to probe one challenge for a weak-form SDE estimator. For each system the last column reports the mean,(std) over $M=1000$ replications of one representative drift parameter (the one most sensitive to the listed challenge), and, below it, the diffusion estimate for the same component. Diffusion entries use the lag-one quadratic-variation form, except the slope form for the Langevin at $\gamma=0.5$. The Cox--Ingersoll--Ross entry lists the drift rate $-\widehat\kappa$ above $\widehat\nu$, and the FitzHugh--Nagumo ratio parameter $\widehat a$ is given as its mode with a $95\%$ Fieller interval. Full drift and diffusion estimates for all nine systems appear in Tables~\ref{tab:drift} and~\ref{tab:sigma}. Although each row is designed to stress a different failure mode, the same estimator under the same default rules recovers every reported parameter with negligible bias. }
\footnotesize
\begin{tabular}{@{}p{2.3cm}p{5.9cm}p{4.3cm}p{3.4cm}@{}}
\toprule
System & Model & Challenge probed & Representative recovery\\
\midrule
Ornstein--Uhlenbeck (OU) & $\dd X=\kappa(\mu-X)\dt+\sigma\dW$ & linear baseline with exact likelihood; calibrates the cost of the weak form ($\tau=1/\kappa$) & $\widehat\kappa=1.497\,(0.126)$ vs $1.5$\newline $\widehat\sigma=0.800\,(0.020)$ vs $0.8$\\
\midrule
Double-well (DW) & $\dd X=(\alpha X-\beta X^3)\dt+\sigma\dW$ & bistability: local relaxation ($\tau=0.5$\,s) vs rare hopping ($\approx13$\,s) & $\widehat\alpha=1.003\,(0.149)$ vs $1$\newline $\widehat\sigma=0.498\,(0.032)$ vs $0.5$\\
\midrule
Stochastic Lorenz-63 (L63) & $\dd x=a(y-x)\dt+\sigma\dd W_1$;\; $\dd y=(x(b-z)-y)\dt+\sigma\dd W_2$;\; $\dd z=(xy-cz)\dt+\sigma\dd W_3$ & chaos: decorrelation ($\tau\approx0.14$\,s) far inside the Lyapunov time; coefficients spanning $2.7$ to $28$ & $\widehat b=28.027\,(0.355)$ vs $28$\newline $\widehat\sigma_y=1.075\,(0.183)$ vs $1$\\
\midrule
Stochastic FitzHugh--Nagumo (FHN) & $\dd V=(V-V^3/3-W+I)\dt+\sigma_V\dd B_1$;\; $\dd W=\epsilon(V+a-bW)\dt+\sigma_W\dd B_2$ & slow--fast excitability; $a,b$ enter only through products with $\epsilon\approx0.08$ & $\widehat a=0.684\,[0.61,0.77]$ vs $0.7$\newline $\widehat\sigma_W=0.302\,(0.038)$ vs $0.3$\\
\midrule
Cox--Ingersoll--Ross (CIR) & $\dd X=\kappa(\mu-X)\dt+\nu\sqrt{X}\,\dW$ & state-dependent diffusion through the diffusion term candidate $h(x)=x$ & $-\widehat\kappa=-2.10\,(0.75)$ vs $-2$\newline $\widehat\nu=1.00\,(0.04)$ vs $1$\\
\midrule
Lorenz-96 (L96) & $\dd x_i=[(x_{i+1}-x_{i-2})x_{i-1}-x_i+F]\dt+\sigma\dd W_i$ & dimension: $40$ and $120$ coupled components & $\widehat F=7.99\,(0.04)$ vs $8$ ($d=120$)\newline $\widehat\sigma=0.501\,(0.002)$ vs $0.5$\\
\midrule
Kuramoto & $\dd\theta_i=[\omega+\tfrac{K}{N}\sum_j\sin(\theta_j-\theta_i)]\dt+\sigma\dd W_i$ & network coupling; trigonometric drift term candidates and their debiasing & $\widehat K=2.01\,(0.10)$ vs $2.0$\newline $\widehat\sigma=0.300\,(0.002)$ vs $0.3$\\
\midrule
Underdamped Langevin & $\dd q=p\,\dt$;\; $\dd p=[q-q^3-\gamma p]\dt+\sigma\dd W$ & degenerate noise: no Brownian forcing on $q$ & $p$ in $\dot q$: $1.00\,(0.03)$ vs $1$\newline $\widehat\sigma_p=1.013\,(0.011)$ vs $1$\\
\midrule
Lorenz-63, observed through $C$ & as above, observed as $Y=CX+\eta$
with non-identity $C$ & indirect observation, handled by the
deconvolution of Remark~\ref{rem:obsC} & $x$ in $\dot y$: $28.01\,(0.67)$ vs $28$\newline $\widehat\sigma_y=0.412\,(0.072)$ vs $0.4$\\
\bottomrule
\label{tab:systems}
\end{tabular}
\end{table*}

\subsection{The nine systems}
The nine systems follow in the order of Table~\ref{tab:systems}, each with its simulation settings. Together they stress the features that challenge a weak-form estimator, namely linear and bistable dynamics, chaos, slow--fast excitability, a state-dependent diffusion, a high-dimensional system, non-polynomial drift term candidates, degenerate (hypoelliptic) noise, and a non-identity observation map, and none of them requires any change to the estimator. All use the same disjoint-window IV/GMM procedure, recover drift and diffusion jointly, and, unless noted, inherit the settings above; the reported figures are estimator averages\,(std) over $M=1000$ replications at $L=20$, and Tables~\ref{tab:drift} and~\ref{tab:sigma} collect every parameter.

\subsubsection{Ornstein--Uhlenbeck}
The scalar OU process $\dd X_t=\kappa(\mu-X_t)\dt+\sigma\dW$ with
$\kappa=1.5,\mu=2,\sigma=0.8$ is simulated at $\Delta t=0.01$,
$N=200{,}000$ ($T=2000$), $R=0.02$. The drift term candidates are $g(x)=[1,x]$, so the true
drift vector is $[\kappa\mu,-\kappa]=[3,-1.5]$, the diffusion term candidate
is $h(x)=[1]$, and the characteristic time is the relaxation time
$\tau=1/\kappa\approx0.67$\,s. Table~\ref{tab:drift} shows
unbiased recovery, with the smallest standard deviations at the intermediate
and short windows. At $L=20$ and $\delta=4\Delta t$ this record gives $K\approx4545$ window
pairs, so $\sqrt K\,\Delta t\approx0.67$, outside the rate condition of
Theorem~\ref{thm:clt}; the quadrature bias that this permits is
nonetheless negligible here, the mean of $\widehat\kappa$ lying within
$0.003$ of the truth against a standard deviation of $0.13$.
Table~\ref{tab:sigma} recovers $\sigma$ within Monte Carlo error.
This linear-Gaussian case admits an exact likelihood through the
Kalman filter, so it calibrates the price of the weak-form route. A Kalman
maximum-likelihood fit~\cite{ShumwayStoffer1982} of the same trajectories
attains the Cram\'er--Rao bound, with standard deviations about three times
smaller than \textsc{WAIVES} (Appendix~\ref{si:baselines}); the gap is the cost of never forming a
likelihood. The computational comparison, by contrast, favors \textsc{WAIVES}, at $13$\,ms per trajectory for $\kappa$, $\mu$, and
$\sigma$ against $0.08$--$0.28$\,s for the maximum-likelihood fits, a
factor of $6$ to $22$ (Appendix~\ref{si:baselines}).

\subsubsection{Double-well (bistable)}
$\dd X_t=(\alpha X_t-\beta X_t^3)\dt+\sigma\dW$ with $\alpha=\beta=1$,
$\sigma=0.5$ has stable equilibria at $\pm1$ and a near-well relaxation time
$\tau=1/(2\alpha)=0.5$\,s; the record's
overall autocorrelation time, $\approx13$\,s, instead reflects rare well
hopping, and it is the local time that governs instrument relevance. The
diffusion term candidate is again $h(x)=[1]$. We use $\Delta t=0.01$,
$N=200{,}000$, $R=0.02$, drift term candidates
$g(x)=[1,x,x^2,x^3]$, with the observation-noise debiasing of
Section~\ref{sec:algorithm} applied to the quadratic and cubic
columns.
Table~\ref{tab:drift} shows the bias inflating once the window exceeds the
relaxation scale ($L=100$ corresponds to $1.0$\,s, twice $\tau$), and
Table~\ref{tab:sigma} recovers $\sigma$ within error.
A strong-form baseline regressing raw Euler increments on the same
drift term candidates~\cite{Brunton2016,Boninsegna2018} is corrupted by an order of
magnitude at this noise level ($\widehat\alpha=11.1$ against $1$) while
WAIVES stays unbiased (Appendix~\ref{si:baselines}),
so the correction is measured against an established equation-discovery
method, not only against weak-form least squares.

\subsubsection{Stochastic Lorenz-63}
\begin{figure}[t]
\centering
\includegraphics[width=\linewidth]{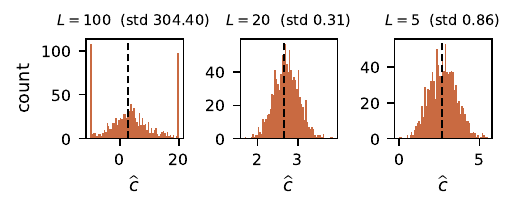}
\caption{Distribution of the Lorenz-63 $\widehat c$ across the $M=1000$ replications, at three window lengths (clipped to $[-10,20]$ for display). At $L=100$ (span $0.2$,s, a substantial fraction of the noise-shortened decorrelation time) the estimate is essentially uninformative, with a tail-dominated sample standard deviation of $\approx306$. At $L=20$ and $L=5$ the distribution concentrates tightly on the true value $c=8/3$. Once the window exceeds the correlation time, the first stage collapses (average $F<1$) and the exactly identified estimator becomes heavy-tailed, so that its average over the replications is governed by a few extreme draws. Shortening the window in such cases restores a tightly concentrated estimator.}
\label{fig:lorenz}
\end{figure}

$\dd x=a(y-x)\dt+\sigma\dd W_1$, $\dd y=(x(b-z)-y)\dt+\sigma\dd W_2$,
$\dd z=(xy-cz)\dt+\sigma\dd W_3$, with $a=10,b=28,c=8/3,\sigma=1$, is chaotic
with Lyapunov time $1/\lambda_{\max}\approx1.1$\,s; the
operative characteristic time is the measured $1/e$ decorrelation
$\tau\approx0.14$\,s, far shorter because the driving noise erases memory
along the attractor. The diffusion term candidate is $h(x)=[1]$ per component. We use $\Delta t=0.002$,
$N=200{,}000$ ($T=400$), $R=0.05$ per component, degree-2 monomial drift term candidates
(10 per component). Table~\ref{tab:drift} shows
the intermediate window $L=20$ is by far the most reliable. Over
$M=1000$ trajectories $\widehat b=28.027$ is within $0.1\%$ of its true
value, while $\widehat a=10.028$ and $\widehat c=2.721$ differ from truth by
$0.3$ and $2.0\%$ (about three and five Monte Carlo standard errors), small
biases consistent with finite-$\Delta t$ discretization at the chaotic
timescales. Here $a$ is read from the $y$ term of the $x$-equation; the $x$
term of the same equation returns $10.093\,(0.377)$, the same within errors,
and carries a $+0.9\%$ finite-$\Delta t$ offset that the $y$ term is free
of. The agreement is an internal consistency check available whenever a
parameter enters more than one term. The long
window's variance increases
sharply once $T_w$ passes $\tau$, as the distribution of the $c$ estimate shows
(Fig.~\ref{fig:lorenz}).

\subsubsection{Stochastic FitzHugh--Nagumo}
\begin{figure*}[t]
\centering
\includegraphics[width=\linewidth]{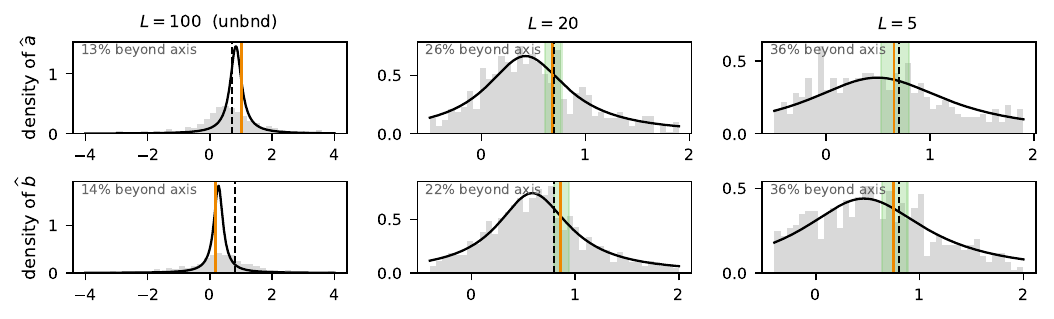}
\caption{Ratio-of-normals inference for the FHN slow parameters $a,b$
($M=1000$). The Marsaglia--Hinkley density (black) matches the Monte Carlo
sampling distribution of the ratio estimator (histogram) in every panel.
A ratio of jointly normal estimates is itself skewed and
heavy-tailed (the gray text gives the mass beyond the axis endpoints). At
$L=20$ and $L=5$ the mode (vertical orange line) lies close to the true value (dashed) with a tight
bounded ensemble-level Fieller interval (green band); at $L=100$ the denominator
$\hat\epsilon$ is indistinguishable from zero, and the density develops Cauchy-like
tails (the tail mass halves with each doubling of the threshold as
$1/r^2$ decay predicts), and the Fieller interval is unbounded, indicating non-identifiability.}
\label{fig:ratio}
\end{figure*}

$\dd V_t=(V_t-V_t^3/3-W_t+I)\dt+\sigma_V\dd B_{1,t}$,
$\dd W_t=\epsilon(V_t+a-bW_t)\dt+\sigma_W\dd B_{2,t}$, with
$I=0.5,a=0.7,b=0.8,\epsilon=0.08,\sigma_V=\sigma_W=0.3$, simulated at
$\Delta t=0.05$, $N=100{,}000$ ($T=5000$), $R=0.05$; drift term candidates
$\{1,v,w,v^2,vw,w^2,v^3\}$, diffusion term candidate $h=[1]$ per
component, and a measured characteristic time $\tau\approx6$\,s set by the
excitable orbit. The slow equation's first stage clears the $F$ threshold at the
operative window (Table~\ref{tab:drift}), and the two depths agree within
uncertainty, so the default depth one applies here as well.
Table~\ref{tab:drift} shows the short and intermediate windows recovering
all four drift parameters, while the longest window destabilizes the slow
$(a,b)$ recovery. The $W$-equation is parameterized as $\epsilon(V+a-bW)$, so $a,b$ are
ratios of estimated coefficients with a near-zero denominator
$\epsilon\approx0.08$. A naive per-trajectory ratio has no finite
moments, so we report the Marsaglia--Hinkley ratio
mode~\cite{Marsaglia2006,Hinkley1969} with a Fieller~\cite{Fieller1954}
$95\%$ interval, both built from the IV sampling covariance
(Fig.~\ref{fig:ratio}). At
$L=20$ and $L=5$ the mode lies at the true value with a bounded interval; at $L=100$ the
Fieller interval is unbounded, a correct report of non-identifiability. Fig.~\ref{fig:ratio} shows the Marsaglia--Hinkley density and
ensemble Fieller intervals behind the $a,b$ entries of
Table~\ref{tab:drift}, at all three window lengths.

\subsubsection{Cox--Ingersoll--Ross (state-dependent diffusion)}
$\dd X_t=\kappa(\mu-X_t)\dt+\nu\sqrt{X_t}\,\dd W_t$, with
$(\kappa,\mu,\nu)=(2,1.5,1)$, $\Delta t=0.01$, $T=100$, $R=0.01$, for
which the diffusion
$\sigma^2(x)=\nu^2 x$ depends on the state itself. With the Feller condition $2\kappa\mu=6\gg\nu^2$
satisfied, the affine drift term candidates $g(x)=[1,x]$,
and the linear diffusion term candidate $h(x)=x$, the drift
coefficients $(\kappa\mu,-\kappa)$ are recovered as $3.14\,(1.09)$ and
$-2.10\,(0.75)$ against truth $(3,-2)$, and the diffusion coefficient as
$\widehat\nu=1.00\,(0.04)$ against the true value of $1$. The multilag estimator \eqref{eq:R-multilag}, applied to each trajectory on its own, returns $\widehat R=0.0100$ against $R=0.01$ on average. The diffusion side thus handles
non-constant $\sigma$ through its diffusion term candidates just as the drift side handles
non-constant $f$. At $L=100$ ($T_w=1.0$, twice $1/\kappa$) a few diverging records inflate the spread (Table~\ref{tab:drift}).

\subsubsection{Lorenz-96 (high dimensional)}
The cyclic system
$\dd x_i=\bigl[(x_{i+1}-x_{i-2})x_{i-1}-x_i+F\bigr]\dt+\sigma\,\dd W_i$ with
forcing $F=8$ and $\sigma=0.5$ ($\Delta t=10^{-3}$, $T=10$,
$R=10^{-3}$), is run at $d=40$ and $d=120$. The pooled
per-component drift term candidates are $[1,\;x_i,\;x_{i-1}x_{i+1},\;x_{i-1}x_{i-2}]$
under cyclic indexing, and the diffusion term candidate is $h=[1]$. The forcing and the three
structural coefficients are recovered tightly at both dimensions
($\widehat F=7.99\,(0.04)$ with structural terms within $0.2\%$ at $d=120$) along
with $\widehat\sigma=0.50$ and $\widehat R=1.00\times10^{-3}$ at both dimensions; at $d=120$ the learned coefficients are more accurate, because the larger system supplies more pooled windows.

\subsubsection{Stochastic Kuramoto network (non-polynomial drift term candidates)}
A network of $N=50$ phase oscillators,
$\dd\theta_i=\bigl[\omega+\tfrac{K}{N}\sum_j\sin(\theta_j-\theta_i)\bigr]\dt
+\sigma\,\dd W_i$, with $\omega=0.5$, $\sigma=0.3$, coupling
$K=2.0$ or $K=0.8$, $\Delta t=0.005$, $T=50$, and $R=10^{-3}$; in this
example only, $N$ counts oscillators and $K$ denotes the coupling, not the moment-system quantities of Section~\ref{sec:iv}. The
per-oscillator drift term candidates are
$[1,\;\overline S_s\cos\theta_i-\overline S_c\sin\theta_i]$, the
mean-field coupling expanded through the order-parameter components
$\overline S_s=\tfrac1N\sum_j\sin\theta_j$ and
$\overline S_c=\tfrac1N\sum_j\cos\theta_j$; the diffusion term candidate is
$h=[1]$. Across both the coherent
and the near-incoherent regime, the natural frequency
and coupling are recovered, $\widehat\omega=0.50\,(0.01)$, with
$\widehat K=2.01\,(0.10)$ and $0.80\,(0.06)$ respectively, $\widehat\sigma\approx 0.30$ and $\widehat R=1.00\times10^{-3}$ in both regimes, confirming that exogeneity and relevance are
properties of the windows, not of the basis.

\subsubsection{Underdamped Langevin (hypoelliptic)}
$\dd q=p\,\dt$, $\dd p=[\,q-q^3-\gamma p\,]\dt+\sigma\,\dd W$, run
at $(\gamma,\sigma)=(0.5,1)$ and $(2,2)$ with $\Delta t=0.005$, $T=100$,
$R=10^{-3}$, the hardest case of the nine. The $q$-equation carries no noise ($\sigma_q=0$), so the diffusion matrix
is rank-deficient and the system is only hypoelliptic. It also
carries two separated timescales, the fast conservative oscillation and the
slow damping time $1/\gamma$, and it is run at the common window $L=20$. The estimates are insensitive to this choice; at $L=50$ the friction estimate is $c_p=-0.51\,(0.25)$, against $-0.51\,(0.23)$ at $L=20$ ($\gamma=0.5$), the kinematic coefficient is more precise at the longer window ($0.012$ against $0.028$), and the $p$-equation terms are slightly more precise at the shorter one. The drift term candidates are $[p]$ for the $q$-equation and
$[q,\;q^3,\;p]$ for the $p$-equation, with diffusion term candidate $h=[1]$ per
component. The kinematic constraint
$\dd q=p\,\dt$ is recovered with coefficient $1.00\,(0.03)$,
the
potential and friction terms of the $p$-equation are recovered
($c_{q^3}=-1.01\,(0.22)$ and $c_p=-0.51\,(0.23)$ at $\gamma=0.5$, and
$c_p=-2.01\,(0.53)$ at $\gamma=2$, against truth $-1$ and $-\gamma$), and
$\widehat\sigma_p$ matches the truth under both diffusion forms, the lag-one quadratic variation \eqref{eq:qv-system} giving $1.00\,(0.01)$ at $\gamma=0.5,\sigma=1$ and $2.00\,(0.02)$ at $\gamma=2,\sigma=2$, and the two-lag slope $1.01\,(0.01)$ and $1.99\,(0.02)$. The multilag estimator \eqref{eq:R-multilag} returns $\widehat R=(1.00,\,1.00)\times10^{-3}$ for $(q,p)$ at $\gamma=0.5$ and $(1.00,\,1.01)\times10^{-3}$ at $\gamma=2$, against $10^{-3}$. The diffusion of $p$ uses the lag-one and lag-two variograms, in which $R$ cancels. These are IV estimates on the full set of drift term candidates. The
degenerate $\sigma_q$ is returned not as zero but as a small, predictable
finite-$\Delta t$ artifact $\approx\sqrt{3\,\E[p^2]\,\Delta t}$, which correctly
flags the component as near-degenerate instead of silently mis-estimating it.

\subsubsection{Lorenz-63 through a non-identity observation map}
This is the case of Remark~\ref{rem:obsC}, and the synthetic counterpart of
the optical-trap analysis of Section~\ref{sec:trap}. The drift and diffusion
are those of the stochastic Lorenz-63 system above
($\bm\sigma=(0.5,0.4,0.55)$), but the state is observed through
\begin{equation}\label{eq:lorenzC}
 C=\tfrac12\begin{pmatrix}1&1&0\\0&1&1\\0&0&1\end{pmatrix},
\end{equation}
taken from \cite{Crisan2022}, with per-sensor noise variance
$R_{y}=10^{-4}$. Here $d_y=d=3$ and $C$ is invertible but far from the
identity. The deconvolution
$\widehat X_{t_n}=(C^\top C)^{-1}C^\top Y_{t_n}$ of Remark~\ref{rem:obsC}
amplifies the sensor noise ($\|C^{-1}\|\approx3$ on the worst
component), so the effective per-state covariance
$R_{\mathrm{eff}}=(C^\top C)^{-1}C^\top R\,C(C^\top C)^{-1}$ has diagonal
$(1.2,0.8,0.4)\times10^{-3}$, an order of magnitude above the per-sensor
level. Sampling and term candidates follow the Lorenz-63 settings above ($\Delta t=10^{-3}$, $N=10^4$ estimation points, $n_{\rm sub}=10$,
meaning the estimation grid retains every tenth simulated step,
the degree-2 monomial drift term candidates and diffusion term candidate $h=[1]$),
all
corrections on, and the
standard settings
$L=20$, $\delta=4\Delta t$, $n_{\text{lags}}=1$.

Over $M=1000$ trajectories (all of which converged) the IV
estimates are $-10.02\,(0.83)$ and $10.00\,(0.53)$ for $x$ and $y$ in $\dot x$
(truth $-10$, $10$); $28.01\,(0.67)$, $-1.01\,(0.43)$ and $-1.001\,(0.019)$
for $x$, $y$ and $xz$ in $\dot y$ (truth $28$, $-1$, $-1$); and
$-2.65\,(0.63)$ and $0.998\,(0.078)$ for $z$ and $xy$ in $\dot z$ (truth
$-2.667$, $1$). The first-stage $F$ averages $64$. The multilag estimator
\eqref{eq:R-multilag} applied to the deconvolved series recovers
$R_{\mathrm{eff}}$ accurately, with mean $\mathrm{diag}( \widehat{R}_{\mathrm{eff}})=(1.20,\,0.79,\,0.39)\times10^{-3}$ over the trajectories. The diffusion, estimated by the drift-corrected quadratic variation
\eqref{eq:qv-system} with the estimated $\widehat R_{\mathrm{eff}}$,
feasible here since $\varrho\approx10$ lies within the operating
range, is recovered as
$\widehat{\bm\sigma}=(0.489,0.412,0.564)$ against $(0.5,0.4,0.55)$, with
standard deviations $(0.095,0.072,0.032)$; the slope form gives
$(0.457,0.436,0.544)$ with standard deviations $(0.275,0.312,0.360)$,
uninformative trajectory by trajectory on this short, deconvolved
record. Every coefficient of a term absent from the true model is estimated without detectable bias, the average estimate across the $23$ such coefficients lying within $1.9$ Monte Carlo standard errors of zero. The accuracy matches that of the directly observed Lorenz-63 experiment. The noise amplification inherent in inverting
$C$ costs precision on the linear rates, with standard deviations of order
$0.5$ to $1$ instead of a few tenths, but neither biases the estimator nor
degrades the quadratic terms $xz$ and $xy$, which remain the best-determined
coefficients in the system. A non-identity observation map is, for this
method, a change of variables and a recalibration of the noise covariance,
not a change of estimator. At $L=5$ the seven coefficients are recovered with about twice the spread of $L=20$, at $L=100$ the deconvolved record is unstable (Table~\ref{tab:drift}), and at the stress level the first stage collapses at every window and depth, on records of $N=2\times10^5$ points as well as $N=10^4$ (Table~\ref{tab:stressrescue}).

For the remaining systems no comparably standard alternative exists, since exact likelihoods for noisily observed records are confined to linear Gaussian models such as OU, and particle or ensemble filters for nonlinear, chaotic, or high-dimensional dynamics require tuning and computation far beyond a single linear solve. By analogy with the
two cases above, we expect a well-tuned likelihood method, where it can
be run, to be more precise than \textsc{WAIVES} and substantially
slower, and strong-form regression to be biased by the same
errors-in-variables mechanism.

\subsection{Drift and diffusion recovery}
\begin{table*}[t]
\centering
\caption{ Drift recovery for all nine systems at the main-analysis noise level (using the fixed variance $R$ per system of Table~\ref{tab:noiselevels}). Entries are the estimator average and standard deviation in parentheses, taken over $M=1000$ single-trajectory estimates ($M=200$ for Lorenz-96), each using its trajectory’s own $\widehat R$, at raw IV depth one, with gap $\delta=4\Delta t$ (depth-two comparisons appear in Table~\ref{tab:stressrescue}). The columns compare the window lengths $L=100$, $20$ (the operative window), and $5$, with true values in the last column. The FitzHugh--Nagumo ratio parameters $a,b$ are reported as the Marsaglia--Hinkley mode with a $95\%$ Fieller interval (unbnd'' means the interval is unbounded, indicating non-identifiability). The last row of each block gives the average first-stage $F$, with the percentage of replications whose own $F$ falls below the weak-instrument threshold $10$ in parentheses. Where the average $F$ falls below $1$, the first stage carries less signal than noise and we regard the inference as unreliable. These cells read n.i.\ (not identified). Note that at depth one the exactly identified estimator becomes heavy-tailed, and at higher depths the estimates collapse toward the least-squares limit, so no stable summary of the replications exists (Appendix~\ref{si:coverage}). System abbreviations follow Table~\ref{tab:systems}. Stress-level recovery appears in Table~\ref{tab:stressrescue}.}
\label{tab:drift}
\footnotesize

\begin{tabular}{llcccc}
\toprule
System & Parameter & $L=100$ & $L=20$ & $L=5$ & True\\
\midrule
\multirow{3}{*}{OU}
 & $\kappa$ & 1.532\,(0.331) & 1.497\,(0.126) & 1.529\,(0.225) & 1.500\\
 & $\mu$ & 1.999\,(0.026) & 2.001\,(0.029) & 2.000\,(0.060) & 2.000\\
 & 1st-stage $F$ & 77\,(0\%) & 4945\,(0\%) & 36560\,(0\%) & ---\\
\midrule
\multirow{3}{*}{DW}
 & $\alpha$ & 1.114\,(0.599) & 1.003\,(0.149) & 1.012\,(0.336) & 1.000\\
 & $\beta$ & 1.109\,(0.565) & 1.004\,(0.134) & 1.011\,(0.289) & 1.000\\
 & 1st-stage $F$ & 6.2\,(89\%) & 586\,(0\%) & 3231\,(0\%) & ---\\
\midrule
\multirow{4}{*}{L63}
 & $a$ & n.i. & 10.028\, (0.265)  & 10.051 \,(0.765) & 10.000\\
 & $b$ & n.i. & 28.027\,(0.355) & 28.215\,(1.158) & 28.000\\
 & $c$ & n.i. & 2.721\,(0.313) & 2.740\,(0.864) & 2.667\\
 & 1st-stage $F$ & $<1$\,(100\%) & 228\,(0\%) & 4286\,(0\%) & ---\\
\midrule
\multirow{5}{*}{FHN}
 & $I$ & n.i. & 0.505\,(0.099) & 0.505\,(0.184) & 0.500\\
 & $\epsilon$ & n.i. & 0.075\,(0.150) & 0.078\,(0.181) & 0.080\\
 & $a$ & n.i. & 0.684\,[0.61,0.77] & 0.655\,[0.53,0.79] & 0.700\\
 & $b$ & n.i. & 0.860\,[0.80,0.94] & 0.749\,[0.64,0.88] & 0.800\\
 & 1st-stage $F$ & $<1$\,(100\%) & 16\,(2\%) & 233\,(0\%) & ---\\
\midrule
\multirow{2}{*}{CIR}
 & $\kappa\mu$ & $-1.46$\,(228.02) & 3.14\,(1.09) & 3.12\,(1.27) & 3.00\\
 & $-\kappa$ & 0.91\,(149.38) & $-2.10$\,(0.75) & $-2.09$\,(0.87) & $-2.00$\\
\midrule
L96 ($d=120$)
 & $F$ (forcing) & 8.00\,(0.04) & 7.99\,(0.04) & 8.03\,(0.09) & 8.00\\
\midrule
\multirow{3}{*}{Kuramoto}
 & $\omega$ & 0.50\,(0.01) & 0.50\,(0.01) & 0.50\,(0.03) & 0.50\\
 & $K$ (coh.) & 2.00\,(0.16) & 2.00\,(0.10) & 2.01\,(0.18) & 2.00\\
 & $K$ (incoh.) & 0.80\,(0.07) & 0.80\,(0.06) & 0.81\,(0.11) & 0.80\\
\midrule
\multirow{4}{*}{Langevin}
 & $p$ in $\dot q$ & 1.00\,(0.01) & 1.00\,(0.03) & 1.01\,(0.11) & 1.00\\
 & $c_{q^3}$ & $-1.00$\,(0.52) & $-1.01$\,(0.22) & $-1.00$\,(0.26) & $-1.00$\\
 & $c_p$ ($\gamma=0.5$) & $-0.53$\,(0.32) & $-0.51$\,(0.23) & $-0.51$\,(0.29) & $-0.50$\\
 & $c_p$ ($\gamma=2$) & $-2.07$\,(1.11) & $-2.01$\,(0.53) & $-2.07$\,(0.56) & $-2.00$\\
\midrule
\multirow{7}{*}{\shortstack[l]{L63\\$Y=CX+\eta$}}
 & $x$ in $\dot x$ & $-8.67$\,(30.57) & $-10.02$\,(0.83) & $-10.22$\,(1.94) & $-10.00$\\
 & $y$ in $\dot x$ & 8.48\,(30.23) & 10.00\,(0.53) & 10.15\,(1.20) & 10.00\\
 & $x$ in $\dot y$ & 28.46\,(14.21) & 28.01\,(0.67) & 28.28\,(1.53) & 28.00\\
 & $y$ in $\dot y$ & $-1.157$\,(15.424) & $-1.012$\,(0.426) & $-1.083$\,(0.952) & $-1.000$\\
 & $xz$ in $\dot y$ & $-1.014$\,(0.463) & $-1.001$\,(0.019) & $-1.009$\,(0.042) & $-1.000$\\
 & $z$ in $\dot z$ & $-0.212$\,(40.889) & $-2.647$\,(0.633) & $-2.618$\,(1.000) & $-2.667$\\
 & $xy$ in $\dot z$ & 0.729\,(4.219) & 0.998\,(0.078) & 1.001\,(0.117) & 1.000\\
\bottomrule
\end{tabular}
\end{table*}
\begin{table}[t]
\centering
\caption{Diffusion recovery at the operative window $L=20$ and the main-analysis noise level (using the fixed variance $R$ per system of Table~\ref{tab:noiselevels}). Entries are the estimator average and standard deviation in parentheses, taken over $M=1000$ single-trajectory estimates ($M=200$ for Lorenz-96), each using its trajectory’s own $\widehat R$, at depth one. QV denotes the lag-one quadratic-variation system \eqref{eq:qv-system}, and slope denotes the multilag slope \eqref{eq:qv-vgram}, replaced by the two-lag slope for the Langevin $p$ component. True values appear in the second column. System abbreviations follow Table~\ref{tab:systems}.}
\label{tab:sigma}
\footnotesize
\begin{tabular}{lccc}
\toprule
System & $\sigma$ (true) & QV \eqref{eq:qv-system} & slope \eqref{eq:qv-vgram}\\\\
\midrule
OU & 0.800 & 0.800\,(0.020) & 0.800\,(0.021)\\
DW & 0.500 & 0.498\,(0.032) & 0.499\,(0.027)\\
FHN ($V$) & 0.300 & 0.309\,(0.037) & 0.296\,(0.032)\\
FHN ($W$) & 0.300 & 0.302\,(0.038) & 0.300\,(0.030)\\
L63 ($x$) & 1.000 & 1.001\,(0.210) & 1.056\,(0.196)\\
L63 ($y$) & 1.000 & 1.075\,(0.183) & 1.249\,(0.210)\\
L63 ($z$) & 1.000 & 1.080\,(0.185) & 1.286\,(0.240)\\
\midrule
CIR ($\nu$) & 1.000 & 1.00\,(0.04) & 0.996\,(0.086)\\
L96 ($d=40$) & 0.500 & 0.502\,(0.005) & 0.501\,(0.005)\\
L96 ($d=120$) & 0.500 & 0.501\,(0.002) & 0.501\,(0.002)\\
Kuramoto ($K=2.0$) & 0.300 & 0.300\,(0.002) & 0.300\,(0.002)\\
Kuramoto ($K=0.8$) & 0.300 & 0.300\,(0.002) & 0.300\,(0.002)\\
Langevin ($p$, $\gamma=0.5$) & 1.000 & 1.001\,(0.012) & 1.013\,(0.011)\\
Langevin ($p$, $\gamma=2$) & 2.000 & 2.000\,(0.024) & 1.993\,(0.018)\\
L63 via $C$ ($x$) & 0.500 & 0.489\,(0.095) & 0.457\,(0.275)\\
L63 via $C$ ($y$) & 0.400 & 0.412\,(0.072) & 0.436\,(0.312)\\
L63 via $C$ ($z$) & 0.550 & 0.564\,(0.032) & 0.544\,(0.360)\\
\bottomrule
\end{tabular}
\end{table}
Table~\ref{tab:drift} reports the drift recovery for every parameter of the nine systems at the three window lengths. The quadratic-variation estimator \eqref{eq:qv-system} recovers the diffusion
magnitude alongside the drift. Table~\ref{tab:sigma} reports
$\widehat\sigma$ at the operative window $L=20$, with all corrections of \eqref{eq:qv-system}
enabled for every system, alongside the multilag slope form of
Section~\ref{sec:iv}. The noise level that the debiasing and
the quadratic-variation forms use is each trajectory's own multilag
estimate \eqref{eq:R-multilag}. Averaged over trajectories, it returns $\widehat R=0.0200$ for the
OU and double-well records against $R=0.02$,
$(0.0500,\,0.0498,\,0.0498)$ for Lorenz-63 and $(0.0500,\,0.0500)$ for
FitzHugh--Nagumo against $0.05$. At the stress level of
Table~\ref{tab:stressrescue} it reproduces the fixed stress variances to
within $0.03\%$ for every component. The slope form recovers every dissipative-model component within
Monte Carlo error. The lag-one form matches it except on the
FitzHugh--Nagumo $V$ component; there $\varrho\approx22$ approaches the
edge of the operating regime and the lag-one estimate is biased upward by
$3\%$. Lorenz-63 lies at $\varrho\approx50$, where the variogram still
recovers $\widehat R=0.0499$ against $0.05$, but the $y$ and $z$
diffusions are biased upward by $7$--$8\%$ in
the lag-one form and $25$--$29\%$ in the slope form (ten or more Monte
Carlo standard errors; Table~\ref{tab:sigma}). The lag-one $x$ is exact and the slope $x$ is within
$6\%$. The excess on the slope reflects drift contamination of the longer
lags at this extreme $\varrho$, which the corrections of
\eqref{eq:qv-system} remove only to leading order. On the dissipative
systems the slope's variance is comparable to the lag-one form.

\subsection{Robustness to observation noise}
Tables~\ref{tab:stressrescue} and~\ref{tab:sigmastress} repeat every drift and diffusion estimate of all nine systems at a
``stress'' noise level, a fixed variance $R_{\mathrm s}$ per component
equal to the square of $15\%$ of the component's range over a record of the
length used in the experiments (Table~\ref{tab:noiselevels}; galleries in Fig.~\ref{fig:noise}). The OU and
double-well estimates stay unbiased, with $1.9$ and $3.7$ times their spread
at the main-analysis noise level. For FitzHugh--Nagumo and Lorenz-63 the
depth-one first stage collapses ($F<1$). The depth rule restores bounded
FitzHugh--Nagumo intervals, while the Lorenz-63 estimates stay centered
near the truth but with spreads too large to be informative (depth-two
$F\approx1$). The diffusion
estimators lose precision as Section~\ref{sec:iv} anticipates, up to the
point where the Lorenz-63 diffusion is no longer identifiable from the
increments.

The resulting choices of span, gap, depth, and slope lags are collected
as a design rule in Appendix~\ref{si:choice}.

\section{Real-data illustrations}
\label{sec:realdata}

The example systems above validated \textsc{WAIVES} against parameters
known by design. We now apply \textsc{WAIVES} to two real data sets to
demonstrate the method's ease of application and its usefulness in
practice. The first is the unsmoothed Atlantic Multidecadal Oscillation (AMO)
index, the area-averaged, detrended North Atlantic sea-surface-temperature
anomaly \cite{Enfield2001,Kaplan1998}. It is an observational record with
no ground truth and no tractable likelihood, where competing model classes
must be fitted and compared, highlighting the need for a fast, assumption-light estimator such as
\textsc{WAIVES}. The second is the calibration of an optical trap~\cite{Gieseler2021}, where the diffusion coefficient is fixed by first principles and the drift is measured independently by standard calibration, so \textsc{WAIVES} can be checked against known physics.

\subsection{North Atlantic sea-surface temperature}
\label{sec:amo}
\begin{figure*}[t]
\centering
\includegraphics[width=\linewidth]{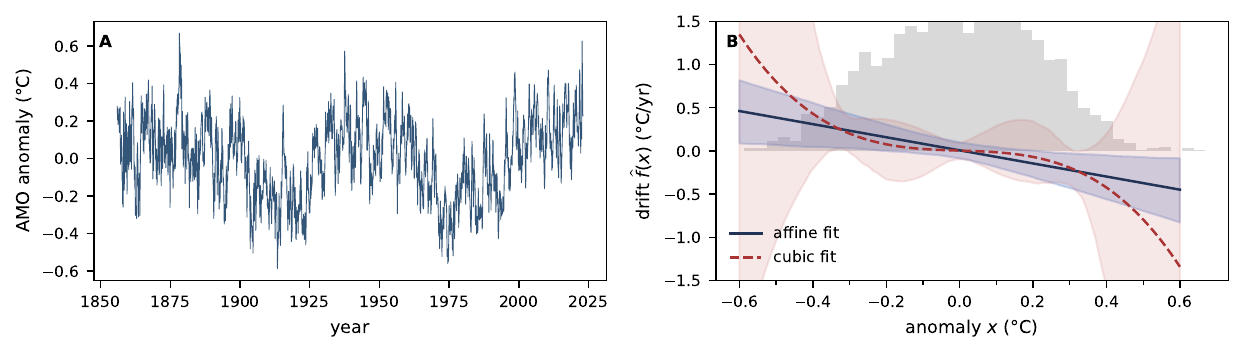}
\caption{\textsc{WAIVES} applied to the unsmoothed AMO index.
(\emph{A}) The deseasonalized monthly series, 1856--2023. (\emph{B}) The
estimated drift at the operative window: the affine fit (solid) with its
$95\%$ window-pair bootstrap band (shaded), the cubic-drift fit (dashed)
with its band (lighter shading), and the empirical distribution of the
observed anomalies (gray histogram). The affine fit's slope is
$-\widehat\kappa$, where the restoring rate $\kappa$ measures how strongly
an anomaly is pulled back toward the long-term mean ($\widehat\kappa = 0.76$ with $95\%$ Wald interval
of $[0.20, 1.31]$.) The bootstrap bands are shown as an additional uncertainty-quantification mode, serving as a diagnostic for unmodeled structure (in the gap and span sweeps of this section, an estimate that wanders outside of its bootstrap bands  would flag a misspecified drift, as the residual seasonal cycle did for the intercept). Bistability, which is defined as two preferred AMO states, would appear as
a cubic drift bending away from the affine fit. However, over the range of observed
states the two fits coincide within their uncertainty bands, and the record cannot
distinguish a single well from a shallow double well.}
\label{fig:amo}
\end{figure*}

Beyond being a canonical long climate series ($N=2005$ monthly values,
1856--2023, $\Delta t=1/12$\,yr), the AMO index is at the center of an
active debate about the stability of the Atlantic meridional overturning
circulation. In that debate, competing stochastic descriptions (monostable
models with a single restoring equilibrium, and bistable or tipping-prone
alternatives) are fitted to closely related sea-surface-temperature
records and compared \cite{Boers2021,Ditlevsen2023,Morr2026}. We use this data set to demonstrate \textsc{WAIVES}, not to enter that debate; the index is a temperature proxy whose relationship to the overturning circulation is itself contested~\cite{BenYami2024,Cotronei2026,Mann2021}.

These data have several noteworthy features, including very low
measurement noise, as expected for an index averaged over an entire
ocean basin (the variogram estimator of Section~\ref{sec:iv} yields
$\widehat R^{1/2}=0.006\,^\circ$C). With the observation error effectively absent, the
dominant bias mechanism is expected to stem from the shared-driving-noise endogeneity, which makes the record a well-suited application for
\textsc{WAIVES}.
In addition, a residual annual cycle persists in these data, even after the standard anomaly processing. This cycle is also visible in the gap-sweep diagnostic plots in Fig.~\ref{fig:amo-si}, where the drift intercept varies with the gap $\delta$ in a systematic,
calendar-dependent pattern. 
Removing the monthly climatology yields an intercept statistically
indistinguishable from zero while leaving the restoring rate essentially
unchanged (Section~\ref{amo:season} and Fig.~\ref{fig:amo-si}).
The relevance diagnostic detects this as well, so all subsequent analyses use the deseasonalized series.

We thus model the deseasonalized anomaly $X_t$ as an SDE with unknown drift and
constant unknown diffusion, beginning with the affine drift term candidates $g(x)=[1,x]$ and
diffusion term candidate $h=[1]$, that
is, the OU model $\dd X_t=-\kappa(X_t-\mu)\dt+\sigma\dW$. Here $1/\kappa$
is the relaxation time, and the mean-reversion rate $\kappa$ controls how
strongly the anomaly is pulled toward its equilibrium. We choose the
window span by sweeping over the
spans of $0.5$--$3$\,yr, from well inside to just beyond the record's
$\approx2.5$\,yr decorrelation time, the comparison that
Proposition~\ref{prop:rel} ties instrument informativeness to. We select the
operative span by the first-stage strength $F$ of Section~\ref{sec:numerics}, which is
maximized at the shortest span of $0.5$\,yr ($F=197$, satisfying $F\ge10$). Confidence intervals are Wald intervals
from the plug-in covariance (Appendix~\ref{si:coverage}).

The resulting estimates are summarized in Fig.~\ref{fig:amo}. At the operative window, the IV estimate of the mean-reversion rate is
$\widehat\kappa=0.76$ (95\% CI
$[0.20,1.31]$)\,yr$^{-1}$. The anomaly relaxes toward
a single equilibrium, statistically indistinguishable from zero, on
a relaxation time of $1/\widehat\kappa\approx1.3$\,yr. The diffusion
estimate is $\widehat\sigma=0.30\,^\circ$C\,yr$^{-1/2}$, with two
independent estimators (the multilag slope \eqref{eq:qv-vgram} and the
quadratic-variation estimator \eqref{eq:qv-system}) agreeing to within
$7\%$.

We check the fitted model in three ways. First, an OU
process with parameters set to these estimates has stationary standard
deviation $\widehat\sigma/\sqrt{2\widehat\kappa}=0.25\,^\circ$C,
close to the observed $0.21\,^\circ$C, the record's sample standard
deviation. Second, the fitted model implies a correlation time near $1.3$\,yr, whereas the record's measured $1/e$ time is $2.5$\,yr, so the index superposes several timescales and the operative-span estimate targets the fast, short-time relaxation. Third, 
repeating the
estimation on simulated OU records of the same length, rate, and
diffusion, with observation noise at the record's scale, confirms that the Wald
interval attains $93\%$ coverage (Section~\ref{amo:synth}).
Weak-form least squares on the same moment system, by contrast, fails to
identify the restoring rate. On the real record its estimates flip sign across the span sweep,
and on matched synthetic records it is driven to zero at every span, as
Proposition~\ref{prop:ols-zero} implies (Section~\ref{amo:synth}).

Finally, we examine the bistability hypothesis directly. Bistability here
means that the system possesses two stable equilibria (two coexisting
climate states), so that noise or slow forcing could tip it from one to the
other. The minimal drift with this property is cubic:
$f(x)=\theta_1x+\theta_3x^3$ with $\theta_1>0$ and $\theta_3<0$ has stable
states at $x_*=\pm\sqrt{\theta_1/|\theta_3|}$ separated by an unstable
equilibrium at zero (a ``double well''). In contrast, the linear drift $-\kappa x$
(with $\kappa>0$) always restores toward the single equilibrium state at
zero. Enlarging the drift term candidates to $[x,x^3]$ therefore allows the data to
discriminate between the two hypotheses, with bistability corresponding to a
negative $\theta_3$ of sufficient magnitude. The statistical inference question is then whether
the record can distinguish such a value from zero. The resulting 95\%
interval, $\theta_3\in[-38,+21]$, admits both hypotheses at the same time.
It contains zero, so a purely linear drift cannot be rejected. However, it
also contains, for example, $\theta_3=-9$, which corresponds to a double
well calibrated to the AMO record itself, with stable states one sample
standard deviation from zero ($\pm0.21\,^\circ$C) and a near-well
relaxation time corresponding to that of the fitted model, $1/\widehat\kappa$
(Fig.~\ref{fig:amo}B, with the derivation in Section~\ref{amo:power}). In fact, stable states anywhere between $0.5$ and
$1.5$ sample standard deviations imply $|\theta_3|$ between roughly $34$
and $4$, all inside the interval. Furthermore, a synthetic power analysis (Section~\ref{amo:power}) indicates that no estimator built
on these moment conditions could resolve a cubic component of that
magnitude from a single 167-year record. This conclusion is also supported
by a GMM model comparison. The linear model passes the Hansen
$J$-test~\cite{Hansen1982} ($J=0.73$, $p=0.39$), with no evidence of
misspecification in its over-identified moment system. Similarly, the
moment-selection criteria of Andrews and Lu~\cite{AndrewsLu2001}, which can
be viewed as the GMM counterparts of BIC and AIC, favor the linear model (Section~\ref{amo:model}). This preference may reflect parsimony rather than evidence against bistability; in synthetic double-well records the same criteria still select the linear model in $99\%$ of replicates, so the comparison has very low power at this record length. The instrumental record, analyzed through this model
class, therefore can neither support nor rule out bistability.

We emphasize that our analysis is complementary to the recent early-warning
analyses of AMOC stability~\cite{Boers2021,Ditlevsen2023}. Those studies
ask
whether the stability of the
currently occupied state has been declining over time as a bifurcation is approached. In contrast, we ask whether the time-averaged short-time drift of the
basin-mean AMO index, over the range of states the record contains, supports
one restoring equilibrium or two. These two questions are complementary, and a single well in the stationary short-time drift can be fully consistent with a slow approach to a fold. Our practical non-identifiability finding may, however, be directly
relevant to that debate: projected tipping times are known to be sensitive
to the assumed form of the drift~\cite{BenYami2024,Morr2026}, and the power
analysis above underlines that records of this
length cannot distinguish the candidate drift structures.

\subsubsection{Data and preprocessing}
\label{amo:data}
We use the unsmoothed, long AMO index from NOAA PSL (Kaplan SST V2 source
\cite{Kaplan1998,Enfield2001}), as monthly values January 1856 through January
2023 ($N=2005$; the series is no longer extended because its source data set
was frozen). The published series is the area-weighted North Atlantic SST
anomaly, linearly detrended by the provider. We do not smooth it, since
smoothing would destroy the quadratic-variation structure on which the
diffusion estimator and the noise-variance estimate rely. Time is measured in
years, $\Delta t=1/12$.

\subsubsection{Residual seasonality and the gap-sweep diagnostic}
\label{amo:season}
\begin{figure}[t]
\centering
\includegraphics[width=\linewidth]{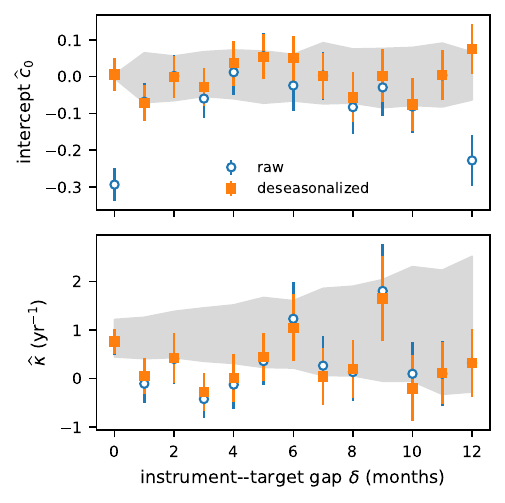}
\caption{The gap sweep as a built-in misspecification check. Sweeping the
instrument--target gap $\delta$ at $L=6$ rotates the calendar phase of the
window pairs (pair span $2L+\delta/\Delta t$ steps) through any residual
annual cycle. A correctly specified, time-homogeneous drift does not
depend on $\delta$. The shaded bands show the central $80\%$ of the same
estimates on matched synthetic records without seasonality, and the error
bars are window-pair bootstrap intervals. Because all markers derive from
the same record, they are strongly correlated across gaps and will tend to
wander together. Thus, the diagnostic is a systematic dependence on
$\delta$, for example a cluster of estimates pulled far from the null at
particular gaps. The error bars, which describe each estimate's own
uncertainty, may likewise extend beyond the band without indicating
misspecification. \emph{Top:} on the raw series (open markers) the drift
intercept $\widehat c_0$ is displaced far outside the null at the
calendar-aligned gaps ($\delta=0,12$ months): at these gaps the pair span
is a whole number of years, so every pair occupies the same calendar
months and the cycle's contribution adds coherently across pairs instead
of averaging out, displacing the intercept. Removing the monthly
climatology (filled markers) reduces the $\delta$-dependence. \emph{Bottom:}
the slope $\widehat\kappa$ is essentially unchanged by deseasonalization
and shows no systematic dependence on $\delta$, with its markers tracking
the null band across the sweep.}
\label{fig:amo-si}
\end{figure}

Every choice of the gap $\delta$ between the instrument and target windows
defines a valid estimator of the same parameters. All such estimators share
one probability limit, but not one finite-sample distribution, because
instrument relevance decays with $\delta$ and the sampling scatter grows
accordingly. The specification comparison is therefore made against a null distribution that carries this gap dependence, the gray band of Fig.~\ref{fig:amo-si}, which at each gap represents the central range of estimates obtained by repeating the entire analysis, including deseasonalization, on records simulated from the fitted model. On the raw series, the drift intercept
departs from this null, varying with $\delta$ in a systematic,
calendar-dependent pattern.
Calendar-month means range from
$-0.035$ to $+0.035\,^\circ$C, with the seasonal decline from September to
December steepest. On the raw series this deterministic component
contaminates the weak-form moments. The pair span is
$(n_{\mathrm{lags}}+1)L+\delta/\Delta t$ observations, so
sweeping the gap $\delta$ rotates the calendar phase of
the window pairs through the annual cycle, and the drift intercept traces
that cycle. At calendar-aligned spans, where every target window covers the
same months, the bump-weighted autumn decline loads coherently onto the
intercept, biasing it to $-0.29\,^\circ$C\,yr$^{-1}$,
six standard errors from zero and far outside the matched
synthetic null
(Fig.~\ref{fig:amo-si}, top). Removing the monthly pattern (subtracting
each calendar month's mean) restores an intercept statistically
indistinguishable from zero at every window and at the operative
configuration ($\widehat c_0=0.01$, s.e.\ $0.05$), within
the synthetic null across the gap sweep. The slope $\widehat\kappa$ is almost unaffected by deseasonalization (Fig.~\ref{fig:amo-si}, bottom), because the annual cycle enters through the level moments, to which the intercept is sensitive, while the relaxation rate is identified by the state-dependent moments, which the cycle leaves nearly untouched. The remaining
scatter of $\widehat\kappa$ across gaps, lag counts, and record halves is
consistent with the null distribution obtained from matched synthetic
records. The observed across-gap spread and the first/second-half
difference lie at the 85th--90th percentile of that null, so we find no
evidence of residual misspecification.

\subsubsection{Estimation settings}
\label{amo:settings}
All settings follow the analysis above, namely the affine drift term candidates $[1,x]$,
enlarged to $[1,x,x^3-3\widehat R\,x]$ for the model comparison; diffusion
term $h=[1]$; bump banks $p\in\{4,6,8\}$; depth $n_{\mathrm{lags}}=1$;
the grid-sum quadrature of Eqs.~(\ref{eq:obs-b})--(\ref{eq:obs-Z}); polynomial debiasing with the variogram
$\widehat R$; Wald intervals from the plug-in covariance
(Appendix~\ref{si:coverage}) for the operative-window
estimates; and a window-pair bootstrap with $4000$ resamples for the
diagnostics at longer spans (span sweep, sliding windows, and
model-comparison null bands). One setting departs from the benchmark conventions, the gap, which is $\delta=0$ here instead of $4\Delta t$ because the record is short and every observation spent on a buffer costs window pairs. At $\delta=0$ the paired windows share only the boundary grid point, where every test function vanishes, so the discrete sums remain disjoint. Furthermore, the gap-sweep diagnostic of the preceding paragraph indicates
that a larger buffer would add little robustness. Finally, the bootstrap resamples window pairs independently; the pairs
are disjoint, and dependence between neighbouring pairs enters only
through the state shared across their common boundary. On AMO-matched
synthetic records this bootstrap is conservative at the longer spans,
where the first stage is weak (coverage $96$--$98\%$ at median $F\le9$),
while the Wald interval stays calibrated down to $F\approx3$
(Appendix~\ref{si:coverage}). Windows of
$L\in\{6,9,12,18,24,36\}$ observations give spans $0.5$--$3$\,yr, bracketing
the record's $1/e$ autocorrelation time of $2.5$\,yr. The operative window
$L=6$ has the strongest first stage, $F=197$ versus $F=11$--$65$ at longer
spans (equivalently, a first-stage $R^2$, the share of regressor variance
the instruments explain, of $0.55$ versus $0.27$--$0.37$), and the smallest
standard error.

\subsubsection{Robustness at the operative window}
\label{amo:robust}
The baseline estimate is $\widehat\kappa=0.76$ (window-pair bootstrap
s.e.\ $0.30$; the standard errors in this subsection are bootstrap ones,
while the interval quoted above is Wald, s.e.\ $0.28$). Varying the test-function bank moves it well within sampling variation, to $0.75\,(0.34)$ with bump powers $\{6,8,10\}$ and $0.72\,(0.23)$ with $\{2,3,4\}$.
Deepening or displacing the instrument instead weakens the estimate
substantially, to $0.07\,(0.34)$ at $n_{\mathrm{lags}}=2$ and
$0.19\,(0.60)$ at gap $\delta=8$ months. Both changes are expected on a
record this short, where the added block or buffer pushes the instrument
outside the system's memory and the first stage collapses
(Proposition~\ref{prop:rel}). Splitting the record gives $1.00\,(0.46)$ on
1856--1939 and $-0.14\,(0.45)$ on 1940--2023. Each of these differences
lies within the across-configuration variation of the matched synthetic
null quantified in the preceding paragraph, at the 85th--90th percentile at
the extremes, and the lower second half is consistent with the
sliding-window view below.

\subsubsection{Synthetic calibration and coverage}
\label{amo:synth}
We simulate $M=800$ OU records with the estimated parameters ($\kappa=0.77$, $\sigma=0.30$, $N=2005$, $\Delta t=1/12$) and noise variance $R=3\times10^{-4}$, about seven times the record's $\widehat R\approx4\times10^{-5}$ (repeating the calibration at the record's own $\widehat R$ changes none of the summaries), apply the identical procedure, including per-record climatology removal, and recover $\widehat\kappa$ with mean $0.81$, median $0.80$, and standard deviation $0.32$. Two conclusions follow. First, the
real-data Wald standard error ($0.28$) is close to the true sampling
standard deviation, and the nominal-95\% Wald interval attains $93\%$
coverage.
Second, the six-point window carries only a small
($\approx5\%$) quadrature bias at this rate, negligible against the
sampling spread. We report the
uncorrected estimate. The weak-form OLS estimate on the same synthetic records is $0.00$
(s.d.\ $0.19$) at every window span from $0.5$ to $3$\,yr, a complete
cancellation, as Proposition~\ref{prop:ols-zero} requires. On the real
record, four of the six spans fall within this zero-centered null, while
the shortest and the two-year spans exceed it ($+0.61$ and $+0.67$).
Proposition~\ref{prop:ols-zero} applies to any stationary diffusion,
regardless of timescale structure, so these excursions are not a
multiscale effect. They are consistent with sampling variability under
the strong dependence of estimates across overlapping spans, or with mild
nonstationarity of the record, not with a recoverable drift signal. The OLS values remain mutually
inconsistent across spans, flipping sign, and identify no rate.

\subsubsection{Weak-form model comparison, linear versus cubic drift}
\label{amo:model}
\textsc{WAIVES} never forms a likelihood, so classical information criteria
are unavailable. The GMM framework supplies the standard substitutes. Taking
the weak-form integrals of $[1,\,x,\,x^3-3\widehat R x]$ over the
instrument window as instruments ($m=3$ moment conditions per test function
and window pair), the linear drift model ($p=2$) is over-identified by one
condition while the cubic model ($p=3$) is exactly identified. Two-step
efficient GMM with window-pair--clustered weighting yields the Hansen
$J$-statistic \cite{Hansen1982}. Because the models are nested and share
instruments and weighting, $J_{\mathrm{lin}}-J_{\mathrm{cub}}
=J_{\mathrm{lin}}$ is the GMM distance test of $\theta_3=0$
\cite{NeweyWest1987}, and the Andrews--Lu moment-selection criteria
\cite{AndrewsLu2001},
$\mathrm{MMSC\text{-}BIC}=J-(m-p)\ln N_w$ and
$\mathrm{MMSC\text{-}AIC}=J-2(m-p)$, evaluated per model, (smaller preferred), play the
roles of BIC and AIC. On the AMO record at the operative window
($N_w=166$ pairs), the cubic-model GMM estimate
$(\theta_1,\theta_3)=(-0.13,-5.9)$ agrees with the exactly identified IV
estimate with the enlarged set of drift term candidates. The linear model passes the specification test
($J=0.73$, one degree of freedom, $p=0.39$); and both criteria prefer the
linear model ($\mathrm{MMSC\text{-}BIC}$ of $-4.4$ versus $0$;
$\mathrm{MMSC\text{-}AIC}$ of $-1.3$ versus $0$). Validation on matched
synthetics (300 replicates per scenario, full estimation procedure, including
deseasonalization) shows the procedure is well calibrated but, at this
record length, weakly powered. Under a true linear drift the nominal-5\%
$J$-test rejects in $5\%$ of replicates and MMSC-BIC selects the cubic
model in $1\%$; under true double wells with $|\theta_3|=9$ (minima at
$\pm1$ standard deviation) and $|\theta_3|=4$ ($\pm1.5$ standard
deviations), the $J$-test rejects in only $3\%$ and $4\%$, and MMSC-BIC
selects the cubic model in $1\%$ at either magnitude. The real-data preference for
the linear drift therefore reflects parsimony at low power, not evidence
against bistability.

\subsubsection{Power analysis for the cubic component}
\label{amo:power}
On synthetic records matched to the AMO scales with a purely linear drift,
the single-record IV estimate of the cubic coefficient is
spread over several tens in magnitude at the operative window. A
double well with minima at $\pm1$ standard deviation ($\pm0.21\,^\circ$C)
and near-well relaxation matching $\widehat\kappa$ requires
$|\theta_3|=\widehat\kappa/(2x_*^2)\approx9$; wells at $\pm1.5$ standard
deviations require $|\theta_3|\approx4$. Both magnitudes are inside the
single-record sampling spread, and the real-data 95\% interval
$[-38,+21]$ contains them together with zero. The cubic point estimate is itself monostable,
and its bootstrap band  contains the affine fit
(Fig.~\ref{fig:amo}B). In this case, the binding constraint is the record length, not the estimator. Distinguishing the record-calibrated $\theta_3=-9$ from zero at the $95\%$ level would require the standard error reduced by a factor of three, and with the $1/\sqrt{T}$ scaling of Theorem~\ref{thm:clt} that means a record roughly ten times longer, on the order of two millennia.

\subsubsection{A sliding-window view of the restoring rate}
\label{amo:sliding}
\begin{figure}[t]
\centering
\includegraphics[width=\linewidth]{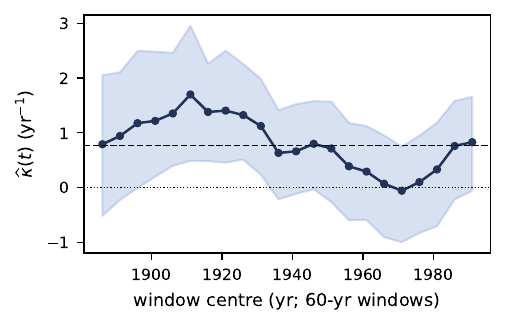}
\caption{Stability of the restoring rate over time from sliding-window
estimates $\widehat\kappa(t)$ on the deseasonalized AMO record, 60-yr
windows stepped by 5\,yr, each estimated at the operative span with a
window-pair bootstrap band ($95\%$). The dashed line marks the
full-record estimate $\widehat\kappa=0.76$. Consecutive windows share
$55$ of their $60$ years, so the estimates are strongly correlated and
the path wanders smoothly (its decadal wiggles are not independent
evidence of change). No trend is resolvable at this record's power (the
variance-weighted trend across window centers is
$-1.1\pm1.2$\,yr$^{-1}$ per century). The record is thus consistent with
a constant restoring rate, and slower drifts in the AMO's persistence can
be neither confirmed nor excluded.}
\label{fig:amo-sliding}
\end{figure}

The analysis above focuses on the short-time drift. The same
analysis nonetheless sheds light on the record's stability history. The
restoring rate $\kappa$ is the quantity whose decline toward zero signals
loss of resilience in recent analyses of closely related AMOC
fingerprints~\cite{Boers2021,Ditlevsen2023}. Near its stable state, the
ramped saddle-node model of ref.~\cite{Ditlevsen2023} linearizes to the
OU form fitted here. Its restoring rate is
$\alpha=2\sqrt{A|\lambda|}$, the same quantity our $\kappa$ estimates,
from the drift, not from variance and autocorrelation. Their index
also differs, the HadISST subpolar-gyre temperature with twice the global
mean removed, and carries a faster rate ($\alpha\approx2$--$3$\,yr$^{-1}$)
than the basin-wide AMO, so magnitudes are not directly comparable. We
therefore estimate
$\kappa$ in sliding 60-yr windows stepped by 5\,yr, each with the operative
span and a window-pair bootstrap (Fig.~\ref{fig:amo-sliding}). The
pointwise estimates fluctuate around the full-record value with wide
intervals. They dip toward zero around the 1960s--70s window centers and
recover thereafter. A variance-weighted trend across window centers is
$-1.1\pm1.2$\,yr$^{-1}$ per century, indistinguishable from zero at this
record's power.
This is a demonstration of applicability, not an adjudication. A 60-yr
window resolves rates of order $1$\,yr$^{-1}$ with roughly $50\%$ relative
uncertainty, so declines of the size debated for the overturning
circulation remain below detectability here, consistent with the power
analysis above. The value of the weak-form route is that it extracts the
drift-based indicator directly, with likelihood-free uncertainty, rather
than through variance and autocorrelation proxies.

\subsection{Calibration of an optical trap}
\label{sec:trap}
\begin{table}[t]
\centering
\caption{Optical-trap calibration on the data of \cite{Gieseler2021} at
$L=20$, $n_{\text{lags}}=2$, and $\delta=8\Delta t$ (Appendix~\ref{si:choice}), showing mean\,(std) across the five records, each fitted separately. The gap exceeds the synthetic default $\delta=4\Delta t$ because
the detector noise here is itself autocorrelated, and the instrument window must
extend beyond its correlation length. Stiffness $\kappa$ is reported in
pN\,$\mu$m$^{-1}$, and diffusion $D$ in $\mu$m$^2$s$^{-1}$ (references are the
Einstein--Smoluchowski $D=k_BT/\gamma$ and the noise-corrected
equipartition
$\kappa_{\mathrm{eq}}^{\mathrm{corr}}=k_BT/(\langle x^2\rangle-\widehat
R_{\mathrm{PSD}})$, while  FORMA appears under both drag conventions). On real
laboratory data, with no known ground truth, \textsc{WAIVES} recovers
stiffness and diffusion in agreement with both the thermodynamic
reference and the established FORMA estimator, and supplies a standard
error for each, confirming that the method's bias correction and
uncertainty quantification carry over from simulation to experiment.}
\label{tab:trap}
\footnotesize
\setlength{\tabcolsep}{4pt}\resizebox{\columnwidth}{!}{\begin{tabular}{lccc}
\toprule
& $2.3$\,mW & $6.0$\,mW & $9.2$\,mW\\
\midrule
$D$ (Einstein--Smoluchowski) & 0.2085 & 0.2085 & 0.2085\\
$\widehat D$ \textsc{WAIVES} slope & 0.208\,(0.005) & 0.213\,(0.003) & 0.213\,(0.006)\\
$\widehat D$ \textsc{WAIVES} lag-one QV & 0.145\,(0.002) & 0.137\,(0.002) & 0.136\,(0.003)\\
$\widehat D$ PSD (Lorentzian) & 0.216\,(0.003) & 0.217\,(0.003) & 0.218\,(0.004)\\
$\widehat D$ FORMA & 0.144\,(0.002) & 0.134\,(0.002) & 0.131\,(0.002)\\
\midrule
$\kappa_{\mathrm{eq}}^{\mathrm{corr}}$ (reference) & 13.63\,(0.33) & 35.04\,(0.58) & 56.61\,(1.05)\\
$\kappa_{\mathrm{eq}}$ raw & 13.61 & 34.92 & 56.28\\
$\widehat\kappa$ \textsc{WAIVES} & 14.07\,(0.60) & 35.99\,(0.74) & 57.44\,(2.28)\\
$\widehat\kappa$ PSD (Lorentzian) & 14.16\,(0.22) & 36.24\,(0.30) & 58.46\,(0.30)\\
$\widehat\kappa$ FORMA, Stokes $\gamma$ & \phantom{0}9.41\,(0.14) & 22.62\,(0.18) & 35.79\,(0.09)\\
$\widehat\kappa$ FORMA, $\gamma=k_BT/\widehat D$ & 13.65\,(0.33) & 35.14\,(0.59) & 56.82\,(1.03)\\
\bottomrule
\end{tabular}}
\end{table}

A dielectric bead held in an optical trap is, in the overdamped limit,
the OU process of Section~\ref{sec:numerics},
\begin{equation}\label{eq:trap-sde}
 \dd X_t=-\frac{\kappa}{\gamma}X_t\dt+\sqrt{2D}\dW.
\end{equation}
It is one of the few measured systems in which both coefficients are
known from physical first principles, with no
time-series estimation. We use the published calibration data of
\cite{Gieseler2021}, a polystyrene bead of radius
$a=1.030\,\mu$m in water at $T=293.15$\,K. Its transverse position is
recorded by a quadrant photodiode at $\Delta t=10^{-5}$\,s, with five
trajectories of $N=10^6$ samples at each of three laser powers.

The physical references follow from measured quantities alone. Stokes' law
determines the drag, $\gamma=6\pi\eta
a=19.415$\,pN\,ms\,$\mu$m$^{-1}$, where $\eta$ is the dynamic viscosity
of water. The Einstein--Smoluchowski relation then gives the diffusion,
$D=k_BT/\gamma=0.2085\,\mu$m$^2$s$^{-1}$. For the stiffness, the
equipartition theorem gives
$\kappa=k_BT/\langle x^2\rangle$, where $\langle x^2\rangle$ is the
stationary variance of the measured bead position. No curve is fitted to
obtain any of these, though all inherit the $\approx2\%$ per kelvin
sensitivity of water's viscosity through the Stokes drag. Because
detector noise adds variance to the measured positions, the plain
equipartition estimate divides by too large a $\langle x^2\rangle$ and
understates the stiffness. We therefore assess stiffness against the
noise-corrected reference
$\kappa_{\mathrm{eq}}^{\mathrm{corr}}=k_BT/(\langle
x^2\rangle-\widehat R)$, with $\widehat R$ estimated from the fitted
noise floor of the power spectrum (Table~\ref{tab:trap}).

We fit \textsc{WAIVES} with drift term candidates $g(x)=[1,x]$ and
constant diffusion term candidate $h=[1]$. The released records are
calibrated one-dimensional $x$ positions, so no deconvolution is
involved. We always report
$\widehat\kappa=\gamma\,\widehat\theta$, with $\widehat\theta$ the
recovered drift-rate coefficient of \eqref{eq:trap-sde}, using the Stokes
$\gamma$. For the diffusion we use the multilag slope form
\eqref{eq:qv-vgram} of Section~\ref{sec:iv}, a choice dictated by the
detector's filtering, which also places these records outside the white-noise observation model \eqref{eq:obs} of the theorems, so that this application tests robustness beyond their assumptions. Quadrant photodiodes low-pass the position signal at
these bandwidths \cite{BergSorensen2006}, which depresses the shortest-lag
increments and biases any estimator built on them. The lag-one variogram
therefore fails to recover $R$ on this detector, and FORMA's diffusion
estimate is low by roughly a third on these records for the same reason. The negative fitted multilag intercept \eqref{eq:R-hat} is what such
filtering produces, whereas added white noise would push it positive,
though we have not fit the detector's transfer function directly. The slope is unaffected by either distortion, and it returns $\widehat D$ within $-0.2\%$, $+2.3\%$, and $+2.1\%$ of the
Einstein value at the three powers (Table~\ref{tab:trap}).
Table~\ref{tab:trap} reports the full results at
the operative window $L=20$. 
\textsc{WAIVES} recovers the stiffness to
within about $3\%$ of the noise-corrected equipartition reference at all
three powers, and agrees to within $1$--$2\%$ with the power-spectral
Lorentzian fit, the accepted standard in this field.

The comparison with FORMA reveals that its drift
coefficient is biased low by $31$--$37\%$ and its diffusion by an almost
identical factor. Its published stiffness is nonetheless accurate,
because FORMA infers the drag self-consistently as $k_BT/\widehat D$
instead of taking it from Stokes' law, and the two biases cancel. That
cancellation costs a drag coefficient inflated $45$--$59\%$ above its
measured value. Table~\ref{tab:trap} reports FORMA
under both conventions, so the cancellation can be seen directly. The disjoint-window construction makes the cancellation unnecessary. The Lorentzian fit is equally accurate on this record, so the gain lies
in scope. The weak-form estimator needs no spectral model, carries over
directly to nonlinear drift term candidates, and returns drift and
diffusion jointly from the same moment system.

Because raising the laser power shortens the relaxation time
$\tau=\gamma/\kappa$ from $1.43$ to $0.35$\,ms at fixed $\Delta t$, this
data set sweeps the ratio $L\Delta t/\tau$ by experimental control, and
so reproduces the window sweep of
Fig.~\ref{fig:variance}, on measured data. At $L=100$
the span crosses $\tau$ as the power is raised ($L\Delta t/\tau=0.70$,
$1.80$, $2.90$) and the estimates degrade as that sweep predicts, to $13.93\,(0.82)$,
$32.71\,(4.79)$, and $18.01\,(21.65)$, the last no longer carrying any
information. $L=20$ is again the operative window.

Because this is the one record in the paper with an external ground
truth, we also use it to check the $F$-triggered instrument depth rule of
Remark~\ref{rem:testfns}. The first-stage $F$ exceeds $10^4$ at either
depth and every power, so this is not a case of weak instruments in the
sampling sense~\cite{StaigerStock1997}. At the stiffest power the
operative span lies at $L\Delta t/\tau\approx0.57$, marginally beyond the
nominal $0.5\tau$ band of Section~\ref{sec:numerics}, and the agreement
below indicates that the band is conservative. Depth one is nonetheless
attenuated by up to $10\%$ at the highest power. The attenuation grows as
the instrument moves further from the target, and matched synthetic records
with white noise show no such effect. The signature therefore points to
residual exogeneity contamination by slowly correlated detector noise.
The depth-consistency check of Remark~\ref{rem:testfns} detects this
failure, which is invisible to the first-stage $F$, and escalates to depth
two, with GMM weighting then suppressing the attenuation at every power. The
slope-based diffusion estimate \eqref{eq:qv-vgram}, robust to short-lag
distortion by construction, stays within $3\%$ of the Einstein value at
either depth.

\section{Discussion}
\label{sec:discussion}

The weak form has gained popularity in deterministic equation learning by
trading a noise-amplifying operation (differentiation) for a
noise-suppressing one (integration). The same trade carries a hidden cost
in the stochastic setting. The smoothing integral that makes the algorithm
robust to measurement noise also entangles the driving Brownian increment
in both the regressor and the residual, resulting in endogeneity that
leaves ordinary least-squares estimators biased and
inconsistent (Section~\ref{sec:endogeneity}). We restore consistency through instrumental variables built from nearby disjoint windows, whose disjointness ensures exogeneity, through independence of Brownian increments over non-overlapping intervals, while their proximity preserves relevance. The resulting estimator, \textsc{WAIVES}, is a consistent,
asymptotically normal, and asymptotically unbiased GMM estimator. While
\textsc{WAIVES} is presented here as the inference layer, it can also be used for  simple model discovery. Given a library of term candidates, it provides unbiased coefficients with calibrated standard errors, so dropping the terms whose coefficients are statistically indistinguishable from zero, and refitting, provides  a feasible and fast model selection, like what was done in Section~\ref{sec:amo} for the linear against the cubic drift.

\textsc{WAIVES} needs no likelihood, which makes it an order of magnitude faster than the Kalman maximum-likelihood fit on OU (Appendix~\ref{si:baselines}) and applicable where likelihoods cannot be evaluated.
We have presented several examples with joint drift and diffusion
identification for diagonal diffusion (constant or entering through
state-dependent diffusion term candidates).
Every estimate reported here is based on a single
trajectory. When several trajectories of the same system are available,
the moment conditions, and the noise covariance, could be pooled across
them, further reducing the standard errors (Appendix~\ref{si:numerics}, examines the noise-covariance pooling).

In terms of limitations, diffusion recovery can be challenging when
measurement noise is large relative to the increment, that is, when the
noise-to-signal ratio $\varrho_i=2R_{ii}/(\sigma_i^2\Delta t)$ of
Section~\ref{sec:iv} is large. In such situations the accuracy of the
diffusion estimators decreases (see Section~\ref{sec:numerics}). Refining $\Delta t$ improves the drift noise condition $\norm{R}_{\mathrm{op}}\Delta t/(\sigma_{\min}^2T_w^2)\to0$ but inflates $\varrho$, so finer sampling makes the diffusion harder to recover~\cite{GloterJacod2001}; with $R$ set by the measurement device, $\Delta t$ is a compromise, resolved by the operating-range diagnostics.

Several extensions are immediate, including a WSINDy-style extension of \textsc{WAIVES}~\cite{MessengerBortz2021}, for model discovery at scale, with a systematic study of sparsification and regularization. 
Similarly, we will extend \textsc{WAIVES} to accommodate partially observed states ($d_y<d$ in Remark~\ref{rem:obsC}),
where some state variable trajectories are completely unobserved
and the instruments must be
built from the observed components alone. Finally, we note that the methodology
presented here is not restricted to SDEs. Any weak-form regression whose
residual is a localized stochastic integral potentially faces the same endogeneity and
can be addressed by the same principles. Specifically, space--time test
functions for SPDEs are the natural next application, and our
current research is developing the corresponding instrumental-variable theory and fast,
likelihood-free algorithms for SPDEs as well.

\section*{Data and code availability}
All data are publicly available. The
unsmoothed AMO index (file \texttt{amon.us.long.data.txt}, January
1856--January 2023) is from NOAA PSL
(\url{https://psl.noaa.gov/data/timeseries/AMO/}), Kaplan SST V2 source
(DOI 10.1029/97JC01736). The optical-trap records are the files
\texttt{Data\_x\_positions\_Exp\_\{I,II,III\}.mat} of the Gieseler et
al.\ tutorial data (DOI 10.1364/AOP.394888, Sec.~3 calibration data set). The code used to generate the tables and figures in the manuscript is archived on Zenodo (DOI 10.5281/zenodo.24369122).

\section*{Acknowledgments}
The authors thank Daniel A.\ Messenger for valuable discussions during the development of this work, and Prof.\ Ian Grooms for his helpful feedback on the analysis of AMO index data. This work was funded in part by the NSF Division Of Environmental
Biology (EEID grant DEB-2109774) Division of Mathematical Sciences (eMB grant
DMS-2527532), NIFA Biological Sciences (grant 2019-67014-29919), and the
NIH-NIGMS Division of Biophysics, Biomedical Technology, and Computational
Bioscience (grant MIRA R35GM149335).
This work utilized the Blanca condo computing resource at the University of
Colorado Boulder. Blanca is jointly funded by computing users and the
University of Colorado Boulder.
Large language models (Claude, Anthropic; ChatGPT, OpenAI) were used under
author direction for manuscript editing, for assistance with analysis and
replication code, and for literature, reference, and internal-consistency
checking. All scientific ideas, derivations, and conclusions are the
authors' own; the authors verified all results, proofs, and references and
take full responsibility for the content of this article.

\bibliographystyle{unsrtnat}
\bibliography{refs}

@inproceedings{KuangLin2026,
  author    = {Kuang, Simon and Lin, Xinfan},
  title     = {Instrumental variables system identification with {$L^p$} consistency},
  booktitle = {Proceedings of the 8th Annual Conference on Learning for Dynamics and Control},
  series    = {Proceedings of Machine Learning Research},
  volume    = {331},
  pages     = {1713--1740},
  year      = {2026},
}

@article{Zeiringer2026,
  author  = {Zeiringer, Thomas and Garnier, Hugues and Horn, Martin},
  title   = {Instrumental variable framework for nonlinear continuous-time system identification},
  journal = {Journal of the Franklin Institute},
  volume  = {363},
  pages   = {108619},
  year    = {2026},
}

@article{Stefanski1989,
  author  = {Stefanski, Leonard A.},
  title   = {Unbiased estimation of a nonlinear function of a normal mean with application to measurement error models},
  journal = {Communications in Statistics---Theory and Methods},
  volume  = {18},
  number  = {12},
  pages   = {4335--4358},
  year    = {1989}
}

@article{Nakamura1990,
  author  = {Nakamura, Tsuyoshi},
  title   = {Corrected score function for errors-in-variables models: methodology and application to generalized linear models},
  journal = {Biometrika},
  volume  = {77},
  number  = {1},
  pages   = {127--137},
  year    = {1990}
}

@book{Carroll2006,
  author    = {Carroll, Raymond J. and Ruppert, David and Stefanski, Leonard A. and Crainiceanu, Ciprian M.},
  title     = {Measurement Error in Nonlinear Models: A Modern Perspective},
  edition   = {2nd},
  publisher = {Chapman \& Hall/CRC},
  address   = {Boca Raton},
  year      = {2006}
}

@article{MessengerBortz2021PDE,
  author  = {Messenger, Daniel A. and Bortz, David M.},
  title   = {Weak {SINDy} for partial differential equations},
  journal = {Journal of Computational Physics},
  volume  = {443},
  pages   = {110525},
  year    = {2021},
}

@article{Reinbold2020,
  author  = {Reinbold, Patrick A. K. and Gurevich, Daniel R. and Grigoriev, Roman O.},
  title   = {Using noisy or incomplete data to discover models of spatiotemporal dynamics},
  journal = {Physical Review E},
  volume  = {101},
  pages   = {010203},
  year    = {2020}
}

@article{MessengerBortz2022,
  author  = {Messenger, Daniel A. and Bortz, David M.},
  title   = {Learning mean-field equations from particle data using {WSINDy}},
  journal = {Physica D: Nonlinear Phenomena},
  volume  = {439},
  pages   = {133406},
  year    = {2022},
}

@article{Minor2026,
author = {Minor, Seth and Elderd, Bret D. and Van Allen, Benjamin and Bortz, David M. and Dukic, Vanja},
title = {Weak-Form Learning for Mean-Field Partial Differential Equations: An Application to Insect Movement},
journal = {SIAM Journal on Life Sciences},
volume = {1},
number = {3},
pages = {444-484},
year = {2026},
eprint = { 
https://doi.org/10.1137/25M1812038
}
}

@article{Ragwitz2001,
  author  = {Ragwitz, Mario and Kantz, Holger},
  title   = {Indispensable finite time corrections for {Fokker}--{Planck} equations from time series data},
  journal = {Physical Review Letters},
  volume  = {87},
  pages   = {254501},
  year    = {2001}
}

@article{Boettcher2006,
  author  = {B{\"o}ttcher, Frank and Peinke, Joachim and Kleinhans, David and Friedrich, Rudolf and Lind, Pedro G. and Haase, Maria},
  title   = {Reconstruction of complex dynamical systems affected by strong measurement noise},
  journal = {Physical Review Letters},
  volume  = {97},
  pages   = {090603},
  year    = {2006}
}

@book{Pavliotis2014,
  author    = {Pavliotis, Grigorios A.},
  title     = {Stochastic Processes and Applications: Diffusion Processes, the {Fokker}--{Planck} and {Langevin} Equations},
  series    = {Texts in Applied Mathematics},
  volume    = {60},
  publisher = {Springer},
  address   = {New York},
  year      = {2014}
}

@article{Friedrich2011,
  author  = {Friedrich, Rudolf and Peinke, Joachim and Sahimi, Muhammad and Tabar, M. Reza Rahimi},
  title   = {Approaching complexity by stochastic methods: From biological systems to turbulence},
  journal = {Physics Reports},
  volume  = {506},
  number  = {5},
  pages   = {87--162},
  year    = {2011}
}

@book{Wooldridge2010,
  author    = {Wooldridge, Jeffrey M.},
  title     = {Econometric Analysis of Cross Section and Panel Data},
  edition   = {2nd},
  publisher = {MIT Press},
  address   = {Cambridge, MA},
  year      = {2010}
}

@article{Hansen1982,
  author  = {Hansen, Lars Peter},
  title   = {Large sample properties of generalized method of moments estimators},
  journal = {Econometrica},
  volume  = {50},
  number  = {4},
  pages   = {1029--1054},
  year    = {1982},
}

@article{HansenScheinkman1995,
  author  = {Hansen, Lars Peter and Scheinkman, Jos{\'e} A.},
  title   = {Back to the future: generating moment implications for continuous-time {M}arkov processes},
  journal = {Econometrica},
  volume  = {63},
  number  = {4},
  pages   = {767--804},
  year    = {1995}
}

@article{BibbySorensen1995,
  author  = {Bibby, Bo Martin and S{\o}rensen, Michael},
  title   = {Martingale estimation functions for discretely observed diffusion processes},
  journal = {Bernoulli},
  volume  = {1},
  number  = {1/2},
  pages   = {17--39},
  year    = {1995}
}

@article{AndrewsLu2001,
  author  = {Andrews, Donald W. K. and Lu, Biao},
  title   = {Consistent model and moment selection procedures for {GMM} estimation with application to dynamic panel data models},
  journal = {Journal of Econometrics},
  volume  = {101},
  number  = {1},
  pages   = {123--164},
  year    = {2001},
}

@article{NeweyWest1987,
  author  = {Newey, Whitney K. and West, Kenneth D.},
  title   = {Hypothesis testing with efficient method of moments estimation},
  journal = {International Economic Review},
  volume  = {28},
  number  = {3},
  pages   = {777--787},
  year    = {1987},
}

@article{MessengerBortz2021,
  author  = {Messenger, Daniel A. and Bortz, David M.},
  title   = {Weak {SINDy}: {Galerkin}-based data-driven model selection},
  journal = {Multiscale Model. Simul.},
  volume  = {19},
  number  = {3},
  pages   = {1474--1497},
  year    = {2021},
  publisher = {SIAM}
}

@article{BortzMessengerDukic2023,
  author  = {Bortz, David M. and Messenger, Daniel A. and Duki\'c, Vanja},
  title   = {Direct estimation of parameters in {ODE} models using {WENDy}:
             Weak-form estimation of nonlinear dynamics},
  journal = {Bull. Math. Biol.},
  volume  = {85},
  number  = {11},
  pages   = {110},
  year    = {2023}
}

@article{Brunton2016,
  author  = {Brunton, Steven L. and Proctor, Joshua L. and Kutz, J. Nathan},
  title   = {Discovering governing equations from data by sparse
             identification of nonlinear dynamical systems},
  journal = {Proc. Natl. Acad. Sci. U.S.A.},
  volume  = {113},
  number  = {15},
  pages   = {3932--3937},
  year    = {2016}
}

@article{SchmidtLipson2009,
  author  = {Schmidt, Michael and Lipson, Hod},
  title   = {Distilling free-form natural laws from experimental data},
  journal = {Science},
  volume  = {324},
  number  = {5923},
  pages   = {81--85},
  year    = {2009}
}

@article{RudyBruntonProctorKutz2017,
  author  = {Rudy, Samuel H. and Brunton, Steven L. and Proctor, Joshua L.
             and Kutz, J. Nathan},
  title   = {Data-driven discovery of partial differential equations},
  journal = {Sci. Adv.},
  volume  = {3},
  number  = {4},
  pages   = {e1602614},
  year    = {2017}
}

@book{BruntonKutz2022,
  author    = {Brunton, Steven L. and Kutz, J. Nathan},
  title     = {Data-Driven Science and Engineering: Machine Learning,
               Dynamical Systems, and Control},
  edition   = {2},
  publisher = {Cambridge University Press},
  year      = {2022}
}

@article{Sargan1958,
  author  = {Sargan, J. D.},
  title   = {The estimation of economic relationships using instrumental
             variables},
  journal = {Econometrica},
  volume  = {26},
  number  = {3},
  pages   = {393--415},
  year    = {1958}
}

@article{Rummel2025,
  author  = {Rummel, Nicholas and Messenger, Daniel A. and Becker, Stephen and Dukic, Vanja and Bortz, David M.},
  title   = {{WENDy} for Nonlinear-in-Parameters {ODE}s},
  journal = {SIAM Journal on Scientific Computing},
  year    = {2026},
  note    = {in press},
  eprint  = {2502.08881},
  archivePrefix = {arXiv},
}

@article{Marsaglia2006,
  author  = {Marsaglia, George},
  title   = {Ratios of normal variables},
  journal = {J. Stat. Softw.},
  volume  = {16},
  number  = {4},
  pages   = {1--10},
  year    = {2006}
}

@article{Hinkley1969,
  author  = {Hinkley, D. V.},
  title   = {On the ratio of two correlated normal random variables},
  journal = {Biometrika},
  volume  = {56},
  number  = {3},
  pages   = {635--639},
  year    = {1969}
}

@article{Fieller1954,
  author  = {Fieller, E. C.},
  title   = {Some problems in interval estimation},
  journal = {J. R. Stat. Soc. B},
  volume  = {16},
  number  = {2},
  pages   = {175--185},
  year    = {1954}
}

@article{Sorensen2004,
  author  = {S{\o}rensen, Helle},
  title   = {Parametric inference for diffusion processes observed at discrete
             points in time: a survey},
  journal = {Int. Stat. Rev.},
  volume  = {72},
  number  = {3},
  pages   = {337--354},
  year    = {2004}
}

@article{Pedersen1995,
  author  = {Pedersen, Asger Roer},
  title   = {A new approach to maximum likelihood estimation for stochastic
             differential equations based on discrete observations},
  journal = {Scand. J. Stat.},
  volume  = {22},
  number  = {1},
  pages   = {55--71},
  year    = {1995}
}

@article{Roberts2001,
  author  = {Roberts, Gareth O. and Stramer, Osnat},
  title   = {On inference for partially observed nonlinear diffusion models
             using the {Metropolis--Hastings} algorithm},
  journal = {Biometrika},
  volume  = {88},
  number  = {3},
  pages   = {603--621},
  year    = {2001}
}

@article{AitSahalia2002,
  author  = {A{\"i}t-Sahalia, Yacine},
  title   = {Maximum likelihood estimation of discretely sampled diffusions:
             a closed-form approximation approach},
  journal = {Econometrica},
  volume  = {70},
  number  = {1},
  pages   = {223--262},
  year    = {2002}
}

@article{BPRF2006,
  author  = {Beskos, Alexandros and Papaspiliopoulos, Omiros and
             Roberts, Gareth O. and Fearnhead, Paul},
  title   = {Exact and computationally efficient likelihood-based estimation
             for discretely observed diffusion processes},
  journal = {J. R. Stat. Soc. B},
  volume  = {68},
  number  = {3},
  pages   = {333--382},
  year    = {2006}
}

@article{Beskos2009,
  author  = {Beskos, Alexandros and Papaspiliopoulos, Omiros and
             Roberts, Gareth O.},
  title   = {{Monte Carlo} maximum likelihood estimation for discretely
             observed diffusion processes},
  journal = {Ann. Stat.},
  volume  = {37},
  number  = {1},
  pages   = {223--245},
  year    = {2009}
}

@article{Golightly2008,
  author  = {Golightly, Andrew and Wilkinson, Darren J.},
  title   = {Bayesian inference for nonlinear multivariate diffusion models
             observed with error},
  journal = {Comput. Stat. Data Anal.},
  volume  = {52},
  number  = {3},
  pages   = {1674--1693},
  year    = {2008}
}

@article{Andrieu2010,
  author  = {Andrieu, Christophe and Doucet, Arnaud and Holenstein, Roman},
  title   = {Particle {Markov} chain {Monte Carlo} methods},
  journal = {J. R. Stat. Soc. B},
  volume  = {72},
  number  = {3},
  pages   = {269--342},
  year    = {2010}
}

@article{Kantas2015,
  author  = {Kantas, Nikolas and Doucet, Arnaud and Singh, Sumeetpal S. and
             Maciejowski, Jan and Chopin, Nicolas},
  title   = {On particle methods for parameter estimation in state-space
             models},
  journal = {Stat. Sci.},
  volume  = {30},
  number  = {3},
  pages   = {328--351},
  year    = {2015}
}

@article{Ionides2006,
  author  = {Ionides, Edward L. and Bret\'{o}, Carles and King, Aaron A.},
  title   = {Inference for nonlinear dynamical systems},
  journal = {Proc. Natl. Acad. Sci. U.S.A.},
  volume  = {103},
  number  = {49},
  pages   = {18438--18443},
  year    = {2006}
}

@article{Beskos2021,
  author  = {Beskos, Alexandros and Crisan, Dan and Jasra, Ajay and
             Kantas, Nikolas and Ruzayqat, Hamza},
  title   = {Score-based parameter estimation for a class of continuous-time
             state space models},
  journal = {SIAM J. Sci. Comput.},
  volume  = {43},
  number  = {4},
  pages   = {A2555--A2580},
  year    = {2021}
}

@article{Billingsley1961,
  author  = {Billingsley, Patrick},
  title   = {The {L}indeberg--{L}\'evy theorem for martingales},
  journal = {Proceedings of the American Mathematical Society},
  volume  = {12},
  number  = {5},
  pages   = {788--792},
  year    = {1961}
}

@article{CraggDonald1993,
  author  = {Cragg, John G. and Donald, Stephen G.},
  title   = {Testing identifiability and specification in instrumental variable models},
  journal = {Econometric Theory},
  volume  = {9},
  number  = {2},
  pages   = {222--240},
  year    = {1993}
}

@incollection{StockYogo2005,
  author    = {Stock, James H. and Yogo, Motohiro},
  title     = {Testing for weak instruments in linear {IV} regression},
  booktitle = {Identification and Inference for Econometric Models: Essays in Honor of Thomas Rothenberg},
  editor    = {Andrews, Donald W. K. and Stock, James H.},
  publisher = {Cambridge University Press},
  address   = {Cambridge},
  pages     = {80--108},
  year      = {2005}
}

@book{Vaart1998,
  author    = {van der Vaart, A. W.},
  title     = {Asymptotic Statistics},
  publisher = {Cambridge University Press},
  address   = {Cambridge},
  year      = {1998}
}

@article{MattinglySH2002,
  author  = {Mattingly, Jonathan C. and Stuart, Andrew M. and Higham, Desmond J.},
  title   = {Ergodicity for {SDEs} and approximations: locally {Lipschitz} vector fields and degenerate noise},
  journal = {Stochastic Processes and their Applications},
  volume  = {101},
  number  = {2},
  pages   = {185--232},
  year    = {2002}
}

@article{ShumwayStoffer1982,
  author  = {Shumway, Robert H. and Stoffer, David S.},
  title   = {An approach to time series smoothing and forecasting using the {EM} algorithm},
  journal = {Journal of Time Series Analysis},
  volume  = {3},
  number  = {4},
  pages   = {253--264},
  year    = {1982}
}

@misc{Morr2026,
  author = {Morr, Andreas and Ben-Yami, Maya and Groenke, Brian and
            Sch{\"o}tz, Christof and Cotronei, Alessandro and
            Myrvoll-Nilsen, Eirik and Bathiany, Sebastian and
            Rypdal, Martin and Boers, Niklas},
  title  = {Extrapolation from historical data cannot reliably predict the
            time of a potential {AMOC} collapse},
  year   = {2026},
  note   = {arXiv preprint, arXiv:2604.20341 [physics.geo-ph]},
  eprint = {2604.20341},
  archivePrefix = {arXiv}
}

@article{Cotronei2026,
  author  = {Cotronei, Alessandro and Myrvoll-Nilsen, Eirik and Rypdal, Martin},
  title   = {Evaluating model uncertainty in critical threshold estimations
             from time series data: Application to the {Atlantic} meridional
             {Overturning} {Circulation}},
  journal = {Frontiers in Climate},
  volume  = {8},
  pages   = {1761461},
  year    = {2026},
}

@article{Crisan2022,
  author  = {Crisan, Dan and Del Moral, Pierre and Jasra, Ajay and
             Ruzayqat, Hamza},
  title   = {Log-normalization constant estimation using the ensemble
             {Kalman--Bucy} filter with application to high-dimensional models},
  journal = {Adv. Appl. Probab.},
  volume  = {54},
  number  = {4},
  pages   = {1139--1163},
  year    = {2022}
}

@misc{Tzen2019,
  author  = {Tzen, Belinda and Raginsky, Maxim},
  title   = {Neural stochastic differential equations: deep latent {Gaussian}
             models in the diffusion limit},
  year    = {2019},
  note    = {arXiv:1905.09883}
}

@inproceedings{Li2020,
  author    = {Li, Xuechen and Wong, Ting-Kam Leonard and
               Chen, Ricky T. Q. and Duvenaud, David},
  title     = {Scalable gradients for stochastic differential equations},
  booktitle = {Proc. 23rd Int. Conf. Artif. Intell. Stat. (AISTATS)},
  series    = {Proc. Mach. Learn. Res.},
  volume    = {108},
  pages     = {3870--3882},
  year      = {2020},
  publisher = {PMLR}
}

@inproceedings{Kidger2021,
  author    = {Kidger, Patrick and Foster, James and Li, Xuechen and
               Oberhauser, Harald and Lyons, Terry},
  title     = {Neural {SDE}s as infinite-dimensional {GAN}s},
  booktitle = {Proc. 38th Int. Conf. Mach. Learn. (ICML)},
  series    = {Proc. Mach. Learn. Res.},
  volume    = {139},
  pages     = {5453--5463},
  year      = {2021},
  publisher = {PMLR},
}

@article{Boninsegna2018,
  author  = {Boninsegna, Lorenzo and N{\"u}ske, Feliks and Clementi, Cecilia},
  title   = {Sparse learning of stochastic dynamical equations},
  journal = {J. Chem. Phys.},
  volume  = {148},
  number  = {24},
  pages   = {241723},
  year    = {2018}
}

@article{Callaham2021,
  author  = {Callaham, Jared L. and Loiseau, Jean-Christophe and
             Rigas, Georgios and Brunton, Steven L.},
  title   = {Nonlinear stochastic modelling with {Langevin} regression},
  journal = {Proc. R. Soc. A},
  volume  = {477},
  number  = {2250},
  pages   = {20210092},
  year    = {2021}
}

@article{Wanner2024,
author = {Wanner, Mathias and Mezi{\'c}, Igor},
title = {On Higher Order Drift and Diffusion Estimates for Stochastic SINDy},
journal = {SIAM Journal on Applied Dynamical Systems},
volume = {23},
number = {2},
pages = {1504-1539},
year = {2024},
}

@book{Soderstrom1983,
  author    = {S\"{o}derstr\"{o}m, Torsten and Stoica, Petre},
  title     = {Instrumental Variable Methods for System Identification},
  series    = {Lect. Notes Control Inf. Sci.},
  volume    = {57},
  publisher = {Springer-Verlag},
  address   = {Berlin},
  year      = {1983}
}

@book{Soderstrom1989,
  author    = {S\"{o}derstr\"{o}m, Torsten and Stoica, Petre},
  title     = {System Identification},
  publisher = {Prentice Hall},
  address   = {Hemel Hempstead, UK},
  year      = {1989}
}

@article{Enfield2001,
  author  = {Enfield, David B. and Mestas-Nu{\~n}ez, Alberto M. and Trimble, Paul J.},
  title   = {The {Atlantic} multidecadal oscillation and its relation to rainfall and river flows in the continental {U.S.}},
  journal = {Geophysical Research Letters},
  volume  = {28},
  number  = {10},
  pages   = {2077--2080},
  year    = {2001},
}

@article{Kaplan1998,
  author  = {Kaplan, Alexey and Cane, Mark A. and Kushnir, Yochanan and Clement, Amy C. and Blumenthal, M. Benno and Rajagopalan, Balaji},
  title   = {Analyses of global sea surface temperature 1856--1991},
  journal = {Journal of Geophysical Research: Oceans},
  volume  = {103},
  number  = {C9},
  pages   = {18567--18589},
  year    = {1998},
}

@article{Boers2021,
  author  = {Boers, Niklas},
  title   = {Observation-based early-warning signals for a collapse of the {Atlantic} {Meridional} {Overturning} {Circulation}},
  journal = {Nature Climate Change},
  volume  = {11},
  pages   = {680--688},
  year    = {2021},
}

@article{Ditlevsen2023,
  author  = {Ditlevsen, Peter and Ditlevsen, Susanne},
  title   = {Warning of a forthcoming collapse of the {Atlantic} meridional overturning circulation},
  journal = {Nature Communications},
  volume  = {14},
  pages   = {4254},
  year    = {2023},
}

@article{Mann2021,
  author  = {Mann, Michael E. and Steinman, Byron A. and Brouillette, Daniel J. and Miller, Sonya K.},
  title   = {Multidecadal climate oscillations during the past millennium driven by volcanic forcing},
  journal = {Science},
  volume  = {371},
  number  = {6533},
  pages   = {1014--1019},
  year    = {2021},
}

@article{Young1970,
  author  = {Young, Peter C.},
  title   = {An instrumental variable method for real-time identification of a noisy process},
  journal = {Automatica},
  volume  = {6},
  number  = {2},
  pages   = {271--287},
  year    = {1970},
}

@book{HallHeyde1980,
  author    = {Hall, Peter and Heyde, Christopher C.},
  title     = {Martingale Limit Theory and Its Application},
  publisher = {Academic Press},
  address   = {New York},
  year      = {1980}
}

@article{Gieseler2021,
  author  = {Gieseler, Jan and Gomez-Solano, Juan Ruben and Magazz{\`u},
             Alessandro and P{\'e}rez Castillo, Isaac and
             P{\'e}rez Garc{\'i}a, Laura and Gironella-Torrent, Marta and
             Viader-Godoy, Xavier and Ritort, Felix and Pesce, Giuseppe and
             Arzola, Alejandro V. and Volke-Sep{\'u}lveda, Karen and
             Volpe, Giovanni},
  title   = {Optical tweezers --- from calibration to applications: a tutorial},
  journal = {Adv. Opt. Photonics},
  volume  = {13},
  number  = {1},
  pages   = {74--241},
  year    = {2021},
}

@article{BergSorensen2006,
  author  = {Berg-S{\o}rensen, Kirstine and Peterman, Erwin J. G. and
             Weber, Tom and Schmidt, Christoph F. and Flyvbjerg, Henrik},
  title   = {Power spectrum analysis for optical tweezers. {II}: Laser
             wavelength dependence of parasitic filtering, and how to achieve
             high bandwidth},
  journal = {Rev. Sci. Instrum.},
  volume  = {77},
  number  = {6},
  pages   = {063106},
  year    = {2006},
}

@article{StaigerStock1997,
  author  = {Staiger, Douglas and Stock, James H.},
  title   = {Instrumental variables regression with weak instruments},
  journal = {Econometrica},
  volume  = {65},
  number  = {3},
  pages   = {557--586},
  year    = {1997}
}

@article{Kunsch1989,
  author  = {K{\"u}nsch, Hans R.},
  title   = {The jackknife and the bootstrap for general stationary observations},
  journal = {Annals of Statistics},
  volume  = {17},
  number  = {3},
  pages   = {1217--1241},
  year    = {1989}
}

@article{BenYami2024,
  author  = {Ben-Yami, Maya and Morr, Andreas and Bathiany, Sebastian and Boers, Niklas},
  title   = {Uncertainties too large to predict tipping times of major {Earth} system components from historical data},
  journal = {Science Advances},
  volume  = {10},
  number  = {31},
  pages   = {eadl4841},
  year    = {2024},
}

@article{GloterJacod2001,
  author  = {Gloter, Arnaud and Jacod, Jean},
  title   = {Diffusions with measurement errors. {I}. Local asymptotic normality},
  journal = {ESAIM: Probability and Statistics},
  volume  = {5},
  pages   = {225--242},
  year    = {2001},
}

@article{NeweyWest1994,
  author  = {Newey, Whitney K. and West, Kenneth D.},
  title   = {Automatic Lag Selection in Covariance Matrix Estimation},
  journal = {The Review of Economic Studies},
  volume  = {61},
  number  = {4},
  pages   = {631--653},
  year    = {1994},
}

@article{Kinal1980,
  author  = {Kinal, Terrence W.},
  title   = {The Existence of Moments of $k$-Class Estimators},
  journal = {Econometrica},
  volume  = {48},
  number  = {1},
  pages   = {241--249},
  year    = {1980}
}

@article{lu2024wcr,
author = {Lu, Liwei and Zeng, Zhijun and Jiang, Yan and Zhu, Yi and Hu, Pipi},
title = {Weak collocation regression method: Fast reveal hidden stochastic
dynamics from high-dimensional aggregate data},
journal = {J. Comput. Phys.},
volume = {502},
pages = {112799},
year = {2024}
}

@article{jiang2026stwcr,
author = {Jiang, Yan and Zeng, Zhijun and Yang, Wuyue and Hong, Liu and
Hu, Pipi and Zhu, Yi},
title = {{STWCR}: Weak collocation regression for revealing hidden
stochastic dynamics from single trajectory data},
journal = {Appl. Math. Comput.},
volume = {521},
pages = {129958},
year = {2026}
}

\clearpage
\appendix
\section{Proofs}
\label{app:proofs}

\subsection{Validity of the disjoint-window instrument}
\label{si:validity}
This appendix proves Proposition~\ref{prop:ols-zero} and then
establishes the two instrument conditions (IV1)--(IV2) for the construction
of Definition~\ref{def:instrument}.

\begin{proof}[Proof of Proposition~\ref{prop:ols-zero}]
Center each $g_j(X)$ and each component $X^{(i)}$ at its stationary mean;
this changes neither $\mathsf b_i$ nor any covariance. Fix $t$ and let $s$
vary. For $s\neq t$ the product rule gives
$\dd_s\bigl(g_j(X_t)X^{(i)}_s\bigr)=g_j(X_t)\,\dd X^{(i)}_s$, since
$g_j(X_t)$ does not vary with $s$; taking expectations and using
stationarity,
$\E[g_j(X_t)\,\dd X^{(i)}_s]=\partial_s\,C_{ji}(t-s)\,\dd s
=-C_{ji}'(t-s)\,\dd s$.
Reversibility and stationarity give
$C_{ji}(-u)=\Cov(g_j(X_0),X^{(i)}_u)=\Cov(g_j(X_u),X^{(i)}_0)=C_{ji}(u)$,
so $C_{ji}$ is even and $C_{ji}'$ is odd on $u\neq0$. By the stochastic
Fubini theorem, justified by the stated moment bounds and the boundedness
of $\phi$ and $C_{ji}'$,
\[
\E[a_j\mathsf b_i]
=-\iint_{[a,b]^2}\phi(t)\phi(s)\,C_{ji}'(t-s)\,\dd s\,\dd t=0,
\]
the integrand being odd under the exchange of $s$ and $t$ on a symmetric
domain. The diagonal $s=t$ carries no contribution, since it has Lebesgue measure zero, $\dd X$ has no atoms in time, and $\E[g_j(X_t)\,\sigma_i(X_t)\,\dd W_t]=0$, where $\sigma_i$ is the $i$th row of $\sigma$. At stationarity
$\E[\mathsf b_i]=\int\phi\,\E[\dd X^{(i)}_t]=0$, so
$\Cov(a_j,\mathsf b_i)=\E[a_j\mathsf b_i]=0$. The moment vector of the
regression of $\mathsf b_i$ on the drift term candidates therefore vanishes, and so does
its population coefficient vector whenever the Gram matrix is invertible.

As for the two special cases, a stationary ergodic scalar diffusion has zero stationary probability flux and is therefore reversible, so the conclusion holds for every choice of drift term candidates; and for $g(x)=x$ in one dimension, $C(u)=\Cov(X_u,X_0)$ is even for any stationary process, so the conclusion for the affine drift holds without reversibility.
\end{proof}

The argument uses only the frozen-$t$ product rule and the evenness of the
cross-covariances, so the proposition extends verbatim to stationary,
time-reversible, square-integrable semimartingales whose cross-covariances are continuously differentiable at all nonzero lags, including jump-diffusions. The identity it discards on the diagonal,
in the scalar case $\E[X_t\,\dd X_t]=-\tfrac12\,\E[\sigma^2(X_0)]\,\dd t$,
follows from $\dd\,\E[X_t^2]=0$ at stationarity, and is  what
makes the adapted one-step regression consistent.

For OU, $\E[A^{2}]=\tfrac{\sigma^{2}}{2\kappa}\iint_{[a,b]^2}
\phi(t)\phi(s)\,e^{-\kappa|t-s|}\dd s\,\dd t >0$ for every nonzero
$\phi$, since the kernel $e^{-\kappa|t-s|}$ is strictly positive
definite, being the Fourier transform of the everywhere-positive Cauchy
density $\kappa/[\pi(\kappa^{2}+\omega^{2})]$. With
$\mathsf b=-\kappa A+\varepsilon$ and $\E[A\mathsf b]=0$, this gives
$\E[A\varepsilon]=\kappa\,\E[A^{2}]>0$.

We now prove the two instrument conditions, Propositions~\ref{prop:exog} and~\ref{prop:rel}, for the construction of
Definition~\ref{def:instrument}. Recall the target residual
$\varepsilon_k=\int_{a_k}^{a_k+T_w}\phi_k(t)\sigma(X_t)\dW$ and the instrument
$Z_{kj}=\int_{a_k'}^{a_k'+T_w}\psi_k(t)g_j(X_t)\dt$ with
$[a_k',a_k'+T_w)\cap[a_k,a_k+T_w]=\varnothing$, the
instrument interval taken half-open so that the possibly shared endpoint
$a_k'+T_w=a_k$ is excluded, and $a_k'+T_w\le a_k$.

\begin{proof}[Proof of Proposition~\ref{prop:exog}]
Write the total residual including observation noise as
$\varepsilon_k^{\mathrm{tot}}=\varepsilon_k+\delta\mathsf b_k-\sum_j\theta_j\,\delta A_{kj}$,
where $\varepsilon_k=\int_{a_k}^{a_k+T_w}\phi_k(t)\sigma(X_t)\dW$ is the process-noise
term and $\delta\mathsf b_k,\delta A_{kj}$ are the contributions of the measurement
errors $\{\eta_n\}_{n\in\mathcal{I}_k}$ to the target-window quadratures.
We show each piece is uncorrelated with $Z_{kj}$.

As for the process noise, let $\Filt_t=\sigma(\{W_s:s\le t\}\cup\{\eta_m:t_m\le t\})$
be the filtration generated by the Brownian path and the errors up to time $t$.
The
exact instrument $Z_{kj}=\int_{a_k'}^{a_k'+T_w}\psi_k(t)\,g_j(X_t)\dt$ is a
functional of the path $\{X_t:a_k'\le t\le a_k'+T_w\}$ only, and its
observable form of Section~\ref{sec:iv} is a function of the observations
$Y_{t_m}=X_{t_m}+\eta_m$ at the instants $t_m\in\mathcal I_k'\subset[a_k',a_k'+T_w)$.
Every time involved is at most $a_k'+T_w$, so $Z_{kj}$ is
$\Filt_{a_k'+T_w}$-measurable. Since $a_k'+T_w\le a_k$ and the filtration grows
with time, $\Filt_{a_k'+T_w}\subseteq\Filt_{a_k}$, so $Z_{kj}$ is
$\Filt_{a_k}$-measurable as well. The integrand $\phi_k(t)\sigma(X_t)$ of
$\varepsilon_k$ is adapted, so $\varepsilon_k$ is an It\^o integral over
$[a_k,a_k+T_w]$ and hence a martingale increment beyond $a_k$,
$\E[\varepsilon_k\mid\Filt_{a_k}]=0$. By the tower property,
\begin{equation}
\E[Z_{kj}\,\varepsilon_k]
=\E\!\big[Z_{kj}\,\E[\varepsilon_k\mid\Filt_{a_k}]\big]
=\E[Z_{kj}\cdot 0]=0.
\label{eq:exog-tower}
\end{equation}

Next we deal with the observation noise. The target perturbations $\delta\mathsf b_k,\delta A_{kj}$
are linear (to first order) in $\{\eta_n\}_{n\in\mathcal{I}_k}$, with
coefficients that depend on the latent path through terms such as
$g_j'(X_{t_n})$, while $Z_{kj}$ depends on $\{W_t\}$ and on
$\{\eta_m\}_{m\in\mathcal{I}_k'}$. Let $\mathcal G_k$ be the
$\sigma$-algebra generated by the latent path $\{X_t\}$ and the
instrument-window errors $\{\eta_m\}_{m\in\mathcal{I}_k'}$. Then $Z_{kj}$
and the perturbation coefficients are $\mathcal G_k$-measurable, while
each $\eta_n$ with $n\in\mathcal{I}_k$ is mean zero and independent of
$\mathcal G_k$, because the index sets $\mathcal{I}_k,\mathcal{I}_k'$ are
disjoint and $\{\eta_n\}$ is i.i.d.\ and independent of $\{W_t\}$.
Conditioning therefore gives
$\E[\delta\mathsf b_k\mid\mathcal G_k]=\E[\delta A_{kj}\mid\mathcal G_k]=0$ (for $\delta A_{kj}$ with the debiasing at the true $R$; the plug-in
$\widehat R$ depends on the target-window errors and is not
$\mathcal G_k$-measurable, which Proposition~\ref{prop:plugin} handles),
and the tower property yields
$\E[Z_{kj}\,\delta\mathsf b_k]=\E\!\big[Z_{kj}\,\E[\delta\mathsf b_k\mid\mathcal G_k]\big]=0$
and likewise $\E[Z_{kj}\,\delta A_{kj'}]=0$. The
suppressed second-order terms are quadratic in the errors with mean
$O(\norm{R})$; the debiased drift term candidates of Section~\ref{sec:algorithm} remove that mean, so
what survives in $\rho_k$ has mean $O(\norm{R}^2)$ beside the
$O(\Delta t)$ quadrature part. Combining with \eqref{eq:exog-tower} gives
the displayed order.
\end{proof}

\begin{proof}[Proof of Proposition~\ref{prop:rel}]
It suffices to show $\E[\Zmat\transpose\Amat]$ is nonsingular, since full
column rank of a $p\times p$ matrix is nonsingularity. Consider a single window
pair and a pair of term indices $(j,j')$. By stationarity and Fubini,
\begin{multline}
\mathbf Q_{jj'}=\E\bigl[Z_{kj}A_{kj'}\bigr]\\
=\int_{a_k'}^{a_k'+T_w}\!\!\int_{a_k}^{a_k+T_w}
 \psi_k(s)\phi_k(t)\,\E\bigl[g_j(X_s)g_{j'}(X_t)\bigr]\dd s\,\dt.
\label{eq:rel-double}
\end{multline}
Center the drift term candidates, $\bar g_j=\E[g_j(X_0)]$, and write
$\E\bigl[g_j(X_s)g_{j'}(X_t)\bigr]=\bar g_j\bar g_{j'}+\Gamma_{jj'}(t-s)$. Collect the
means in $\bar{\bm g}=(\bar g_1,\dots,\bar g_p)\transpose$.
Both test functions are nonnegative, and every separation
$|t-s|$ arising in \eqref{eq:rel-double} lies in $[\delta,\,\delta+2T_w]$. For
any $\bm\zeta\ne\bm 0$ the quadratic form
$\bm\zeta\transpose\mathbf Q\,\bm\zeta=\sum_{j,j'}\zeta_j\zeta_{j'}\mathbf Q_{jj'}$
is therefore
\begin{multline*}
\bm\zeta\transpose\mathbf Q\,\bm\zeta
\\
=\!\int_{a_k'}^{a_k'+T_w}\!\!\int_{a_k}^{a_k+T_w}\!
\psi_k(s)\phi_k(t)\bigl[(\bm\zeta\transpose\bar{\bm g})^2
+\bm\zeta\transpose \bm\Gamma(t-s)\,\bm\zeta\bigr]\dd s\,\dt\\
\;\ge\; m\norm{\bm\zeta}^2\mu_\psi\mu_\phi,
\end{multline*}
with $\mu_\psi=\int\psi_k>0$ and $\mu_\phi=\int\phi_k>0$, so
$\mathbf Q$ is positive definite.

Davydov's covariance inequality, which is
valid for $\beta$-mixing processes, bounds each entry by
$|\Gamma_{jj'}(u)|\le 8\,\beta(u)^{1-2/\nu}\,
\|g_j(X_0)\|_\nu\|g_{j'}(X_0)\|_\nu$, so $\bm\Gamma(u)$ vanishes as
$u\to\infty$, exponentially fast for the geometrically ergodic systems
of Remark~\ref{rem:mixing}. The covariances of the drift term candidates are therefore
appreciable for separations $u\lesssim\tau$, where $\tau$ is the characteristic time, and negligible beyond. The
positivity-within-memory hypothesis is therefore a short-separation
condition. It holds whenever $\delta+T_w\lesssim\tau$ for systems whose
covariances of the drift term candidates decay without sign changes, and it can fail at
particular lags for strongly oscillatory dynamics, which is where
relevance degrades empirically.

Conversely, once $\delta+T_w\gg\tau$ the fluctuating part is $O(\beta)$ and
only the rank-one constant part survives, so the matrix becomes ill-conditioned
and (IV1) degrades.
\end{proof}

\begin{remark}
Proposition~\ref{prop:rel} is the theoretical basis of the empirical
window-selection principle in the main text. Instruments are strong when
$T_w\lesssim\tau$ and weak otherwise. The sufficient condition can be checked directly when correlations decay
monotonically within the window geometry; for oscillatory systems such as
FitzHugh--Nagumo and Lorenz-63 it can fail at some separations, and
nonsingularity of $\mathbf Q$ is then the operative hypothesis, verified
empirically through the first-stage $F$, which is its sample test.
\end{remark}

\subsection{Consistency and asymptotic normality}
\label{si:clt}
\begin{proof}[Proof of Theorem~\ref{thm:cons}]
The IV estimator admits the exact error decomposition (from
\eqref{eq:iv-error})
\begin{equation}
\thetahat_{\IV}-\param
=\Big(\tfrac1K\Zmat\transpose\Amat\Big)^{-1}
 \Big(\tfrac1K\Zmat\transpose\bm{\varepsilon}^{\mathrm{tot}}\Big).
\label{eq:cons-decomp}
\end{equation}
We treat the two factors in turn.

Consider first the denominator of
\eqref{eq:cons-decomp}. The summands $Z_{kj}A_{kj'}$ are functions of the path over
disjoint window pairs of a stationary $\beta$-mixing process, so they form a
stationary mixing array with the same mixing coefficients. Their common mean is
$\E[Z_{1j}A_{1j'}]$, finite under the moment assumptions. Under $\sum_{u\ge1}\beta(u)^{1-2/\nu}<\infty$, Davydov's inequality makes the autocovariances of this array summable, so $\Var(\tfrac1K\sum_k Z_{kj}A_{kj'})=O(1/K)$; the array changes with $\Delta t$, its entries being quadratures of the path, but its mixing coefficients are those of the path at the same time separations and do not depend on $\Delta t$, so the bound holds uniformly along the joint limit, and a law of large numbers for $\beta$-mixing arrays gives
$\tfrac1K\Zmat\transpose\Amat\toprob\mathbf Q$, the per-pair mean matrix
$\mathbf Q=\E[Z_1A_1\transpose]$, which is invertible by
Proposition~\ref{prop:rel}.

On the other hand, for the numerator, by
exogeneity (Proposition~\ref{prop:exog}) each summand
$Z_k\varepsilon_k^{\mathrm{tot}}$ has mean zero, so
$\E[\tfrac1K\Zmat\transpose\bm{\varepsilon}^{\mathrm{tot}}]=0$. Its variance has
three sources, each vanishing under the stated rates, namely (i) the averaging of $K$
weakly dependent mean-zero terms contributes $O(1/K)$; (ii) replacing the
continuous-time integrals by their data quadratures incurs an $O(\Delta t)$
bias per window, whose average over the $K$ windows is itself $O(\Delta t)$ and
vanishes as $\Delta t\to0$ with no condition on $K$ (the rate $K\Delta t^2\to0$ is
needed only once the sum is scaled by $\sqrt K$ in the CLT) controlled in aggregate by $K\Delta t^2\to0$; and (iii) the
observation-noise contribution scales as $\norm{R}_{\mathrm{op}}\Delta t/(\sigma_{\min}^2T_w^2)$ by the
noise-ratio bound \eqref{eq:noise-ratio}, which holds entrywise for
matrix $R$ with these norms. Hence
$\tfrac1K\Zmat\transpose\bm{\varepsilon}^{\mathrm{tot}}\toprob 0$.

Applying the continuous-mapping and Slutsky~\cite{Vaart1998} theorems to
\eqref{eq:cons-decomp}, a convergent invertible matrix times a vanishing
vector, gives $\thetahat_{\IV}-\param\toprob 0$.
\end{proof}

\begin{proof}[Proof of Theorem~\ref{thm:clt}]

Scaling \eqref{eq:cons-decomp} by $\sqrt K$ gives
$\sqrt K(\thetahat_{\IV}-\param)
=(\tfrac1K\Zmat\transpose\Amat)^{-1}\,\tfrac1{\sqrt K}\Zmat\transpose
\bm{\varepsilon}^{\mathrm{tot}}$, and the first factor converges in
probability to $\mathbf Q^{-1}$ as in the consistency proof. For the scaled
sum $\bm S_K=\tfrac1{\sqrt K}\sum_k\bm\xi_k$ with
$\bm\xi_k=Z_k\varepsilon_k^{\mathrm{tot}}$, we use the
martingale-difference property. Order the window pairs in time and let
$\mathcal G_k=\Filt_{b_k}$ be the $\sigma$-algebra generated by the
Brownian path and the measurement errors up to the end of pair $k$. Then
$\mathcal G_{k-1}=\Filt_{b_{k-1}}\subseteq\Filt_{a_k}$, because
$b_{k-1}\le a_k$, and $\bm\xi_k$ is $\mathcal G_k$-measurable. Split the total residual as
$\varepsilon_k^{\mathrm{tot}}=\tilde\varepsilon_k+\rho_k$, where
$\tilde\varepsilon_k$ collects the process-noise integral $\varepsilon_k$
and the observation-noise terms, and $\rho_k$ is the quadrature remainder
of Proposition~\ref{prop:exog}, of mean $O(\Delta t)$. The proof of that
proposition gives $\E[\tilde\varepsilon_k\mid\Filt_{a_k}]=0$ exactly.
Indeed $\varepsilon_k$ is a martingale increment beyond $a_k$, and the
observation-noise terms are functions of the errors in the target window.
Those errors are independent of $\Filt_{a_k}$ and of the path, and they
have mean zero given the path once the drift term candidates are debiased; this holds
exactly for Gaussian errors and the polynomial terms of degree at most
three and trigonometric terms used here.
For other terms the debiasing leaves a remainder of mean
$O(\norm{R}^2)$ in $\rho_k$, and the rate $\sqrt K\norm{R}^2\to0$ is then
required as well. Write $\bm\xi_k=\tilde{\bm\xi}_k+Z_k\rho_k$ with
$\tilde{\bm\xi}_k=Z_k\tilde\varepsilon_k$. The tower property, followed
by pulling the $\Filt_{a_k}$-measurable factor $Z_k$ out of the inner
conditional expectation, gives
\begin{align}
\E[\tilde{\bm\xi}_k\mid\mathcal G_{k-1}]
&=\E\big[\E[Z_k\tilde\varepsilon_k\mid\Filt_{a_k}]\,\big|\,
\mathcal G_{k-1}\big]\notag\\
&=\E\big[Z_k\,\E[\tilde\varepsilon_k\mid\Filt_{a_k}]\,\big|\,
\mathcal G_{k-1}\big]=\bm 0.
\end{align}
Thus $\{\tilde{\bm\xi}_k,\mathcal G_k\}$ is a martingale-difference
sequence. The remainder sum $\tfrac1{\sqrt K}\sum_kZ_k\rho_k$ has mean
$O(\sqrt K\,\Delta t)$ and variance $O(\Delta t^2)$, so it vanishes in
probability under the rate $K\Delta t^2\to0$ of the consistency theorem. It remains to prove the central limit
theorem for $\tilde{\bm S}_K=\tfrac1{\sqrt K}\sum_k\tilde{\bm\xi}_k$,
whose covariance $\bm\Sigma=\Var(\tilde{\bm\xi}_1)$ agrees with
$\Var(Z_1\varepsilon_1^{\mathrm{tot}})$ up to $O(\Delta t)$. The
sequence is strictly stationary, since each $\tilde{\bm\xi}_k$ is the same
fixed functional of the path over pair $k$ and the pairs are translates of
one another by the stride $2T_w+\delta$ of a stationary process. It is
ergodic, since $\beta(u)\to0$ makes the underlying shift mixing, hence
ergodic. It is square integrable with $\bm\Sigma$ finite by assumption.
For such sequences the martingale central limit theorem
applies~\cite{Billingsley1961,HallHeyde1980}. The sequence is indexed by $\Delta t$ as well as by $k$; the theorem is stated for arrays, its conditions below hold for every $\Delta t$, and $\bm\Sigma(\Delta t)\to\bm\Sigma$ as the window quadratures converge to their continuous-time integrals, so it applies along the joint limit. Its conditional-variance
condition,
$\tfrac1K\sum_k\E[\tilde{\bm\xi}_k\tilde{\bm\xi}_k\transpose\mid\mathcal
G_{k-1}]\toprob\bm\Sigma$, follows from the ergodic theorem applied to
the stationary sequence
$\E[\tilde{\bm\xi}_k\tilde{\bm\xi}_k\transpose\mid\mathcal G_{k-1}]$,
whose mean is $\bm\Sigma$ by the tower property. Its conditional
Lindeberg condition~\cite{HallHeyde1980} holds because the nonnegative
averages $\tfrac1K\sum_k\E[\|\tilde{\bm\xi}_k\|^2\mathbf
1\{\|\tilde{\bm\xi}_k\|>\epsilon\sqrt K\}\mid\mathcal G_{k-1}]$ have
unconditional mean
$\E[\|\tilde{\bm\xi}_1\|^2\mathbf 1\{\|\tilde{\bm\xi}_1\|>\epsilon\sqrt K\}]\to0$
by dominated convergence, so they vanish in probability. Hence
$\tilde{\bm S}_K\todist\mathcal N(\bm0,\bm\Sigma)$, so does $\bm S_K$,
and Slutsky's theorem gives
$\sqrt K(\thetahat_{\IV}-\param)\todist
\mathcal N(\bm0,\mathbf Q^{-1}\bm\Sigma\mathbf Q^{-\top})$.

For the efficiency claim, the GMM estimator \eqref{eq:gmm} with weight $\Wmat$
has asymptotic covariance
$(\mathbf Q\transpose\Wmat\mathbf Q)^{-1}\mathbf Q\transpose\Wmat\bm\Sigma\Wmat
\mathbf Q(\mathbf Q\transpose\Wmat\mathbf Q)^{-1}$. Minimizing over $\Wmat$ (a
standard generalized least-squares argument) yields the optimum at
$\Wmat\propto\bm\Sigma^{-1}$, with covariance
$(\mathbf Q\transpose\bm\Sigma^{-1}\mathbf Q)^{-1}$, the smallest asymptotic
covariance attainable with the given moment conditions~\cite{Hansen1982}.
\end{proof}

\begin{proof}[Proof of Proposition~\ref{prop:plugin}]
The proof is a routine delta-method argument. The
estimator is a continuously differentiable function of $\widehat R$,
$\|\widehat R-R\|=O_p(N^{-1/2})$ by a law of large numbers for the squared
increments of the stationary mixing path, and a weight $\Wmat\toprob\bm\Sigma^{-1}$ moves the GMM estimator by only
$o_p(K^{-1/2})$.
\end{proof}

\begin{remark}\label{rem:mixing}
The stationarity and mixing hypothesis above is justified system by system. For the models in Section~\ref{sec:numerics}, the OU diffusion is geometrically ergodic. The double-well and FitzHugh--Nagumo drifts are dissipative at infinity with nondegenerate noise, so a Lyapunov and minorization argument gives geometric ergodicity as well. Stochastic Lorenz-63 with additive nondegenerate noise falls inside the framework
of~\cite{MattinglySH2002}, which was built for this class. Cox--Ingersoll--Ross is geometrically ergodic, the stochastic Kuramoto network evolves on a compact state space with elliptic
noise and is uniformly ergodic, Lorenz-96 is similar to the Lorenz-63 model, and the
hypoelliptic underdamped Langevin equation is also covered by the same work of~\cite{MattinglySH2002}. Geometric ergodicity implies exponential
$\beta$-mixing of the stationary law, and hence the summability condition assumed above. We make no claim about sharp mixing constants. The noise-ratio condition is stated with the infimum $\sigma_{\min}$ and so presumes nondegenerate noise. It does not cover the Cox--Ingersoll--Ross process, whose diffusion vanishes at the boundary, or the position component of the hypoelliptic Langevin equation, for which Appendix~\ref{si:numerics} reports the numerical evidence instead.
\end{remark}

\section{Practical instrument and window selection}
\label{si:choice}

\subsection{Operative window}
The main text selects $L$ empirically. However, for the stationary
OU model every quantity entering the estimator is Gaussian, so the window dependence can be computed in closed form, with no simulation. Consider the drift-slope coefficient with one bump
$\varphi(u)=(u(1-u))^p$ per window, span $T_w=L\Delta t$, gap $\delta$, rate
$\kappa$, stationary variance $\sigma^2/2\kappa$, and noise variance $R$ at spacing $\Delta t$. Write $\mu=\int_0^1\varphi$, $\nu=\int_0^1\varphi^2$ and
$\chi=\int_0^1(\varphi')^2$ for the bump's shape constants, and
\begin{equation*}
n(\lambda)=\int_0^1\!\!\int_0^1\varphi(u)\varphi(w)\,e^{-\lambda|u-w|}\,du\,dw
\end{equation*}
for its correlation function within one window, which satisfies
$n(\lambda)=\mu^2\bigl(1-d_1\lambda+O(\lambda^2)\bigr)$ where $d_1$ is defined below. Since $\varphi/\mu$ is the
Beta$(p+1,p+1)$ density, its variance $v^2$ and mean absolute difference
$d_1=\E|u-w|$ are $v^2=1/[4(2p+3)]$, and $d_1=2\,[(2p+1)!]^4/[(p+1)\,(4p+3)!\,(p!)^4]$. Then, we have the following result for the OU model.

\begin{proposition}\label{prop:window}
As $T\to\infty$ with $K\asymp T/(2T_w+\delta)$ window pairs, the exactly
identified estimator obeys
$\sqrt K(\thetahat_{\IV}-\param)\todist\mathcal N(0,V(T_w,\delta))$ with
$V=\E[Z^2]\,\E[(\varepsilon^{\mathrm{tot}})^2]/\E[ZA]^2$, and
\begin{multline*}
V(T_w,\delta) \approx \frac{4\kappa^2\,e^{2\kappa(\delta+a_\ast T_w)}}
{\sigma^4\mu^4T_w^4}
\Bigl[\frac{\sigma^2}{2\kappa}\,T_w^2\,n(\kappa T_w)+R\Delta t\,\nu T_w\Bigr]\\
\times\Bigl[\sigma^2\nu T_w+\kappa^2R\Delta t\,\nu T_w
+R\Delta t\,\chi\,q(p,L)/T_w\Bigr],
\end{multline*}
where $a_\ast\in(0,1]$ is shape dependent with
$a_\ast=1-v^2\kappa T_w+O((\kappa T_w)^3)$, and $q(p,L)\ge1$ is a
discretization factor with $q\to1$ as $L/p\to\infty$. With
$\rho=\chi q\,R\Delta t/(\nu\sigma^2)$ and $a_0=1-d_1/2$, the span that
minimizes $\Var(\thetahat)=V/K$ solves, to leading order in $\kappa T_w$
and $\delta/T_w$, $\kappa a_0T_w\,(T_w^2+\rho)=\rho$, so that
$\kappa T_w^*=(\kappa^2\rho/a_0)^{1/3}$ when $\kappa^2\rho\ll1$ and
$\kappa T_w^*\to1/a_0$ when $\kappa^2\rho\gg1$.
\end{proposition}

\begin{proof}
All three moments, $\E[Z^2]$, $\E[(\varepsilon^{\mathrm{tot}})^2]$ and $\E[ZA]^2$ are double integrals of the stationary kernel
$\frac{\sigma^2}{2\kappa}\,e^{-\kappa|t-s|}$ against the bump and its translate, and the substitution $s=a_k-\delta-T_w+T_wu$,
$t=a_k+T_ww$ maps each window to $[0,1]$. In the cross moment the two
windows are separated by $\delta+T_w+T_w(w-u)$, so $\E[ZA]=(\sigma^2/2\kappa)T_w^2e^{-\kappa(\delta+T_w)}
\iint\varphi(u)\varphi(w)e^{\kappa T_w(u-w)}\,du\,dw$; the double integral
is $\mu^2(1+v^2\kappa^2T_w^2+O((\kappa T_w)^4))$, the odd terms vanishing
by the symmetry $u\leftrightarrow w$, which gives
$(\sigma^2/2\kappa)\mu^2T_w^2e^{-\kappa(\delta+a_\ast T_w)}$. For the instrument second moment
the same kernel gives $\E[Z^2]=\frac{\sigma^2}{2\kappa} T_w^2m(\kappa T_w)+R\Delta t\nu T_w(1+O(\kappa T_w))$, where the term $R\Delta t\,\nu T_w$ is added by the observation noise independence across samples.
The total residual combines the It\^o isometry with the noise in the
response, $\E[(\varepsilon^{\mathrm{tot}})^2]
=\sigma^2\nu T_w+R\Delta t\,\chi\,q/T_w+\kappa^2 R \Delta t\,\nu T_w$. The first term is just $\E[\varepsilon_k^2]$. The second term is $\Var(\delta\mathsf b_k)$ which
carries $\|\phi'\|^2\propto\chi/T_w$ and, on $L\lesssim 2p$ grid points, an
additional factor $q(p,L)\ge1$ from resolving the bump derivative
discretely. The third term is $\theta^2 \Var(\delta A_k)$. Because $\hat\theta - \theta=\sum_k Z_k \varepsilon_k^\mathrm{tot} / \sum_{k} Z_k A_k$ and the summands over window pairs are one-dependent, the
martingale CLT of the main theorems applies and the per-pair variance is $V=\E[Z^2]\,\E[(\varepsilon^{\mathrm{tot}})^2]/\E[ZA]^2$, which gives the displayed equation in the proposition. The minimizer follows by
differentiating
$\log\Var(\thetahat)=2a_\ast\kappa T_w+\log\bigl(n(\kappa T_w)/\mu^2\bigr)
+\log(1+\rho/T_w^2)$ up to a constant, the first two terms having initial
slope $(2-d_1)\kappa=2\kappa a_0$.
\end{proof}

\begin{figure}[!tb]
\centering
\includegraphics[width=\linewidth]{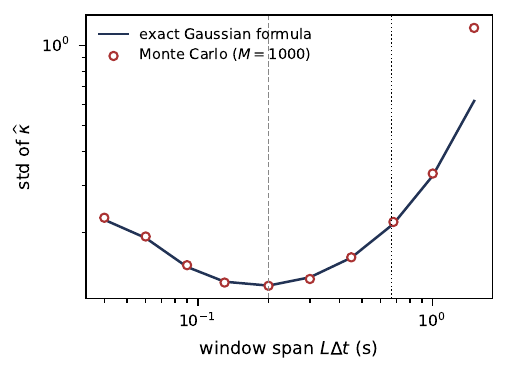}
\caption{Exact window dependence of the estimator standard deviation for
the OU benchmark. The curve evaluates the discrete Gaussian moment formula
of Proposition~\ref{prop:window} with no simulation, with the markers indicating
Monte Carlo standard deviations of the window sweep ($M=1000$) from
Fig.~\ref{fig:variance}. The dashed vertical line marks the operative
window $T_w=20\Delta t$, while the dotted line marks the relaxation time
$1/\kappa$. The formula matches the Monte Carlo points except at the longest span, more than twice $1/\kappa$, where the sample standard deviation exceeds the asymptotic value, and explains the window dependence that motivates the span rule, including the sharp loss of precision as $T_w$ approaches $1/\kappa$.}
\label{fig:wintheory}
\end{figure}

Evaluating the exact discrete Gaussian moments numerically, with no Monte
Carlo and no asymptotics, reproduces the measured standard-deviation curve
of Fig.~\ref{fig:variance} across the sweep except at its longest span, beyond twice the relaxation time (Fig.~\ref{fig:wintheory}).
The constants $a_\ast$, $a_0$ and $q$ are shape dependent, so the rule the
main text extracts, to take $L\Delta t$ a modest fraction of $\tau$ and
refine by the first-stage $F$, is calibrated empirically, not from
the proposition alone. For nonlinear systems $\kappa$ is replaced by the
empirical decorrelation rate of the relevant covariances of the drift term candidates, which is
what the first-stage $F$ measures from data.

\subsection{Depth}
The depth $n_{\text{lags}}$ is the number of past blocks that form the
instrument. Each added block lies one span further back, so its correlation
with the target window carries another factor $e^{-\kappa T_w}$, and a
second block adds little when the first one already predicts the regressors
well. That predictive power is what the first stage measures. It is the
regression of the regressors $\Amat$ on the instruments $\Zmat$ of
Section~\ref{sec:numerics} (summarized by its $F$ statistic). Two distinct failure modes call for depth two. When the first stage is weak in the sampling sense the exactly identified system inherits
weak-instrument bias toward the least-squares limit ($F<10$ here). When a residual violation of exogeneity contaminates the adjacent-block
moment, for example slowly correlated measurement noise on a real detector,
$F$ stays large while the depth-one estimate drifts. Comparing the two
depths exposes it, and the over-identified GMM weighting empirically suppresses the
contaminated direction (Section~\ref{sec:realdata}). The operational form of both is the depth rule of
Remark~\ref{rem:testfns}, with the threshold $F\ge10$ taken from the
weak-instrument literature~\cite{StaigerStock1997, StockYogo2005}. Where the average $F$ falls below $1$ the first stage carries less
signal than noise, no depth repairs it, and we regard the inference as
unreliable. The tables mark such columns n.i.\ (not identified).
Where depth two leaves $F$ between $1$ and $10$, depths three and four can be
tried, and in the stress experiments of Table~\ref{tab:stressrescue} they raise $F$ by
at most a factor of $1.6$, each added block lying a span further back and predicting
less.

\subsection{The first-stage statistic}
Writing the first-stage regression $\Amat=\Zmat\Pi+V$ column-wise, with
$\Pi$ the first-stage coefficient matrix and $V$ its residual matrix, the
reported statistic is the Cragg--Donald
$F=\lambda_{\min}\bigl(\widehat\Sigma_{VV}^{-1/2}\Amat^{\!\top}
P_{\Zmat}\Amat\,\widehat\Sigma_{VV}^{-1/2}\bigr)/m$,
where $\lambda_{\min}$ is the smallest eigenvalue,
$P_{\Zmat}=\Zmat(\Zmat^{\!\top}\Zmat)^{-1}\Zmat^{\!\top}$ is the
projection of \eqref{eq:tsls}, $m$ is the number of instrument columns,
and $\widehat\Sigma_{VV}$ is the first-stage residual
covariance~\cite{CraggDonald1993}. It is computed on one representative
test function per pair, so the effective sample units are the $N_w$
window pairs, and deterministic columns (such as the constant term) are excluded. For a
single column it reduces to the familiar first-stage $F$.

\subsection{Design rule}
The theorems of Appendix~\ref{si:clt} assume the window, gap, depth and diffusion response are fixed in advance, whereas the rule below picks them from the data. We do not prove that this step is harmless; the coverage study of Appendix~\ref{si:coverage} runs the full rule and shows that it costs little. The experiments of Section~\ref{sec:numerics} are designed according to the following choices, whose linear-case theory is given above.
\begin{description}
\item[{\small Span:}] Take $L\Delta t\in[0.1\tau,\,0.5\tau]$ for the
 characteristic time $\tau$, refined by maximizing the first-stage $F$;
 this reproduces $L=20$ for every system of Section~\ref{sec:numerics}.
 In practice $\tau$ is unknown and is replaced by a pilot estimate, the
 empirical autocorrelation time of the record, after which the
 $F$-maximization runs over a small grid of $L$; the span sweeps of the
 two applications of Section~\ref{sec:realdata} each illustrate one of them.
 The window length plays the role of the test-function support radius in
 \textsc{WSINDy} and \textsc{WENDy}, where it is chosen adaptively from
 the data's spectrum~\cite{MessengerBortz2021PDE}; an analogous automated
 rule for the stochastic setting, in which instrument relevance rather
 than smoothing governs the choice, is left to future work.
\item[{\small Gap:}] Keep $\delta$ at a few multiples of $\Delta t$, longer than the correlation length of the measurement noise yet far below $\tau$. The default $\delta=4\Delta t$ costs a variance factor $e^{2\delta/\tau}\approx1$ ($1.07$ to $1.17$ on OU, double-well, Lorenz-63, and FitzHugh--Nagumo), while correlated measurement noise calls for a longer buffer, as on the filtered detector of Section~\ref{sec:trap} (gap $8\Delta t$).

\item[{\small Depth:}] Use $n_{\text{lags}}=1$; raise it to two, with GMM
 weighting, when $F<10$ or the two depths disagree beyond their standard
 errors.
\item[{\small Slope lags:}] Take $j_{\max}\le L/2$ with
 $j_{\max}\Delta t\lesssim0.1\tau$; the default is $j_{\max}=8$.
\end{description}

\section{Simulation protocol and additional diagnostics}
\label{si:numerics}
\begin{table}[!tb]
\centering
\caption{Observation-noise levels for each system. Each row lists the observed component, the fixed noise
variance $R$ of the main analyses, the component's stationary standard
deviation (SD, computed from one long simulation after burn-in), and
$\sqrt R$ as a percentage of that SD. The last two columns give the stress
level values of Tables~\ref{tab:stressrescue} and~\ref{tab:sigmastress}: the
variance $R_{\mathrm{s}}$, fixed per component as the square of $15\%$ of
the component's range (where the range is taken over one reference record), and $\sqrt{R_{\mathrm{s}}}$
as a percentage of the same SD. The Kuramoto row is for $\sin\theta_1$ with
the noise applied to the phase. The observed Lorenz-63 value is per sensor.
FHN stands for FitzHugh--Nagumo, CIR for Cox--Ingersoll--Ross, and  Langevin for underdamped
Langevin. }
\label{tab:noiselevels}
\footnotesize
\setlength{\tabcolsep}{3pt}
\resizebox{\columnwidth}{!}{\begin{tabular}{@{}llccccc@{}}
\toprule
 &  & & Stationary  & $\sqrt R/\mathrm{SD}$ & stress $R_{\mathrm{s}}$ & $\sqrt{R_{\mathrm{s}}}/\mathrm{SD}$\\
System & Component & $R$ &  SD & \% & & \% \\
\midrule
OU & $X$ & $0.02$ & $0.47$ & $29.9$ & $0.300$ & $117$\\
DW & $X$ & $0.02$ & $0.91$ & $15.6$ & $0.288$ & $59$\\
L63 & $x$ & $0.05$ & $7.97$ & $2.8$ & $35.0$ & $74$\\
FHN & $V$ & $0.05$ & $1.59$ & $14.0$ & $0.698$ & $53$\\
CIR & $X$ & $0.01$ & $0.63$ & $16.0$ & $0.499$ & $112$\\
L96 & $x_1$ & $10^{-3}$ & $3.73$ & $0.9$ & $5.22$ & $61$\\
Kuramoto & $\sin\theta_1$ & $10^{-3}$ & $0.73$ & $4.4$ & $0.090$ & $41$\\
Langevin & $q$ & $10^{-3}$ & $1.04$ & $3.1$ & $0.338$ & $56$\\
L63 via $C$ & $Y_1$ & $10^{-4}$ & $8.18$ & $0.1$ & $32.3$ & $69$\\
\bottomrule
\end{tabular}}
\end{table}

\subsection{Simulation protocol}
Each trajectory's observation-noise variance $\widehat R$ is estimated
from that trajectory alone by the multilag variogram. When several
trajectories of the same system are available, $\widehat R$ can instead be
pooled across them, which changes the reported summaries only within
Monte Carlo error. In paired comparisons over $100$ seeds on OU, the
coherent Kuramoto network, and Lorenz-96, a single trajectory determines
$\widehat R$ to within $0.6\%$ (OU) and $0.4\%$ (the other two); the
drift estimates and the multilag slope form of
the diffusion are identical under the two choices, and the lag-one
weak-form diffusion estimate, the one form that subtracts $2\widehat R$
within each window, moves individual trajectories by about half its
standard deviation while its mean and standard deviation change by less
than $0.002$.

In the synthetic examples all trajectories are simulated by Euler--Maruyama
on a grid $n_{\mathrm{sub}}$ times finer than the observation step and then
subsampled ($n_{\mathrm{sub}}=5$ for OU, the double-well, Lorenz-63, FitzHugh--Nagumo and the Kuramoto network; $n_{\mathrm{sub}}=10$ for Cox--Ingersoll--Ross, the underdamped
Langevin, the observed Lorenz-63, and Lorenz-96), with $M=1000$ independent
trajectories per model, run in parallel.
Measurement noise $\eta_n\sim\mathcal{N}(0,R)$ is then added to the
simulated states. Reported means and standard deviations are computed across
the 1000 per-trajectory estimates.

\subsection{Debiasing of the term candidates}
With $Y=X+\eta$ and $\eta\sim\mathcal N(0,R)$ independent of the state, the
odd moments of $\eta$ vanish, so expanding the powers gives
$\E[Y^2\mid X]=X^2+R$, $\E[Y^3\mid X]=X^3+3XR$, and in general only the even
moments of the noise survive, contributing powers of $R$. For drift term candidates $x^2$ and $x^3$ we therefore apply the corrections of
Section~\ref{sec:algorithm}, $\widehat{x^2}\mapsto\widehat{Y^2}-\widehat R$ (and $\widehat{x_ix_j}\mapsto\widehat{Y_iY_j}-\widehat R_{ij}$ for cross terms, which is why the multilag construction of Section~\ref{sec:iv} returns the full matrix) and
$\widehat{x^3}\mapsto\widehat{Y^3}-3\widehat R\,\widehat Y$. Without these
corrections the higher-order columns carry an $O(R)$ bias that
contaminates the drift estimate independently of the endogeneity correction.
The same principle extends to any term candidate $g$. Evaluating $g$ at noisy
arguments blurs it, in that the expectation over the noise, $\E[g(Y)\mid X]$, is the
convolution of $g$ with the density of $\eta$, so $g$ is smoothed in its state
argument on the scale $\smash{\widehat R^{1/2}}$ and the observed values
$g(Y_n)$ scatter around this blurred function instead of around $g$. This
smoothing is distinct from the smoothing performed by the test functions,
which averages over \emph{time} and suppresses the fluctuations of the noise,
whereas the blur is a mean shift, proportional to $R$ and to the curvature of $g$
and identical at every sample, which does not vanish under the temporal
averaging of the weak form. The debiased entry $\widetilde g$ is the
deconvolution, the function whose blur is $g$.
Since $\E[\cos(\omega Y)\mid X]=e^{-\frac12\omega^2 R}\cos(\omega X)$ and
$\E[\sin(\omega Y)\mid X]=e^{-\frac12\omega^2 R}\sin(\omega X)$,
trigonometric terms are eigenfunctions of the blur and are debiased by the multiplicative factor
$e^{\frac12\omega^2\widehat R}$. More generally the debiased entry is
$\widetilde g(Y)=(e^{-\frac12\widehat R\,\partial^2}g)(Y)$, the inverse
heat semigroup applied to $g$, computed via the Taylor expansion
$\sum_k(-\frac12\widehat R)^k\partial^{2k}g/k!$, which truncates exactly
for polynomials and recovers the polynomial rules above. For term candidates that are neither polynomial nor trigonometric, rational or saturating drifts for instance, the deconvolution is ill posed, since the blur damps high frequencies by the factor $e^{-\frac12\omega^2 R}$ that its inverse must amplify; the correction is then applied numerically by Gauss--Hermite quadrature or a low-order Taylor expansion truncated at the variance scale $\widehat R$, which regularizes the inverse by discarding the amplified high-frequency content.

\subsection{IV versus GMM}
At
the default depth $n_{\mathrm{lags}}$ equals one, the moment system is exactly identified and GMM
coincides with IV. When the depth rule selects two blocks, the
reported estimates are two-step GMM. On systems with strong first stages the two variants differ by far less than one Monte
Carlo standard deviation.

\subsection{Ensemble size and computational cost}
In our synthetic examples, the ensemble size $M=1000$ is a reporting choice, not a requirement. Setting
$M=200$, for example, reproduces every conclusion of the main text, with operative-window
table entries agreeing to a few percent. An ensemble as small as $M=5$ already displays the estimator's accuracy (as in the optical trap example). Once the weak-form integrals are formed, one drift recovery is a single linear solve. It takes
tens of milliseconds per trajectory on one CPU core in our reference
implementation; a measured comparison with the Kalman maximum-likelihood
baseline appears in Appendix~\ref{si:baselines}.

\subsection{Off-diagonal noise entries}
The scalar series $S^{\pm}_{t_n}=Y^i_{t_n}\pm Y^j_{t_n}$ carries noise with
variance $R_{ii}+R_{jj}\pm2R_{ij}$, so fitting its scalar variogram
$v^{\pm}_k=\frac{1}{N-k}\sum_{n}(S^{\pm}_{t_{n+k}}-S^{\pm}_{t_n})^2$ on
$(1,k,k^2)$ and halving the intercept yields $\widehat R_{\pm}$ estimating
$R_{ii}+R_{jj}\pm2R_{ij}$, whence
\begin{align}\label{eq:R-polarization}
 \widehat R_{ij}=
\frac{\widehat R_{+}-\widehat R_{ii}-\widehat R_{jj}}{2}
 =\frac{\widehat R_{ii}+\widehat R_{jj}-\widehat R_{-}}{2}
 =\frac{\widehat R_{+}-\widehat R_{-}}{4}.
\end{align}
The symmetric form is preferred for two reasons. It needs neither diagonal
entry, and the fits behind $\widehat R_+$ and $\widehat R_-$ share the same
drift and diffusion nuisance functionals, so their errors are positively
correlated and partially cancel in the difference. Specializing the
diagonal fit to $n_{\mathrm v}=2$ ($n_{\mathrm v}$
counts the variogram lags of Section~\ref{sec:iv}) recovers the classical lag-1/lag-2 form
$\widehat R_{ii}=\tfrac12(2v_1-v_2)$, which returns zero, or a clipped
negative value, as soon as the drift bias dominates.

\subsection{Derivation and defaults of the multilag slope estimator}
By \eqref{eq:Vk-expansion}, $\E[S_{k,i}(j)]$ is quadratic in the lag $j$,
and $\sigma_i^2\Delta t$ is its linear coefficient. Extracting that coefficient is a fixed linear operation on the $S_{k,i}(j)$, whose weights $(w_j)$ in \eqref{eq:qv-vgram} form the linear-coefficient row of the least-squares $(1,j,j^2)$ design over $j=2,\dots,j_{\max}$. The fit
starts at $j=2$, so the most distorted lag never enters. We use
$j_{\max}=8$, which requires $L>j_{\max}$; the choice is examined with
the other design choices in the main text and in Section~\ref{si:slope}.

\subsection{Sensitivity to the number of variogram lags}
On the OU benchmark the estimates are stable across
variogram lag counts $n_{\mathrm v}\in\{3,\dots,10\}$. The recovered $\widehat R$ stays within $0.1\%$ of
the truth at every $n_{\mathrm v}$, with spread shrinking from $2.3\times10^{-4}$ at
$n_{\mathrm v}=3$ to $1.0\times10^{-4}$ at $n_{\mathrm v}=10$ against $R=0.02$, and
$\widehat\sigma$ stays within $0.3\%$ of $0.8$ throughout ($M=400$). The
default $n_{\mathrm v}=6$ lies in the flat region.

\subsection{Stress noise level}
Tables~\ref{tab:stressrescue} and~\ref{tab:sigmastress} report every system
at the stress noise level of Table~\ref{tab:noiselevels}, a fixed variance
per component equal to the square of $15\%$ of the component's range over a
record of the length used in the experiments. Each trajectory 
estimates its own $\widehat R$, which recovers the stress variances to
within $0.03\%$ for every component of OU, the double-well, Lorenz-63 and FitzHugh--Nagumo. At this level
the FitzHugh--Nagumo and Lorenz-63 first stages collapse at depth one
(average $F<1$), and the depth rule selects two instrument
blocks at the unchanged window and gap, whereas the OU and double-well first
stages remain strong ($F=3005$ and $66$) and depth one is the reported
estimator; the tables show depths one to four for every system. Table~\ref{tab:sigmastress} lists the stress-level ratio $\varrho_{\mathrm s}$ next to the
diffusion estimates; where it exceeds $10^4$ (the chaotic systems) the increments carry no
information on $\sigma$ and the cells read n.i.

The $F$ rows of Table~\ref{tab:stressrescue} also separate
what the first stage does and does not control. Where essentially every
replication falls below the threshold, the estimates acquire the predicted
attenuation toward the least-squares limit, with the FitzHugh--Nagumo
current coming in $24\%$ low at depth two, about ten Monte Carlo standard
errors, and the per-trajectory distributions develop heavy tails, as in the
Cox--Ingersoll--Ross column at $L=100$ ($F=2.4$, with $97\%$ of
replications below $10$). In the columns marked n.i.\ the failure is complete, with point estimates orders of magnitude from the truth ($\widehat b$ in the thousands against $b=28$), as the ratio mechanism of Appendix~\ref{si:coverage} predicts. Where no replication falls below the threshold, the estimates are centered with reliable standard errors, but a large $F$ does not promise precision. The Cox--Ingersoll--Ross spreads at the operative window ($F=108$, none below $10$) remain wide because at this noise level ($\sqrt{R_{\mathrm{s}}}$ above the component's stationary SD, Table~\ref{tab:noiselevels}) the record itself carries little information, and the $L=5$ column pairs a larger $F$ ($287$) with a wider spread still.
The threshold $F\ge10$ marks the regime in which per-trajectory
inference is valid, with Wald coverage of $93$--$95\%$ at $F\approx3$
and $90\%$ at $F\approx1$ (Appendix~\ref{si:coverage}); precision is read
from the standard errors it validates.

\begin{table*}[!t]
\centering
\caption{Drift recovery at the stress noise level of
Table~\ref{tab:noiselevels} (the per-component variance $R_{\mathrm{s}}$,
the square of $15\%$ of the component's range). Layout and conventions
follow Table~\ref{tab:drift}; the $L=100$ and $L=5$ columns report raw IV
at depth one, while the operative window $L=20$ shows all four depths: raw
IV at depth one, and two-step GMM with two, three, and four instrument
blocks at depths two to four. The depth rule (Remark~\ref{rem:testfns})
raises the depth to two when the depth-one $F$ falls below $10$ or the two
depths disagree beyond their standard errors. The last row of each block
gives the average first-stage $F$, with the percentage of replications
whose own $F$ falls below $10$ in parentheses. Note that where the average $F$ falls
below $1$ we regard the inference as unreliable and the cells read
n.i.\ (see the caption of Table~\ref{tab:drift} and
Appendix~\ref{si:coverage}).}
\label{tab:stressrescue}
\footnotesize
\setlength{\tabcolsep}{3pt}\begin{tabular}{llccccccc}
\toprule
 & & & \multicolumn{4}{c}{$L=20$, depth} & & \\
\cmidrule(lr){4-7}
System & Parameter & $L=100$ & one & two & three & four & $L=5$ & True\\
\midrule
\multirow{3}{*}{OU}
 & $\kappa$ & 1.534\,(0.354) & 1.494\,(0.241) & 1.503\,(0.247) & 1.495\,(0.243) & 1.526\,(0.246) & 1.511\,(0.951) & 1.500\\
 & $\mu$ & 1.999\,(0.028) & 2.001\,(0.052) & 2.000\,(0.052) & 2.001\,(0.050) & 1.996\,(0.052) & 3.346\,(41.507) & 2.000\\
 & 1st-stage $F$ & 72\,(0\%) & 3005\,(0\%) & 1522\,(0\%) & 1016\,(0\%) & 762\,(0\%) & 6179\,(0\%) & ---\\
\midrule
\multirow{3}{*}{DW}
 & $\alpha$ & 1.158\,(3.087) & 0.974\,(0.548) & 0.995\,(0.499) & 0.997\,(0.499) & 0.967\,(0.477) & 1.019\,(2.908) & 1.000\\
 & $\beta$ & 1.152\,(2.881) & 0.976\,(0.500) & 1.004\,(0.454) & 1.001\,(0.453) & 0.973\,(0.434) & 1.008\,(2.588) & 1.000\\
 & 1st-stage $F$ & 3.2\,(99\%) & 66\,(0\%) & 41\,(0\%) & 29\,(0\%) & 23\,(0\%) & 52\,(0\%) & ---\\
\midrule
\multirow{4}{*}{L63}
 & $a$ & n.i. & n.i. & 9.901\, (9.303) & 9.494 \,(7.727)  & 9.730 \,(6.755)  & n.i. & 10.000\\
 & $b$ & n.i. & n.i. & 27.748\,(15.932) & 27.992\,(13.622) & 27.024\,(11.749) & n.i. & 28.000\\
 & $c$ & n.i. & n.i. & 2.200\,(16.319) & 1.757\,(10.608) & 1.983\,(9.111) & n.i. & 2.667\\
 & 1st-stage $F$ & $<1$\,(100\%) & $<1$\,(100\%) & 1.0\,(100\%) & 1.6\,(100\%) & 1.6\,(100\%) & $<1$\,(100\%) & ---\\
\midrule
\multirow{5}{*}{FHN}
 & $I$ & n.i. & n.i. & 0.382\,(0.379) & 0.327\,(0.336) & 0.321\,(0.311) & 0.633\,(5.714) & 0.500\\
 & $\epsilon$ & n.i. & n.i. & 0.079\,(0.583) & 0.068\,(0.480) & 0.098\,(0.396) & $-1.391$\,(50.863) & 0.080\\
 & $a$ & n.i. & n.i. & 0.683\,[0.44,1.12] & 0.557\,[0.27,0.92] & 0.738\,[0.55,0.98] & 0.283\,[unbnd] & 0.700\\
 & $b$ & n.i. & n.i. & 0.737\,[0.49,1.22] & 0.941\,[0.67,1.56] & 0.522\,[0.33,0.73] & 0.489\,[unbnd] & 0.800\\
 & 1st-stage $F$ & $<1$\,(100\%) & $<1$\,(100\%) & 1.3\,(100\%) & 1.3\,(100\%) & 1.3\,(100\%) & 2.0\,(100\%) & ---\\
\midrule
\multirow{3}{*}{CIR}
 & $\kappa\mu$ & 4.02\,(67.64) & 3.07\,(2.08) & 3.23\,(2.09) & 3.26\,(2.03) & 3.10\,(2.04) & 3.19\,(6.80) & 3.00\\
 & $-\kappa$ & $-2.63$\,(43.33) & $-2.06$\,(1.39) & $-2.15$\,(1.38) & $-2.20$\,(1.34) & $-2.05$\,(1.35) & $-2.13$\,(4.46) & $-2.00$\\
 & 1st-stage $F$ & 2.4\,(97\%) & 108\,(0\%) & 55\,(0\%) & 37\,(0\%) & 28\,(0\%) & 287\,(0\%) & ---\\
\midrule
\multirow{2}{*}{L96 ($d=120$)}
 & $F$ (forcing) & 7.99\,(0.37) & 8.21\,(1.63) & 7.96\,(1.79) & 8.03\,(1.94) & 7.90\,(1.79) & 8.20\,(9.47) & 8.00\\
 & 1st-stage $F$ & 891\,(0\%) & 20194\,(0\%) & 11091\,(0\%) & 7543\,(0\%) & 5849\,(0\%) & 15999\,(0\%) & ---\\
\midrule
\multirow{5}{*}{Kuramoto}
 & $\omega$ & 0.50\,(0.01) & 0.50\,(0.05) & 0.50\,(0.05) & 0.50\,(0.05) & 0.50\,(0.05) & 0.49\,(0.21) & 0.50\\
 & $K$ (coh.) & 2.01\,(0.20) & 2.00\,(0.41) & 2.02\,(0.39) & 2.02\,(0.39) & 2.00\,(0.39) & 1.99\,(2.17) & 2.00\\
 & 1st-stage $F$ & 758\,(0\%) & 7528\,(0\%) & 4250\,(0\%) & 2879\,(0\%) & 2159\,(0\%) & 6732\,(0\%) & ---\\
 & $K$ (incoh.) & 0.81\,(0.08) & 0.80\,(0.22) & 0.80\,(0.23) & 0.80\,(0.22) & 0.81\,(0.22) & 0.77\,(1.09) & 0.80\\
 & 1st-stage $F$ & 3193\,(0\%) & 25518\,(0\%) & 14279\,(0\%) & 9666\,(0\%) & 7240\,(0\%) & 25441\,(0\%) & ---\\
\midrule
\multirow{6}{*}{Langevin}
 & $p$ in $\dot q$ & 1.00\,(0.16) & 1.01\,(0.50) & 1.03\,(0.51) & 0.99\,(0.50) & 0.99\,(0.49) & 0.99\,(2.42) & 1.00\\
 & $c_{q^3}$ & $-1.01$\,(0.66) & $-1.04$\,(0.81) & $-0.98$\,(0.77) & $-1.01$\,(0.77) & $-0.97$\,(0.76) & $-1.05$\,(5.21) & $-1.00$\\
 & $c_p$ ($\gamma=0.5$) & $-0.52$\,(0.39) & $-0.50$\,(0.74) & $-0.46$\,(0.72) & $-0.51$\,(0.74) & $-0.47$\,(0.73) & $-0.62$\,(3.40) & $-0.50$\\
 & 1st-stage $F$ & 5.6\,(90\%) & 96\,(0\%) & 56\,(0\%) & 39\,(0\%) & 30\,(0\%) & 53\,(0\%) & ---\\
 & $c_p$ ($\gamma=2$) & $-2.05$\,(1.24) & $-2.01$\,(1.26) & $-2.04$\,(1.24) & $-1.95$\,(1.20) & $-1.96$\,(1.26) & $-2.15$\,(5.18) & $-2.00$\\
 & 1st-stage $F$ & 4.7\,(98\%) & 61\,(0\%) & 39\,(0\%) & 27\,(0\%) & 21\,(1\%) & 30\,(2\%) & ---\\
\midrule
\multirow{8}{*}{\shortstack[l]{L63\\$Y=CX+\eta$}}
 & $x$ in $\dot x$ & n.i. & n.i. & n.i. & n.i. & n.i. & n.i. & $-10.00$\\
 & $y$ in $\dot x$ & n.i. & n.i. & n.i. & n.i. & n.i. & n.i. & 10.00\\
 & $x$ in $\dot y$ & n.i. & n.i. & n.i. & n.i. & n.i. & n.i. & 28.00\\
 & $y$ in $\dot y$ & n.i. & n.i. & n.i. & n.i. & n.i. & n.i. & $-1.000$\\
 & $xz$ in $\dot y$ & n.i. & n.i. & n.i. & n.i. & n.i. & n.i. & $-1.000$\\
 & $z$ in $\dot z$ & n.i. & n.i. & n.i. & n.i. & n.i. & n.i. & $-2.667$\\
 & $xy$ in $\dot z$ & n.i. & n.i. & n.i. & n.i. & n.i. & n.i. & 1.000\\
 & 1st-stage $F$ & $<1$\,(100\%) & $<1$\,(100\%) & $<1$\,(100\%) & $<1$\,(100\%) & $<1$\,(100\%) & $<1$\,(100\%) & ---\\
\bottomrule
\end{tabular}
\end{table*}

\begin{table*}[!t]
\centering
\caption{Diffusion recovery at the stress noise level of Table~\ref{tab:noiselevels} at $L=20$, with the conventions of Table~\ref{tab:sigma}, at depths one to four. The column $\varrho_{\mathrm s}=2R_{\mathrm s}/(\sigma^2\Delta t)$ gives the noise-to-signal ratio of the increments at the stress level (Section~\ref{sec:iv}). For the Lorenz-63 observed through $C$ it is computed on the deconvolved record, and for Cox--Ingersoll--Ross the state-dependent $\sigma^2(x)$ is evaluated at the mean $\mu$. Note that where $\varrho_{\mathrm s}$ exceeds $10^4$ the diffusion signal lies four orders of magnitude below the noise floor and the increments carry essentially no information on $\sigma$. We label these cells as n.i.\ (not identified).}
\label{tab:sigmastress}
\scriptsize
\setlength{\tabcolsep}{2.5pt}\resizebox{\textwidth}{!}{\begin{tabular}{lcccccccccc}
\toprule
 & & & \multicolumn{2}{c}{depth one} & \multicolumn{2}{c}{depth two} & \multicolumn{2}{c}{depth three} & \multicolumn{2}{c}{depth four}\\
\cmidrule(lr){4-5}\cmidrule(lr){6-7}\cmidrule(lr){8-9}\cmidrule(lr){10-11}
System & $\sigma$ (true) & $\varrho_{\mathrm s}$ & QV & slope & QV & slope & QV & slope & QV & slope\\
\midrule
OU & 0.800 & 94 & 0.738\,(0.345) & 0.763\,(0.250) & 0.766\,(0.315) & 0.764\,(0.245) & 0.734\,(0.334) & 0.757\,(0.248) & 0.751\,(0.324) & 0.764\,(0.245)\\
DW & 0.500 & 231 & 0.444\,(0.348) & 0.448\,(0.304) & 0.469\,(0.354) & 0.437\,(0.301) & 0.453\,(0.346) & 0.450\,(0.309) & 0.477\,(0.353) & 0.443\,(0.296)\\
FHN ($V$) & 0.300 & 310 & 1.110\,(5.703) & 0.281\,(0.248) & 0.410\,(0.301) & 0.281\,(0.242) & 0.369\,(0.298) & 0.279\,(0.249) & 0.371\,(0.292) & 0.285\,(0.244)\\
FHN ($W$) & 0.300 & 457 & 0.844\,(4.905) & 0.287\,(0.296) & 0.370\,(0.351) & 0.317\,(0.289) & 0.363\,(0.341) & 0.296\,(0.293) & 0.353\,(0.344) & 0.310\,(0.294)\\
L63 ($x$) & 1.000 & $3.5\times10^{4}$ & n.i. & n.i. & n.i. & n.i. & n.i. & n.i. & n.i. & n.i.\\
L63 ($y$) & 1.000 & $6.4\times10^{4}$ & n.i. & n.i. & n.i. & n.i. & n.i. & n.i. & n.i. & n.i.\\
L63 ($z$) & 1.000 & $4.8\times10^{4}$ & n.i. & n.i. & n.i. & n.i. & n.i. & n.i. & n.i. & n.i.\\
\midrule
CIR ($\nu$) & 1.000 & 66 & 0.92\,(0.80) & 0.882\,(0.689) & 0.90\,(0.80) & 0.958\,(0.683) & 0.82\,(0.77) & 0.886\,(0.691) & 0.75\,(0.77) & 0.936\,(0.684)\\
L96 ($d=40$) & 0.500 & $6.5\times10^{4}$ & n.i. & n.i. & n.i. & n.i. & n.i. & n.i. & n.i. & n.i.\\
L96 ($d=120$) & 0.500 & $6.4\times10^{4}$ & n.i. & n.i. & n.i. & n.i. & n.i. & n.i. & n.i. & n.i.\\
Kuramoto ($K=2.0$) & 0.300 & 400 & 0.280\,(0.111) & 0.465\,(0.042) & 0.280\,(0.111) & 0.465\,(0.042) & 0.280\,(0.111) & 0.465\,(0.042) & 0.280\,(0.111) & 0.465\,(0.042)\\
Kuramoto ($K=0.8$) & 0.300 & 400 & 0.274\,(0.111) & 0.461\,(0.042) & 0.274\,(0.111) & 0.461\,(0.042) & 0.274\,(0.111) & 0.461\,(0.042) & 0.274\,(0.111) & 0.461\,(0.042)\\
Langevin ($p$, $\gamma=0.5$) & 1.000 & 258 & 0.948\,(0.762) & 1.031\,(0.849) & 0.946\,(0.762) & 1.031\,(0.849) & 0.947\,(0.762) & 1.031\,(0.849) & 0.946\,(0.762) & 1.031\,(0.849)\\
Langevin ($p$, $\gamma=2$) & 2.000 & 120 & 1.791\,(1.012) & 1.771\,(1.176) & 1.790\,(1.012) & 1.771\,(1.176) & 1.789\,(1.012) & 1.771\,(1.176) & 1.788\,(1.012) & 1.771\,(1.176)\\
L63 via $C$ ($x$) & 0.500 & $1.9\times10^{6}$ & n.i. & n.i. & n.i. & n.i. & n.i. & n.i. & n.i. & n.i.\\
L63 via $C$ ($y$) & 0.400 & $1.4\times10^{6}$ & n.i. & n.i. & n.i. & n.i. & n.i. & n.i. & n.i. & n.i.\\
L63 via $C$ ($z$) & 0.550 & $2.0\times10^{5}$ & n.i. & n.i. & n.i. & n.i. & n.i. & n.i. & n.i. & n.i.\\
\bottomrule
\end{tabular}}
\end{table*}

\section{Standard errors, interval coverage, and Normality}
\label{si:coverage}

\begin{table}[!tb]
\centering
\caption{Calibration of the plug-in (Newey--West)
standard errors and of the Wald intervals at each benchmark's operative
record length $T$ (second column, with the number of replications $M$).
The SE ratio is the mean plug-in standard error divided by the sampling
standard deviation of the estimates across replications, so a value of
one indicates a calibrated standard error. When a range is provided, that range is the range of the SE ratios for the listed
parameters. Wald coverage is the fraction of replications whose nominal
$95\%$ interval contains the truth. (For calibration, note that a binomial Monte Carlo standard
error at $M=1000$ for $p=0.95$ equals $0.7$ percentage points.) Across all four systems the plug-in standard errors match the sampling variability to within a few percent, while the Wald intervals attain near-nominal coverage.}
\label{tab:coverage}
\footnotesize
\setlength{\tabcolsep}{3pt}\resizebox{\columnwidth}{!}{\begin{tabular}{lllcc}
\toprule
System & $T$, $M$ & Parameters & SE ratio & Wald coverage\\
\midrule
OU & $2000$, $1000$ & $\kappa$ & $0.99$ & $94.3\%$\\
 & $2000$, $1000$ & $\mu$ & $1.00$ & $95.3\%$\\
DW & $2000$, $1000$ & four drift term candidates & $0.98$--$0.99$ & $94.1$--$94.3\%$\\
L63 & $400$, $1000$ & 7 nonzero terms & $0.96$--$1.02$ & $93.2$--$95.5\%$\\
FHN & $5000$, $1000$ & 7 nonzero terms & $0.95$--$1.01$ & $94.0$--$96.1\%$\\
\bottomrule
\end{tabular}}
\end{table}

\begin{figure*}[!t]
\centering
\includegraphics[width=\linewidth]{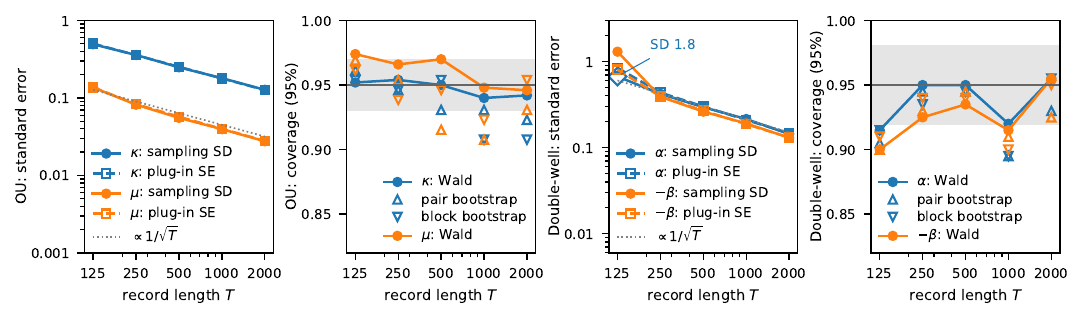}
\caption{Plug-in (Newey--West) standard errors and Wald-interval coverage as the record length $T$ grows, at $L=20$: OU in the left pair of panels ($M=500$), the double-well in the right pair ($M=200$). In the first panel of each pair, the sampling standard deviation across replications (solid) is compared with the mean plug-in standard error (dashed) and with the $T^{-1/2}$ reference rate (dotted). For the double-well at $T=125$, where a few extreme records dominate the sample standard deviation, we use  the robust spread (IQR$/1.349$) instead of the sample standard deviation, and indicate that by the open diamond.  In the second panel of each pair, the coverage of nominal $95\%$ Wald intervals (lines) is compared with that of the pair bootstrap (upward open triangles) and the moving-block bootstrap (downward open triangles). The shaded band is the binomial band $0.95 \pm 1.96\sqrt{0.95\cdot 0.05/M}$ for an empirical coverage at the nominal level of $95\%$, computed using the full $M$. For OU, where the bootstrap was run on $130$ of the $500$ replications, the triangles carry about twice the Monte Carlo error of the band. The plug-in standard error closely tracks the sampling variability at the $T^{-1/2}$ rate, while the Wald intervals hold near-nominal coverage at every record length $T$, as corroborated by the pair bootstrap.}
\label{fig:coverage}
\end{figure*}

\begin{figure*}[!t]
\centering
\includegraphics[width=\textwidth]{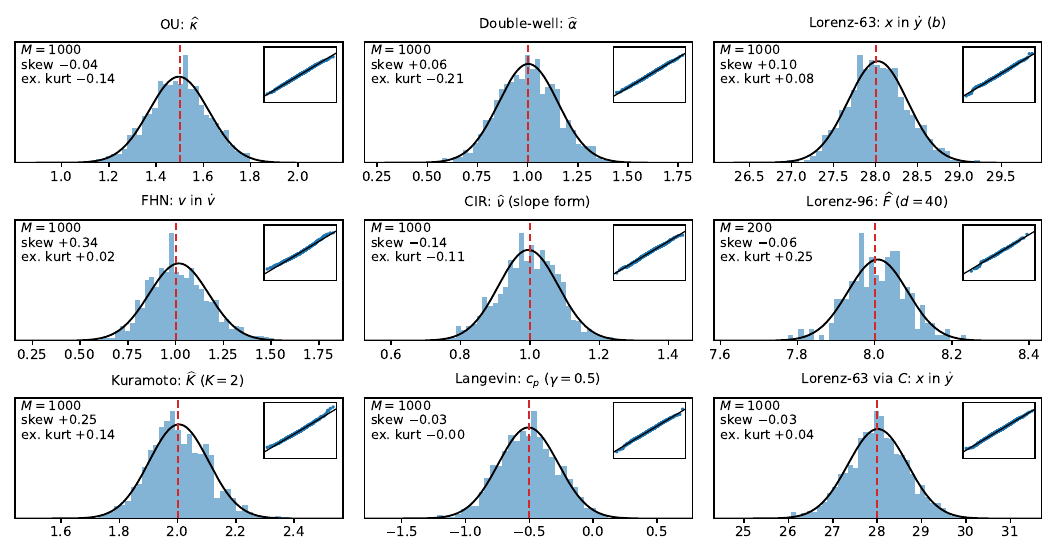}
\caption{Sampling distributions of single-trajectory estimates for each
of the nine systems: histogram of the $M$ estimates of a representative
parameter, fitted Normal density (black), true value (red dashed), and
Normal QQ plot (inset), with skewness and excess kurtosis. The Langevin
panel shows the raw IV estimate, the FitzHugh--Nagumo panel shows the
fast-equation coefficient with the heaviest tails, and the CIR panel shows
the slope-form diffusion estimate. Although Theorem~\ref{thm:clt} is an asymptotic result, the sampling distributions are already close to Gaussian at these finite record lengths, with the printed skewness and excess kurtosis indicating minor remaining departures. This approximate normality is what justifies the Wald intervals of Table~\ref{tab:coverage}.}
\label{fig:hist9}
\end{figure*}

\subsection{Plug-in standard errors}
At depth one the estimator~\eqref{eq:iv-estimator} is exactly identified,
and its error decomposes over window pairs,
$\widehat{\bm\theta}-\bm\theta=G^{-1}\sum_{k=1}^{K}g_k$, where
$G=\sum_k Z_k^\top A_k$ collects the instrument--regressor products of
pair $k$ and $g_k=Z_k^\top\varepsilon_k$ is that pair's score. By
Theorem~\ref{thm:clt} the population scores form a martingale-difference
sequence, so they are uncorrelated across pairs whatever the system's
memory. In finite samples, correlation can enter through the quadrature
remainder $\rho_k$, through evaluating the scores at $\widehat{\bm\theta}$,
and, on real data, through correlated measurement noise. The plug-in
covariance therefore includes a long-run (Newey--West) term as a safeguard;
on the benchmarks it changes the standard errors little (on the AMO
operative window, $0.284$ with it and $0.289$ without),
\begin{align*}
\widehat{\mathrm{Var}}(\widehat{\bm\theta})
={}&G^{-1}\Bigl[\textstyle\sum_k \hat g_k\hat g_k^\top\\
&\textstyle{}+\sum_{\ell=1}^{q}\bigl(1-\tfrac{\ell}{q+1}\bigr)
\sum_k\bigl(\hat g_k\hat g_{k+\ell}^\top+\hat g_{k+\ell}\hat g_k^\top\bigr)\Bigr]
G^{-\top},
\end{align*}
with residual scores $\hat g_k$, Bartlett weights~\cite{NeweyWest1987},
and the automatic bandwidth $q=\lfloor4(K/100)^{2/9}\rfloor$~\cite{NeweyWest1994}
(e.g.\ $q=9$ for OU at $T=2000$). Each equation is treated separately;
derived quantities use the delta method, and ratios use Fieller intervals
as in the main text. The computation needs only the per-pair moments
already formed by the estimator.

\subsection{Design}
For OU ($M=500$) and the double-well ($M=200$) each replication is
simulated once at $T=2000$, and its first $T\in\{125,250,500,1000,2000\}$
time units give nested records of increasing length; estimates are
independent across replications at every $T$. All runs use $L=20$, the default gap $\delta=4\Delta t$, depth one, and each record's own $\widehat R$. For Lorenz-63 and FitzHugh--Nagumo the check is made at the operative record length
($M=1000$; $T=400$ for Lorenz-63 and $T=5000$ for FitzHugh--Nagumo,
whose fast-equation and slow-equation model coefficients are assessed
directly, as these, and not the derived ratios $a,b$, are the
asymptotically Normal quantities). As a comparison, pair-bootstrap
intervals (independent resampling of window pairs) and moving-block
intervals (blocks of $q$ pairs~\cite{Kunsch1989}) are computed with $B=200$ resamples on a
subset of replications.

\subsection{Results}
Fig.~\ref{fig:coverage} shows that the mean plug-in standard error tracks
the sampling standard deviation across replications at every record length
from $T=250$ to $2000$, and that both decay at the $T^{-1/2}$ rate. Wald
coverage of nominal $95\%$ intervals lies within or near the Monte Carlo
band at every length. Table~\ref{tab:coverage} summarizes the four systems at their operative lengths, where the ratio of mean plug-in standard error to sampling standard deviation lies between $0.95$ and $1.02$ for every parameter and Wald coverage between $93.2\%$ and $96.1\%$, within or near the Monte Carlo band in every case. For the $23$ drift term candidates of
Lorenz-63 whose true value is zero, the Wald interval contains zero in
$94.7\%$ of replications on average, so a Wald test for including a term
is also calibrated. The bootstrap intervals agree with the Wald intervals
within Monte Carlo error throughout and do not improve on them; the
moving-block bootstrap undercovers slightly at the longest records ($91\%$ for the OU rate at $T=1000$ and $2000$).

At $T=125$ the double-well's sampling standard deviation is dominated by
a few of the $200$ records (errors of $22$, $5$, and $4$ in $\widehat\alpha$) whose
trajectories visit both wells only a few times. These records report
correspondingly large plug-in standard errors, so coverage is only mildly affected ($92\%$ for $\alpha$ and $90\%$ for $\beta$, with a Monte Carlo standard error of $1.5$ points); a very large standard error on a short record is a signal of
weak identification and should be read as such.

\subsection{Probability limits, not finite-sample means}
The analysis rests on probability limits and limiting distributions rather
than on finite-sample means for a reason specific to instrumental variables.
In the classical linear IV model, with fixed instruments and Gaussian errors, the number of finite moments of the IV estimator grows with the degree of over-identification~\cite{Kinal1980}, and the exactly identified estimator is heavy-tailed, behaving as a ratio whose denominator has positive density at zero. That model does not describe \textsc{WAIVES} in full, since its
instruments are random and its errors are stochastic integrals, but the same
ratio mechanism is present in the depth-one estimator, which is exactly
identified. With a strong instrument the tail mass is negligible, and the
sample means and standard deviations reported in the tables are stable. In the deterministic output-error setting, Kuang and
Lin~\cite{KuangLin2026} confront the same pathology by regularizing the
estimator, through singular-value clipping and instrument truncation. We
instead leave the estimator unregularized, report probability limits, and
rely on over-identification with two-step GMM, which lightens the tails at depths beyond one and has standard asymptotic guarantees.  As
the instrument weakens they become erratic, as in the Lorenz-63 entries at
$L=100$ in Table~\ref{tab:drift} (first-stage $F<1$),
and medians and quantiles become the more reliable
summaries~\cite{StaigerStock1997}. Escalating to depth two over-identifies
the system, which in the classical model lightens the tails. The ratio parameters of FitzHugh--Nagumo are
summarized by Marsaglia modes and Fieller intervals for the same reason.

\subsection{Normality}
Fig.~\ref{fig:hist9} shows the sampling distribution of a representative
parameter for each of the nine systems. All are close to Normal, with
skewness and excess kurtosis within about $\pm0.5$. The exception is the
FitzHugh--Nagumo fast equation, whose model coefficients show mild right
skew and excess kurtosis near $1$, and whose $v$ coefficient is biased
upward by $2\%$ (about three and a half Monte Carlo standard errors), consistent with
its modest depth-one first stage ($F\approx16$); its Wald coverage is
nevertheless nominal.

\subsection{Recommendation}
For single-trajectory inference we recommend Wald intervals from the
plug-in covariance above. They follow directly from the asymptotic
normality of Section~\ref{si:clt}, cost a single pass over the per-pair
moments, and are calibrated on linear, bistable, slow--fast, and chaotic
dynamics; the bootstrap adds computation without improving coverage.
The condition is instrument strength, not record length. On
synthetic records matched to the AMO application (Section~\ref{sec:amo};
$M=1000$), Wald coverage stays between $93\%$ and $95\%$ at spans
leaving as few as $41$ window pairs (median first-stage $F\approx3$), and
falls to $90\%$ only at $F\approx1$, where the instrument identifies
essentially nothing; the bootstrap over-covers there ($97\%$). The
strength rule $F\ge10$ of the main text therefore also secures the Wald
intervals.

\section{Comparison with likelihood baselines}
\label{si:baselines}

\begin{table}[!tb]
\centering
\caption{Likelihood baselines for OU (true $\kappa=1.5$) and the double-well (true
$\alpha=\beta=1$), on the same $M=1000$ trajectories as
Table~\ref{tab:drift}, with entries corresponding to the estimator average\,(std). For OU,
the Kalman filter gives the exact likelihood: MLE-NM and MLE L-BFGS-B are
its maximizers over $(\kappa,\mu,\sigma)$ under two optimizers, and CRB is
the Cram\'er--Rao bound of that likelihood, noted as a standard deviation.
For the double-well, there is no exact likelihood available, and the comparison is done using least-squares optimization on the strong form of the equation. The last column is the median
wall-clock time per trajectory on one core, covering drift, slope diffusion, $\widehat R$, and first-stage $F$ for \textsc{WAIVES}, the MLE-NM and L-BFGS-B fits of $(\kappa,\mu,\sigma)$ for the Kalman MLE, and the least-squares solve for the strong
form. While \textsc{WAIVES} is roughly ten times faster, it does pay a price of larger standard errors (up to a factor of three) compared to its exact-likelihood counterparts, when they exist. Where no exact likelihood exists, as is the case with the double-well example, the comparably fast alternative is the strong-form least squares. At nearly the same cost per trajectory, the strong form is biased by an order of magnitude ($\widehat\alpha\approx11$ against $\alpha=1$) while \textsc{WAIVES} remains unbiased.}
\label{tab:baselines}
\footnotesize
\setlength{\tabcolsep}{4pt}\resizebox{\columnwidth}{!}{\begin{tabular}{llccc}
\toprule
System & Method & $\widehat\kappa$ or $\widehat\alpha$ & $\widehat\beta$ & time/traj.\\
\midrule
\multirow{3}{*}{OU}
 & WAIVES ({\footnotesize $L=20$}) & $1.497\,(0.126)$ & -- & $13$\,ms\\
 & MLE-NM & $1.504\,(0.041)$ & -- & $0.28$\,s\\
 & MLE L-BFGS-B & $1.504\,(0.041)$ & -- & $0.08$\,s\\
 & CRB (std) & $(0.041)$ & -- & --\\
\midrule
\multirow{2}{*}{DW}
 & WAIVES ({\footnotesize $L=20$}) & $1.003\,(0.149)$ & $1.004\,(0.134)$ & $31$\,ms\\
 & Strong-form & $11.10\,(0.33)$ & $10.94\,(0.24)$ & $26$\,ms\\
\bottomrule
\end{tabular}}
\end{table}

Two baselines locate WAIVES against likelihood methods where those are
feasible. For the noisy OU benchmark the exact likelihood
is available through the Kalman filter, since the sampled process is a
linear Gaussian state space model~\cite{ShumwayStoffer1982}. Maximizing it
per trajectory gives the efficient benchmark, and the Cram\'er--Rao bound
follows from the Fisher information of the same likelihood.
Table~\ref{tab:baselines} compares, over the same $M=1000$ trajectories,
WAIVES, the Kalman maximum-likelihood estimator, and the Cram\'er--Rao bound. The Kalman MLE standard deviation
is the same as the Cram\'er--Rao bound. WAIVES, on the other hand, incurs a factor of about $3.1$ in standard deviation on the rate $\kappa$ at the operative window, the efficiency price of the likelihood-free route, while remaining as unbiased as the MLE.
The computational comparison, by contrast, favors \textsc{WAIVES},
and Table~\ref{tab:baselines} reports it alongside the statistical one. Both
timings cover the same three parameters $\kappa$, $\mu$, and $\sigma$. For
\textsc{WAIVES} these come from the drift system with its IV solve, from
the multilag slope estimator of the diffusion, and the first-stage test $F$. It takes 13\,ms per trajectory (the median). One evaluation of the Kalman likelihood takes 1.6\,ms. The evaluation runs the steady-state filter, with the gain
from iterating the scalar Riccati recursion to its fixed point. All $N$
innovations then come from one call to SciPy's compiled recursive filter,
\texttt{scipy.signal.lfilter}. The maximum-likelihood fit runs Nelder--Mead (MLE-NM)
over $(\log\kappa,\mu,\log\sigma)$, starting from $\kappa=1.2$,
$\sigma=0.7$, and the sample mean of the record for $\mu$. Its tolerances are $10^{-6}$ on
the parameters and $10^{-8}$ on the negative log-likelihood. The fit needs 180
evaluations, or 0.28\,s, about 20 times the cost of \textsc{WAIVES}. A
quasi-Newton search (in particular L-BFGS-B with finite-difference
gradients) needs 52 evaluations, or 0.08\,s, still six
times the cost of \textsc{WAIVES}. It uses SciPy's default tolerances, $2.2\times10^{-9}$ relative on the
objective and $10^{-5}$ on the projected gradient. Timings are
medians over the $M=1000$ trajectories on one core of an Intel Xeon Platinum 8592+
(a Blanca node of the University of Colorado Research Computing cluster), with Python 3.8.5, NumPy 1.24.3, SciPy 1.10.1, and single-threaded BLAS.

For a nonlinear case with no exact likelihood we use the double-well
benchmark and compare WAIVES against a strong-form least-squares baseline in the style
of~\cite{Brunton2016,Boninsegna2018}, regressing Euler increments
$\Delta Y_n/\Delta t$ on the same drift term candidates evaluated at $Y_n$. Measurement noise enters that regression on both sides. The
$O(R/\Delta t)$ errors-in-variables terms of the differentiated noisy
record then corrupt the recovered coefficients by an order of magnitude at
the benchmark noise level. This is an endogeneity of the strong form,
mirroring the weak-form endogeneity of Section~\ref{sec:endogeneity}. Smoothing mitigates but does
not remove the effect, which is the documented noise sensitivity that
weak-form methods were built to avoid. WAIVES removes both endogeneities at
once (see Table~\ref{tab:baselines} second block).

\end{document}